\documentclass[twocolumn,preprintnumbers,superscriptaddress,nofootinbib,aps,prd,floatfix]{revtex4}

\usepackage{amsmath,amssymb}
\usepackage{graphicx} %loads the graphicx.sty package 
\usepackage{epstopdf} %loads the epstopdf.sty package 
\usepackage{slashed}
\usepackage{subfigure}
\usepackage{color}
\usepackage{multirow,footmisc}
\usepackage{hyperref}
\usepackage{enumitem}
\usepackage{lipsum}
\usepackage{enumitem}
\usepackage{bm}

\usepackage{footmisc,ulem}

\usepackage{array}
\newcolumntype{L}[1]{>{\raggedright\let\newline\\\arraybackslash\hspace{0pt}}m{#1}}
\newcolumntype{C}[1]{>{\centering\let\newline\\\arraybackslash\hspace{0pt}}m{#1}}
\newcolumntype{R}[1]{>{\raggedleft\let\newline\\\arraybackslash\hspace{0pt}}m{#1}}

\newcommand{\be}{\begin{eqnarray*}}
\newcommand{\ee}{\end{eqnarray*}}

\newcommand{\bee}{\begin{eqnarray}}
\newcommand{\eee}{\end{eqnarray}}
\newcommand{\beeq}{\begin{equation}}
\newcommand{\eeeq}{\end{equation}}

\renewcommand{\vec}{\bf}

\begin{document}

%%%%%%%%%%%%%%%%%%%%%%%%%%%%%%%%%%%%%%%%%%%%%%%%%%%%%%%%%%
\title{Higgs coupling measurements at the LHC}
%%%%%%%%%%%%%%%%%%%%%%%%%%%%%%%%%%%%%%%%%%%%%%%%%%%%%%%%%%
\begin{abstract}
  Due to the absence of tantalising hints for new physics during the
  LHC's Run 1, the extension of the Higgs sector by dimension six
  operators will provide the new phenomenological standard for
  searches of non-resonant extensions of the Standard Model. Using all
  dominant and subdominant Higgs production mechanisms at the LHC, we
  compute the constraints on Higgs physics-relevant dimension six
  operators in a global and correlated fit. We show in how far these
  constraints can be improved by new Higgs channels becoming
  accessible at higher energy and luminosity, both through inclusive
  cross sections as well as through highly sensitive differential
  distributions. This allows us to discuss the sensitivity to new
  effects in the Higgs sector that can be reached at the LHC if direct
  hints for physics beyond the SM remain elusive. We discuss the
  impact of these constraints on well-motivated BSM scenarios.
\end{abstract}
%%%%%%%%%%%%%%%%%%%%%%%%%%%%%%%%%%%%%%%%%%%%%%%%%%%%%%%%%%
%
%
\author{Christoph Englert} %\email{christoph.englert@glasgow.ac.uk}
\affiliation{SUPA, School of Physics and Astronomy, University of
  Glasgow,\\Glasgow G12 8QQ, UK\\[0.1cm]}
\author{Roman Kogler} %\email{}
\affiliation{Institut f\"ur Experimentalphysik, Universit\"at Hamburg,
  22761 Hamburg, Germany\\[0.1cm]}
\author{Holger Schulz} %\email{holger.schulz@durham.ac.uk}
\affiliation{Institute for Particle Physics Phenomenology, Department
  of Physics,\\Durham University, Durham DH1 3LE, UK\\[0.1cm]}
\author{Michael Spannowsky} %\email{michael.spannowsky@durham.ac.uk}
\affiliation{Institute for Particle Physics Phenomenology, Department
  of Physics,\\Durham University, Durham DH1 3LE, UK\\[0.1cm]}

\preprint{IPPP/15/66, DCPT/15/132}

\maketitle

%%%%%%%%%%%%%%%%%%%%%%%%%%%%%%%%%%%%%%%%
\section{Introduction}
\label{sec:intro}
%%%%%%%%%%%%%%%%%%%%%%%%%%%%%%%%%%%%%%%%
Since the Higgs boson's discovery in 2012~\cite{Aad:2012tfa,Chatrchyan:2012ufa}, ATLAS and
CMS have quickly established a picture of
consistency with the Standard Model (SM) expectation of the Higgs
sector~\cite{Aad:2013wqa,Chatrchyan:2013lba}. By now, a multitude of constraints have been
formulated across many dominant and subdominant Higgs production
modes~\cite{atlascms}. All these measurements, as well as the absence
of a direct hint for new physics from exotics searches, seem to
suggest that the scale of new physics is well separated from the
electroweak scale. This motivates\footnote{Note, however, that current
  Higgs measurements still allow for models with light degrees of
  freedom, see e.g. \cite{King:2014xwa}.} the extension of the Higgs
sector by dimension six operators~\cite{Buchmuller:1985jz,Hagiwara:1986vm,Giudice:2007fh,Grzadkowski:2010es,Contino:2013kra}
\begin{equation}
  \label{eq:hfit}
  {\cal{L}}_{\text{Higgs}} = {\cal{L}}_{\text{Higgs}}^{\text{SM}} +
  \sum_i { c_i \over \Lambda^2} O_i 
\end{equation}
to capture new interactions beyond the Standard Model (BSM) in 
a model-independent way - within the generic limitations of effective
field theories. Constraints on these operators from a series of Run 1 and other
measurements have been provided~\cite{Azatov:2012bz,Corbett:2012ja,Corbett:2012dm,Espinosa:2012im,Plehn:2012iz,Carmi:2012in,Peskin:2012we,Dumont:2013wma,Djouadi:2013qya,Lopez-Val:2013yba,Englert:2014uua,Ellis:2014dva,Ellis:2014jta,Falkowski:2014tna,Corbett:2015ksa,Buchalla:2015qju,Aad:2015tna,Berthier:2015gja}.

A question that arises at this stage in the LHC programme is the
ultimate extent to which we will be able to probe the presence of such
interactions. Or asked differently: what are realistic estimates of Wilson coefficient
constraints that we can expect after Run 2 or the high luminosity
phase if direct hints for new physics will remain elusive? With a
multitude of additional Higgs search channels as well as differential
measurements becoming available, the complexity of a fit of the
relevant dimension six operators becomes immense.

It is the purpose of this work to provide these estimates. Using the
{\sc{Gfitter}}~\cite{Flacher:2008zq, Baak:2011ze, Baak:2012kk, Baak:2014ora} 
and {\sc{Professor}}~\cite{Buckley:2009bj} frameworks, we construct
predictions of fully-differential cross sections, evaluated to the
correct leading order expansion in the dimension six extension
$\hbox{d}\sigma=\hbox{d}\sigma^{\text{SM}} +
\hbox{d}\sigma^{\{O_i\}}/\Lambda^2$. We derive constraints on the
Wilson coefficients in a fit of the dimension six operators relevant
for the Higgs sector, inputting a multitude of present as well as projections
of future LHC Higgs measurements.

This paper is outlined as follows. In Sec.~\ref{sec:framework} we
introduce our approach in more detail. In particular, we discuss the
involved Higgs production and decay processes and review our
interpolation methods in the dimension six operator space, as well as
introduce the key elements of our fit procedure. 
In Sec.~\ref{sec:statistics} we give an overview of the statistical setup used. 
Our results using LHC Run 1 data are compared 
to existing and related work in Sec.~\ref{sec:results8}. 
This sets the stage for the extrapolation 
to 14 TeV LHC centre-of-mass energy in Sec.~\ref{sec:analysis}, where we 
detail the assumptions made when extrapolating to higher 
luminosities. Our results are presented in Sec.~\ref{sec:results14}, where we 
give estimates of the sensitivity that can be expected at the LHC for the operators
considered in this work. 
An example on how the EFT constraints can be used in the context of a well-defined 
BSM model is given in Sec.~\ref{sec:interpretation}.
We give a discussion of our results and
conclude in Sec.~\ref{sec:conc}.

Throughout this work we will use the so-called strongly-interacting
light Higgs basis~\cite{Giudice:2007fh} adopting the ``bar notation''
(this choice is not unique and can be related to other
bases~\cite{Falkowski:2015wza}), and constrain deviations from the SM
with leading order electroweak precision.  A series of publications
have extended the dimension six framework to next-to-leading
order~\cite{Passarino:2012cb,Jenkins:2013zja,Jenkins:2013sda,Jenkins:2013wua,Alonso:2013hga,Hartmann:2015oia,Ghezzi:2015vva,Hartmann:2015aia,Englert:2014cva,Grober:2015cwa}.
The impact of these modified electroweak corrections can in
principle be large in phase space regions where SM electroweak
corrections are known to be sizable and should be treated on a
case-by-case basis. However, this is not the main objective of this
analysis and we consider higher order electroweak effects beyond the scope
of this work.

%%%%%%%%%%%%%%%%%%%%%%%%%%%%%%%%%%%
\section{Framework and Assumptions}
\label{sec:framework}
%%%%%%%%%%%%%%%%%%%%%%%%%%%%%%%%%%%

We perform a global fit within a well defined Higgs boson EFT framework assuming SM gauge and 
global symmetries and a SM field content. We focus on the phenomenology of the Higgs boson that can be cast into narrow width approximation calculations, 
\begin{equation}
\label{eq:nw}
\sigma(pp\to H \to X) = \sigma (pp\to H) {\text{BR}}(H\to X)\,.
\end{equation}
Therefore, we can divide the simulation of the underlying dimension six
phenomenology into production and decay of the Higgs boson. We discuss
our approach to these parts in the following.

We consider the set of operators known as the strongly-interacting
light Higgs Lagrangian in bar convention (for details see
Refs.~\cite{Contino:2013kra,Giudice:2007fh,Buchalla:2015wfa,Buchalla:2014eca})
\begin{widetext}
\begin{equation}
\label{eq:silh}
\begin{split} {\cal L}_{\text{SILH}} = &\frac{\bar
c_H}{2v^2}\partial^\mu \left( H^\dagger H \right) \partial_\mu \left(
H^\dagger H \right)
    + \frac{\bar c_T}{2v^2}\left (H^\dagger {\overleftrightarrow {
D^\mu}} H
    \right)  
    \left(   H^\dagger{\overleftrightarrow D}_\mu H\right) 
    - \frac{\bar c_6\lambda}{v^2}\left( H^\dagger H \right)^3  
    \\ 
    &+ \left( \frac{\bar c_{u,i} y_{u,i}}{v^2}H^\dagger H  {\bar
        u}_L^{(i)} H^c u^{(i)}_R +{\text{h.c.}}\right)  
  + \left( \frac{\bar c_{d,i} y_{d,i}}{v^2}H^\dagger H {\bar d}_L^{(i)} H d^{(i)}_R+{\text{h.c.}}\right) 
    \\
    &+\frac{i\bar c_Wg}{2m_W^2}\left( H^\dagger  \sigma^i \overleftrightarrow
      {D^\mu} H \right )( D^\nu  W_{\mu \nu})^i  
    +\frac{i\bar c_Bg'}{2m_W^2}\left( H^\dagger  \overleftrightarrow {D^\mu}
      H \right )( \partial^\nu  B_{\mu \nu})  
    \\ 
    & +\frac{i\bar c_{HW} g}{m_W^2}
    (D^\mu H)^\dagger \sigma^i(D^\nu H)W_{\mu \nu}^i
    +\frac{i\bar c_{HB}g'}{m_W^2}
    (D^\mu H)^\dagger (D^\nu H)B_{\mu \nu}
    \\
    &   +\frac{\bar c_\gamma {g'}^2}{m_W^2}H^\dagger H
    B_{\mu\nu}B^{\mu\nu} 
    +\frac{\bar c_g g_S^2}{m_W^2}H^\dagger H
    G_{\mu\nu}^a G^{a\mu\nu}\,.
\end{split} 
\end{equation}
\end{widetext}
While this basis is not complete~\cite{Buchalla:2014eca,Alonso:2013hga}, it is sufficient for the 
purposes of this paper.
In particular we assume flavour-diagonal dimension six effects and in order to directly reflect the oblique correction subset of LEP
measurements of $S,T$ we decrease the number of degrees of freedom in the fit by
identifying (see also~\cite{Ellis:2014jta,Contino:2013kra,Giudice:2007fh,Barbieri:2004qk})
\begin{equation}
  \bar c_T=0\,,\quad \bar c_W + \bar c_B =0\,.
\end{equation}
We do not include anomalous triple gauge vertices to our fit~\cite{Eboli:2010qd,Corbett:2013pja,Ellis:2014jta,Falkowski:2015jaa}.

%%%%%%%%%%%%%%%%%%%%%%%%%%%%%%%%%%%%%%%%%%%%
\begin{figure*}[!t]
\subfigure[~\label{fig:suba} Comparison of $HZ$ and $H+{\text{j}}$ production.]{\includegraphics[height=0.27\textwidth]{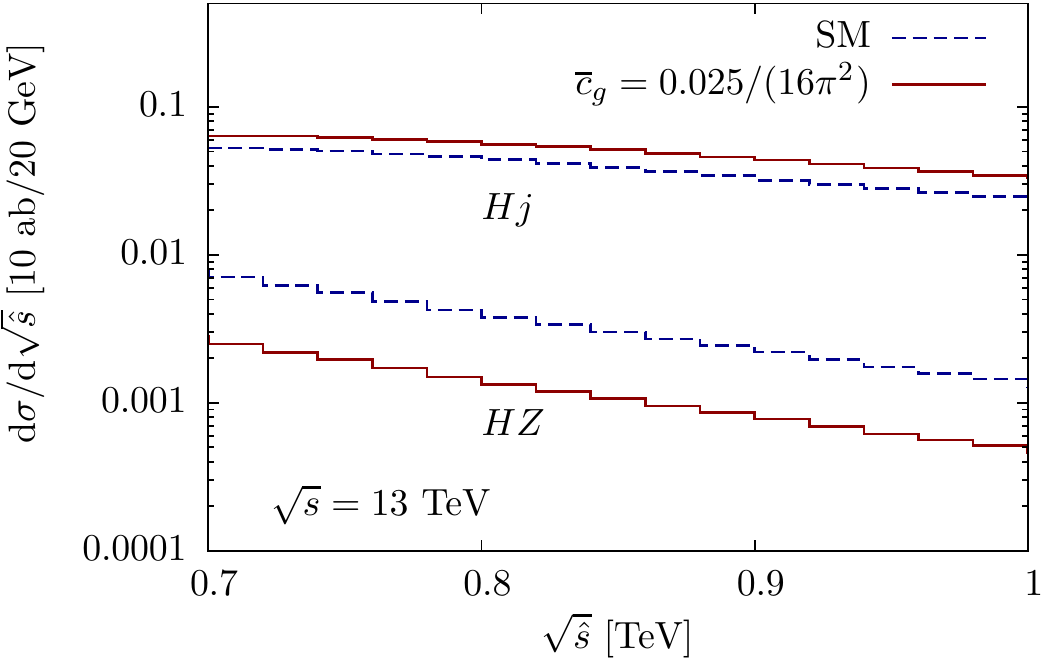}}\hspace{1cm}
\subfigure[~\label{fig:subb} Correlation of Higgs transverse momentum and partonic centre-of-mass energy (at tree-level) for a typical $2\to 2$ process (here $pp\to HZ$ in the SM).]{\includegraphics[height=0.27\textwidth]{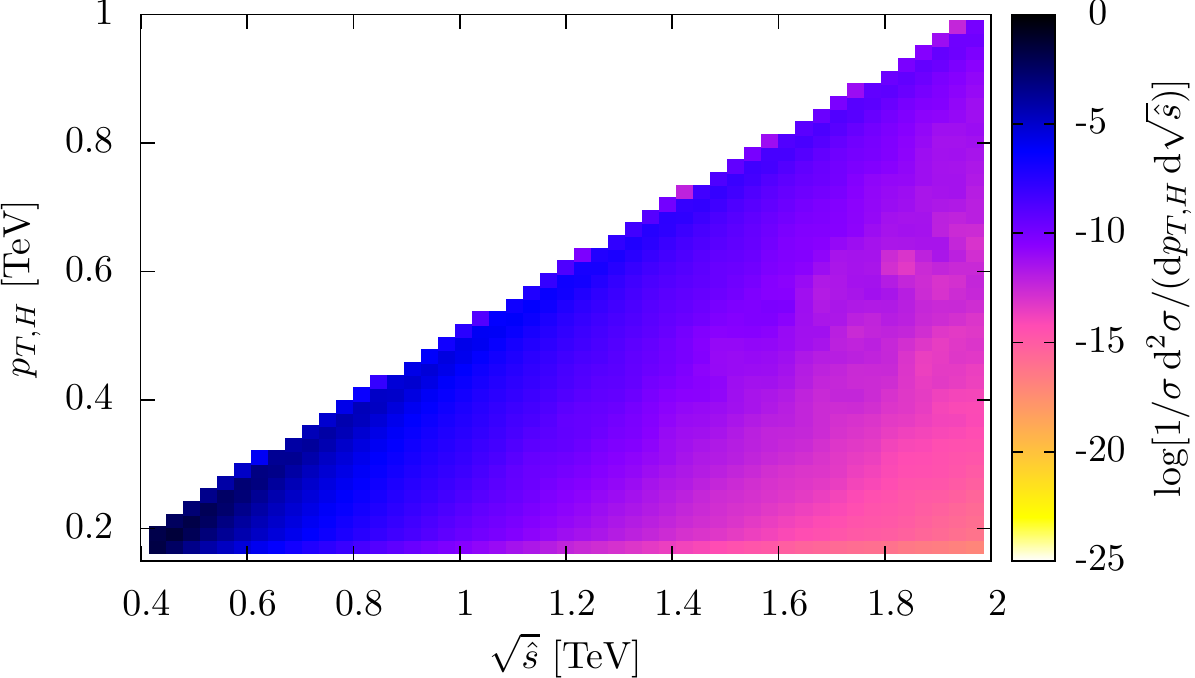}}
\vspace{-0.3cm}
\caption{Comparison of parton-level $pp\to HZ$ and $pp\to H+\text{j}$ for large partonic centre-of-mass energy $\sqrt{\hat s}$ and a particular value of $\bar c_g$. The Higgs branching ratios are rescale to have the correct SM signal strength in gluon fusion, leading to normalisation differences or enhancements at large momentum transfers. Normalisation effects from flat K factors as well as different acceptances are included to illustrate the relative importance of different production modes as signal channels. \label{fig:comp}}
\end{figure*}
%%%%%%%%%%%%%%%%%%%%%%%%%%%%%%%%%%%%%%%%%%%%

\subsection{Higgs Production and Decay}

To simulate the Higgs boson phenomenology, we employ the
  narrow width approximation
\begin{multline}
 \sigma^{\text{d=6}}(pp\to H \to X) \\ = \sigma (pp\to H,\{c_i\}) {\text{BR}}(H\to X,\{c_i\})
\end{multline}
where we linearise both parts in the Wilson coefficients. This
factorisation is motivated from being able to extract the Higgs as the
pole of the full amplitude, which is possible to all orders of
perturbation theory \cite{Passarino:2010qk}. We detail the production
and decay parts in the following.

\subsubsection*{Production}

For the production we rely on an implementation of dimension six
operators analogous to \cite{Alloul:2013naa}, which we have cross
checked and introduced in \cite{Englert:2014ffa}. The Monte-Carlo
integration of the Higgs production processes is performed with a
modified version of {\sc{Vbfnlo}} \cite{Arnold:2012xn} that interfaces {\sc{FeynArts}}, {\sc{FormCalc}}, and {\sc{LoopTools}} \cite{Hahn:1998yk,Hahn:2000kx} using a model file output by {\sc{FeynRules}} \cite{Christensen:2008py,Alloul:2013bka,Alloul:2013jea} and we only
consider ``genuine" dimension six effects that arise from the
interference of the dimension six amplitude with the SM. Writing
\begin{equation}
{\cal{M}}={\cal{M}}_{\text{SM}}+{\cal{M}}_{d=6}\,,
\end{equation} 
we obtain a squared matrix element of the form
\begin{equation}
  |{\cal{M}} |^2 = | {\cal{M}}_{\text{SM}} |^2 + 2\, \text{Re} \{
  {\cal{M}}_{\text{SM}}  {\cal{M}}_{d=6}^\ast \} + {\cal{O}}(1/\Lambda^4) \,,
\end{equation}
and we consistently neglect the dimension eight contributions that
arise from squaring the dimension six effects. Similar to higher order
electroweak or QCD calculations, the differential cross sections are
not necessarily positive definite in this expansion, but negative bin
entries provide a means to judge the validity of the Wilson
coefficient and the dimension six approach in general.

For parameter choices close to the SM, including $|{\cal{M}}_{d=6}|^2$
is typically not an issue and the parameters $c_i^2$ are often
numerically negligible for inclusive observables such as signal
strengths. However, to obtain an inclusive measurement, we marginalise
over a broad range of energies at the LHC and a positive theoretical
cross section might be misleading as momentum dependencies of some
dimension six operators violate a naive scaling $c_i E^2/\Lambda^2
<c_i^2 E^4/\Lambda^4$ in the tails of momentum-dependent
distributions, where $E$ denotes the respective and process-relevant
energy scale. For this reason, we choose to calculate cross sections
to the exact order $\sim 1/\Lambda^2$ and later reject Wilson
coefficient choices that lead to a negative differential cross section
for integrated bins of a given LHC setting when this part of the phase
space is resolved; such negative cross sections signal bigger
contributions of the $d=6$ terms than we expect in the SM, and we
cannot justify limiting our analysis to dimension six operators if new
physics becomes as important as the SM in observable phase space
regions. This provides a conservative tool to gauge the validity of
our approach, but care has to be taken by interpreting the results
when connecting to concrete physics scenarios. In strongly interacting
scenarios, it can be shown that the squared $d=6$ terms are important,
for small Wilson coefficients they are negligible. The latter avenue
should be kept in mind for our results.

\subsubsection*{Included production modes and operators}

We consider the production modes $pp\to H$, $pp \to H+{\text{j}}$, $pp
\to t\bar t H$, $p p \to WH$, $p p \to ZH$ and $p p\to H+2{\text{j}}$
(via gluon fusion and weak boson fusion) in a fully differential
fashion by including the differential Higgs transverse momentum
distributions to setting constraints. As we demonstrate, including
energy-dependent differential information whenever possible, is key to
setting most stringent constraints on the dimension six extension by
including the information of the distributions' shapes beyond the
total cross section, especially when probing blind directions in the
signal strength, as shown in Fig.~\ref{fig:suba}. Note that for the
underlying $2\to 2$ and $2\to 3 $ processes in the regions of detector
acceptance, the Higgs transverse momentum is highly correlated with
the relevant energy scales that probe the new interactions,
Fig.~\ref{fig:subb}, and therefore is a suitable observable to include
in this first step towards a fully-differential Higgs fit.  Expanding
the cross sections to linear order in the Wilson coefficients as done
in this work is not a mere technical twist, but allows us to obtain a
description of the high-$p_T$ cross sections within our
approximations.

The operator $(H^\dagger H)^3$ and off-shell Higgs production in the
EFT
framework~\cite{Azatov:2014jga,Cacciapaglia:2014rla,Englert:2014ffa}
deserve additional comments. Dihiggs production is the only process
which provides direct sensitivity to $\bar c_6$~\cite{Goertz:2014qta}
and factorises from the global fit, at least at leading order. Hence,
the $\bar c_6$ can be separated from the other directions to good
approximation. While Higgs pair production process can serve to lift
$y_t$-degeneracies in the dimension six
extension~\cite{Azatov:2015oxa,Dolan:2015zja}, the sensitivity to
$\bar c_6$ is typically small when we marginalise over $\bar
c_{u3}$. The latter can be constrained either in $pp \to \bar t t H$,
$pp\to ZZ$ in the Higgs off-shell
regime~\cite{Azatov:2014jga,Cacciapaglia:2014rla,Englert:2014ffa,Langenegger:2015lra,Dawson:2014ora}
or $pp\to H+{\text{j}}$
\cite{Harlander:2013oja,Banfi:2013yoa,Grojean:2013nya,Buschmann:2014twa},
however only the former of these processes provides direct sensitivity
to $\bar c_{u3}$ without significant limitations due to
marginalisation over the other operator directions. Current recast
analyses place individual constraints in the range of $|\bar
c_{u3}|\lesssim 5$ and $|\bar c_g|\lesssim 10^{-3}$
\cite{Grojean:2013nya} for the 8 TeV data set.

While the expected sensitivity to $pp\to HH(+{\text{jets}})$ still
remains experimentally vague at this stage in the LHC
programme~\cite{ATL-PHYS-PUB-2014-019,Barr:2013tda}, the potential to
observe $pp\to \bar t t H$ is consensus. We therefore do not include
$pp\to HH$ to our projections and also omit off-shell Higgs boson
production, since experimental efficiencies during the LHC high
luminosity phase will significantly impact the sensitivity in these
channels. We leave a more dedicated discussion of these channels to
future work~\cite{toapp}.

Due to the small Yukawa couplings of first and second generation
quarks and leptons, we limit ourselves to modified top-Higgs and
bottom-Higgs couplings throughout and neglect modifications of the
lepton-Higgs system too.  An overview of the tree-level sensitivity of
the production channels considered in this work is given in
Tab.~\ref{tab:sens}.

\subsubsection*{Branching ratios}

\begin{figure}[!t]
\centering
\includegraphics[width=0.4\textwidth]{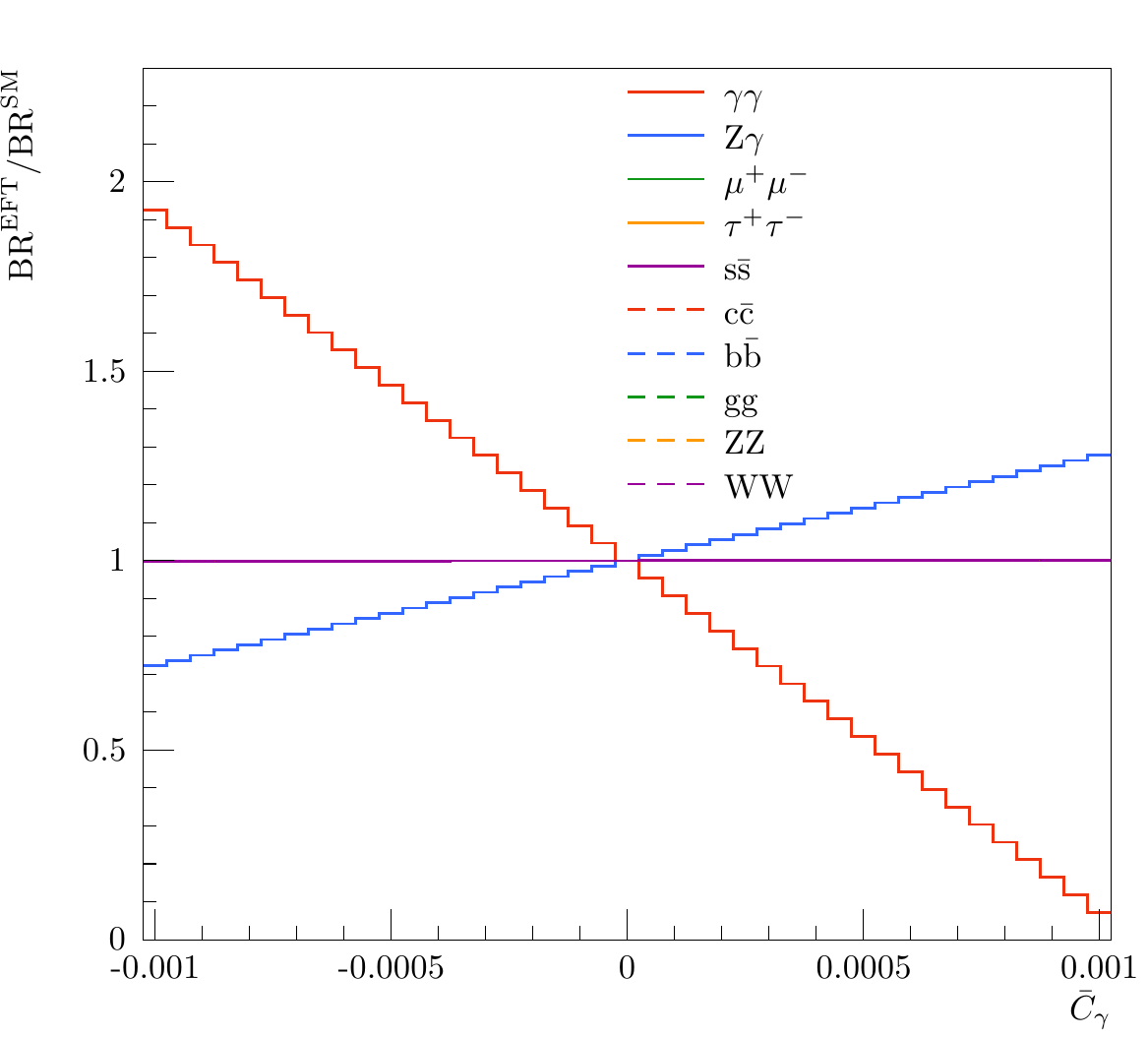}
\caption{Variation of Higgs branching ratios as function of $\bar{c}_\gamma$
normalized to the SM values. 
\label{fig:br_cgam}}
\end{figure}

For the branching ratios, we rely on {\sc{eHdecay}} to
  include the correct Higgs branching ratios in the dimension six
  extended Standard Model, which is detailed in~\cite{Contino:2014aaa}
  and implements a linearisation in the Wilson coefficients. 
  The branching ratios are therefore sensitive to all Wilson coefficients
  affecting single Higgs physics. 
  An example for the variation of the branching ratios as function of
  $\bar{c}_\gamma$ is shown in Fig.~\ref{fig:br_cgam}.

  We sample a broad range of dimension six parameter choices and
  interpolate them using the {\sc{Professor}} method detailed in the
  appendix~\ref{sec:prof}. This also allows us to identify already at
  this stage a reasonable Wilson coefficient range with a
  positive-definite Higgs decay phenomenology that limits the validity
  (i.e. the positive definiteness) of our narrow width
  approximation. We find an excellent interpolation of the
  {\sc{eHdecay}} output (independent of the interpolated sample's size
  and choice) and we typically obtain per mille-level accuracy of the
  Higgs partial decay widths and branching ratios, which is precise
  enough for the limits we can set. Interpolation using
  {\sc{Professor}} is key to performing the fit in the high
  dimensional space of operators and observables in a very fast and
  accurate way.

%%%%%%%%%%%%%%%%%%%%%%%%%%%%%%%%%%%%%%%%%%%%
%%%%%%%%%%%%%%%%%%%%%%%%%%%%%%%%%%%%%%%%%%%%

 %%%%%%%%%%%%%%%%%%%%%%%%%%%%%%%%%%%%%%%%%%%%%%%%%%%%%%%
\begin{table}[bt]
\begin{tabular}{|l|c|}
\hline
production process & included sensitivity \\
\hline
\hline
$pp\to H$ &  \multirow{4}{*}{$\bar c_g, \bar c_{u3}, \bar c_H$}\\
$pp \to H+\text{j}$ & \\
$pp \to H+2\text{j}$ (gluon fusion)& \\
$pp \to t\bar tH$ & \\
\hline
$pp \to VH$ & \multirow{2}{*}{$\bar c_{W}, \bar c_B, \bar c_{HW}, \bar c_{HB}, \bar c_{\gamma}, \bar c_H$ }\\
$pp \to H+2\text{j}$ (weak boson fusion) & \\
\hline
\end{tabular}
\caption{Tree-level sensitivity of the various production mechanisms. \label{tab:sens}}
\end{table}
%%%%%%%%%%%%%%%%%%%%%%%%%%%%%%%%%%%%%%%%%%%%%%%%%%%%%%%

%%%%%%%%%%%%%%%%%%%%%%%%%%%%%%%%%%%%%%
\section{Statistical Analysis}
\label{sec:statistics}
%%%%%%%%%%%%%%%%%%%%%%%%%%%%%%%%%%%%%%

The {\sc{Gfitter}} software is used for calculating the likelihood and
the estimation of the sensitivity of data.
The data include correlated experimental systematic uncertainties, 
implemented in the form of a covariance matrix. 
These uncertainties either come from measurements at the LHC 
or from pseudo-data, as explained in detail below. 
The theoretical uncertainties are taken into account as nuisance parameters. 
The negative log-likelihood $-2 \ln \mathcal{L}= \chi^2$ is constructed as
\begin{equation}
\chi^2 = (\bm{x} - \bm{t}(c_i, \delta_k))^T \bm{V}^{-1} (\bm{x} - \bm{t}(c_i, \delta_k)) 
+ \sum_k \delta_k^2
\label{eq:chi2}
\end{equation}
where $\bm{x}$ denotes the vector of measurements, $\bm{t}(\bar{c}_i, \delta_k)$ the 
theory predictions depending on the Wilson coefficients $c_i$ 
and theoretical uncertainties implemented as nuisance parameters 
$\delta_k$ and $\bm{V}$ is the covariance matrix. The covariance matrix
is given by 
\begin{equation}
\bm{V} = \bm{V}_{\text{stat}} + \bm{V}_{\text{syst}},
\end{equation}
consisting of an uncorrelated statistical part $\bm{V}_{\text{stat}}$
and a correlated experimental systematic uncertainty $\bm{V}_{\text{syst}}$. 
The sensitivity on the coefficients $\bar{c}_i$ 
is calculated by scanning the likelihood as function of a given
set of Wilson coefficients. For each point of this scan 
the likelihood is profiled over the other dimension six coefficients 
and the nuisance parameters $\delta_k$ corresponding to the theoretical uncertainties. 
The profiling is performed by a multi-dimensional fit.
Each of these fits include up to 32 free parameters.
Out of those, 24 are due to nuisance parameters of the 
theoretical uncertainties and the other 8 parameters are 
the dimension six coefficients themselves.

%%%%%%%%%%%%%%%%%%%%%%%%%%%%%%%%%%%%%%
\section{Results for Run 1}
\label{sec:results8}
%%%%%%%%%%%%%%%%%%%%%%%%%%%%%%%%%%%%%%
In the following we will evaluate the status of the effective
Lagrangian~Eq.~\eqref{eq:silh} in light of available Run 1 analyses. Similar
analyses have been performed by a number of groups~, see
e.g.~\cite{Ellis:2014jta,Englert:2014uua,Corbett:2015ksa}. Comparing
the above fit-procedure to these results not only allows us to
validate the highly non-trivial fitting procedure against other
approaches, but also to extend these results by including additional
measurements which have become available in the meantime.
We include Run 1 experimental analyses using 
{\sc{HiggsSignals}}~v1.4~\cite{Bechtle:2013xfa, Bechtle:2014ewa}, based on 
{\sc{HiggsBounds}}~v4.2.1~\cite{Bechtle:2008jh, Bechtle:2011sb, Bechtle:2013wla, Bechtle:2015pma},
which calculates $\chi^2$ given in Eq.~\eqref{eq:chi2} taking into account experimental 
and theoretical correlations, as well as signal acceptances.

Specifically, we include the following analyses. Higgs decays to bosons 
have been measured in the channels
$H\to \gamma\gamma$~\cite{Aad:2014eha, Khachatryan:2014ira}, 
$H\to ZZ^{(*)} \to 4l$~\cite{Aad:2014eva, Chatrchyan:2013mxa} and
$H\to WW^{(*)} \to 2l 2\nu$~\cite{ATLAS:2014aga, Aad:2015ona, Chatrchyan:2013iaa, CMS-PAS-HIG-13-017}. 
These analyses have sensitivity to the gluon-fusion, $H+2{\text{j}}$ and $VH$ production modes. 
The coupling to leptons has been probed in the 
$H\to \tau^+\tau^-$ channel~\cite{Aad:2015vsa, Chatrchyan:2014nva}, 
with some evidence for 
$H\to b\bar{b}$ in $VH$ production~\cite{Aad:2014xzb, Chatrchyan:2013zna} 
and a search for $H\to \mu^+ \mu^-$~\cite{Khachatryan:2014aep}.
The coupling to top quarks has been addressed through $t\bar{t}H$
production in the $H\to b\bar{b}$ decay~\cite{Aad:2015gra, Khachatryan:2014qaa}
and in leptonic decays, sensitive to the
$H\to ZZ^{(*)}$, $H\to WW^{(*)}$ and $H\to \tau^+\tau^-$ 
channels~\cite{Aad:2015iha, Khachatryan:2014qaa}.
This results in a total of 78 measurements included in the fit.
All measurements used are listed in appendix~\ref{sec:run1meas}, 
together with the values of $\mu$, the uncertainties and details on the signal acceptances.
Correlations between the measurements are introduced due to the acceptance 
of a given experimental measurement to
a number of production and decay modes and the overall luminosity 
measurement.  
Also, the theoretical uncertainties from the normalisation  
of the signal strength measurements to the SM prediction, as 
included in the experimental results, are taken to be fully 
correlated among the experimental measurements~\cite{Bechtle:2013xfa, Bechtle:2014ewa}.
Correlations due to theory uncertainties in the calculations with dimension six effects
are included as well.

%%%%%%%%%%%%%%%%%%%%%%%%%%%%%%%%%%%%%%
\begin{figure*}[p]
 \begin{center}
  \includegraphics[width=0.44\textwidth]{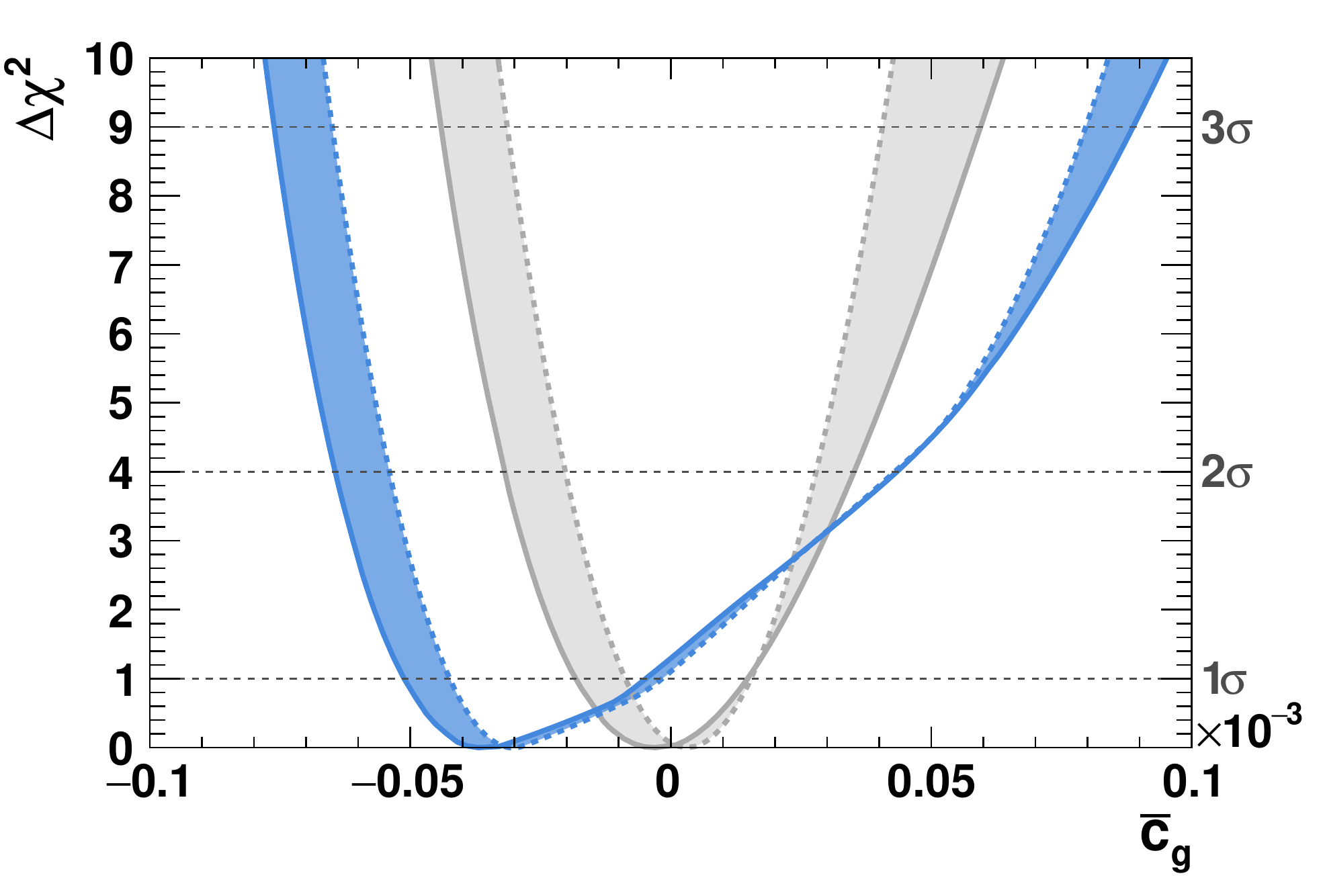}
  \hspace{0.7cm}
  \includegraphics[width=0.44\textwidth ]{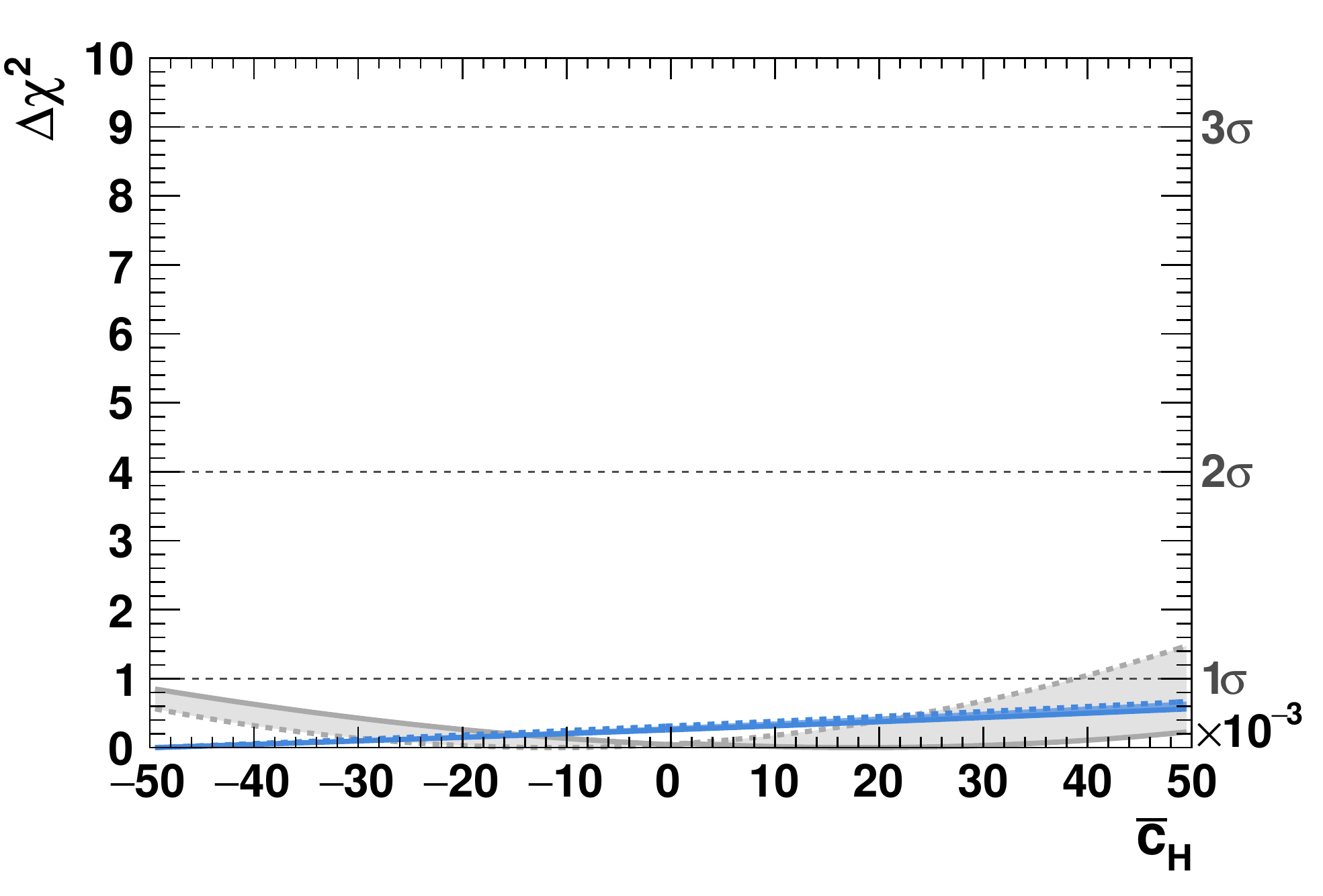}\\[0.2cm]
  \includegraphics[width=0.44\textwidth ]{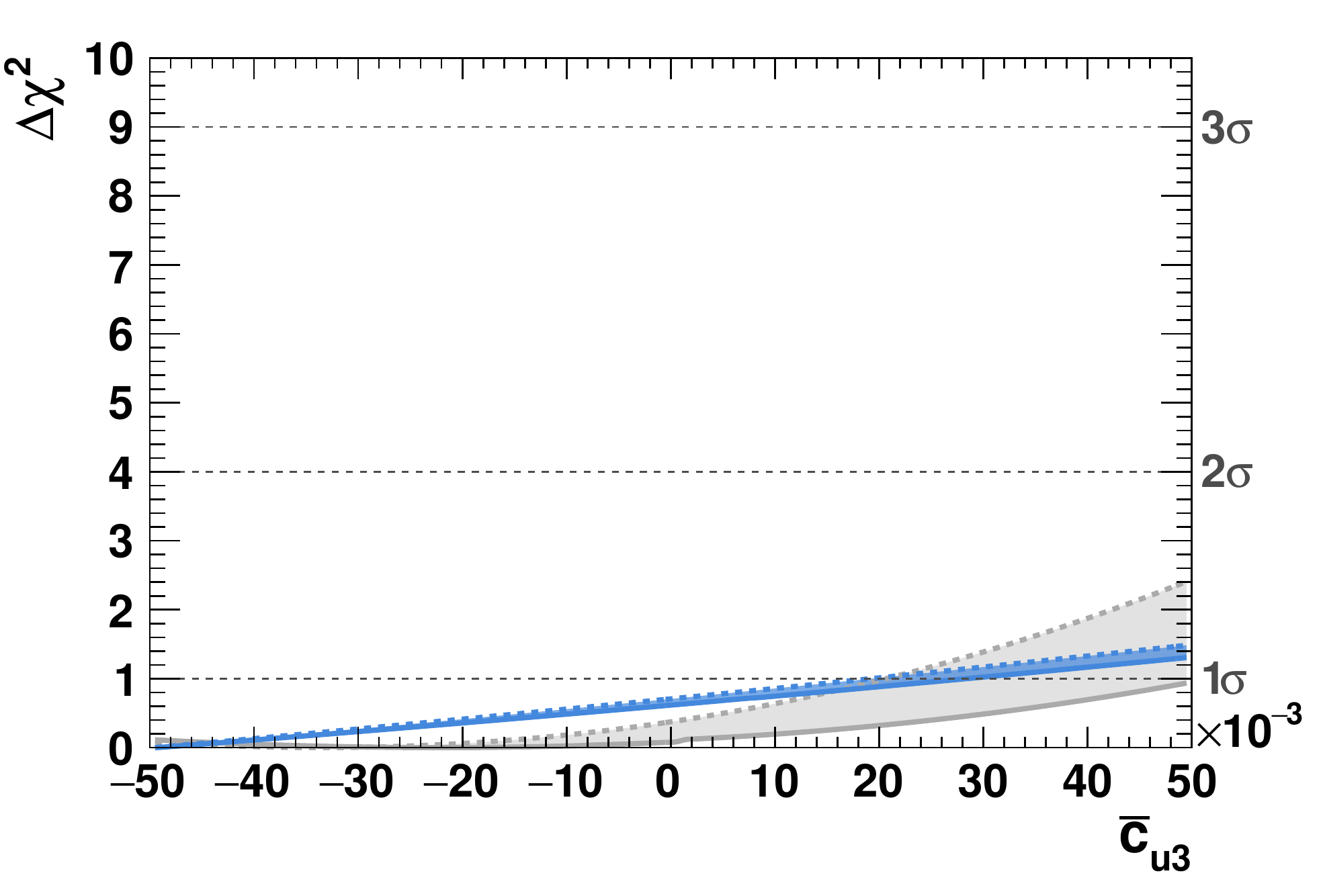}
  \hspace{0.7cm}
  \includegraphics[width=0.44\textwidth]{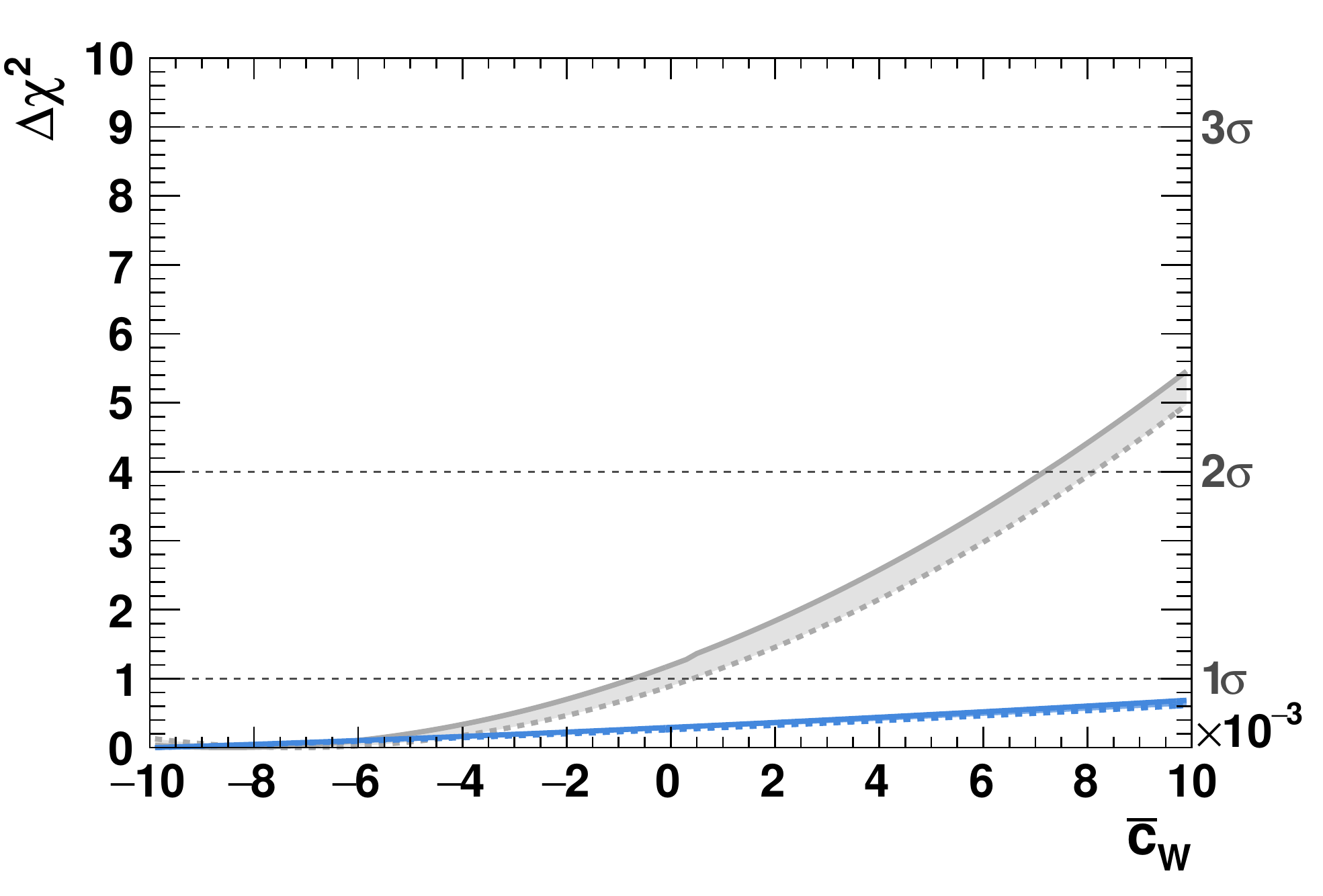} \\[0.2cm]
  \includegraphics[width=0.44\textwidth]{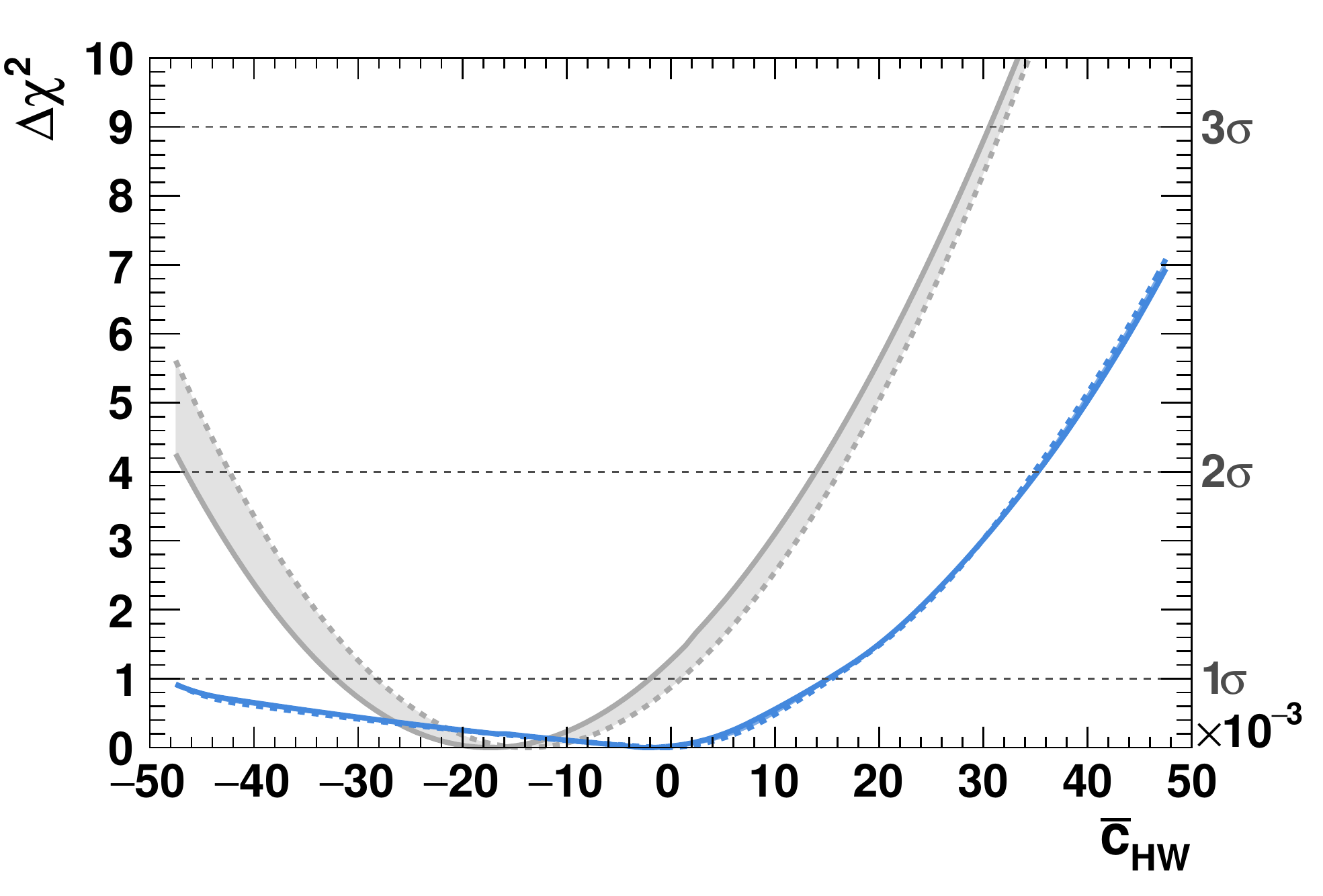}
  \hspace{0.7cm}
  \includegraphics[width=0.44\textwidth]{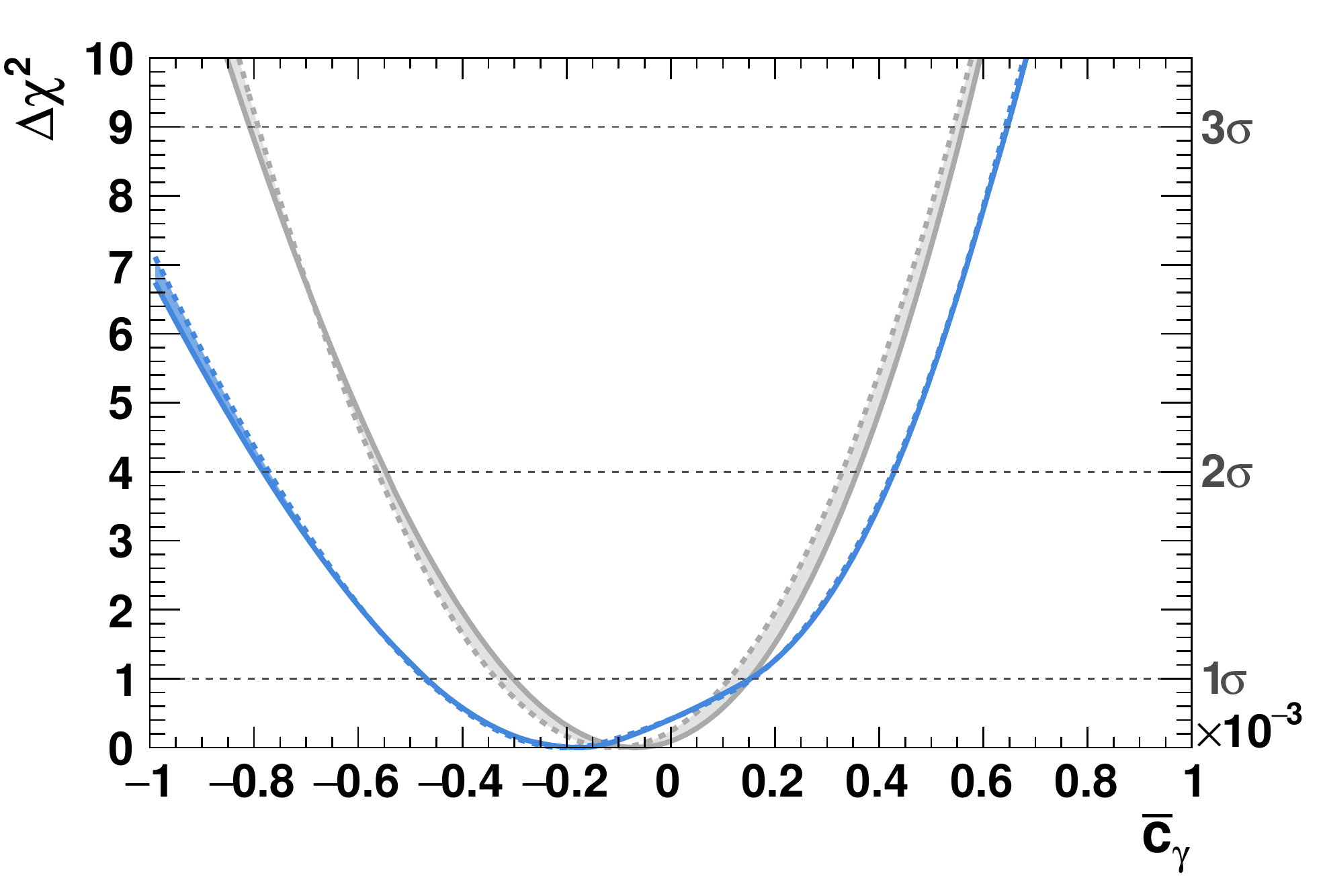}\\[0.2cm]
  \includegraphics[width=0.44\textwidth]{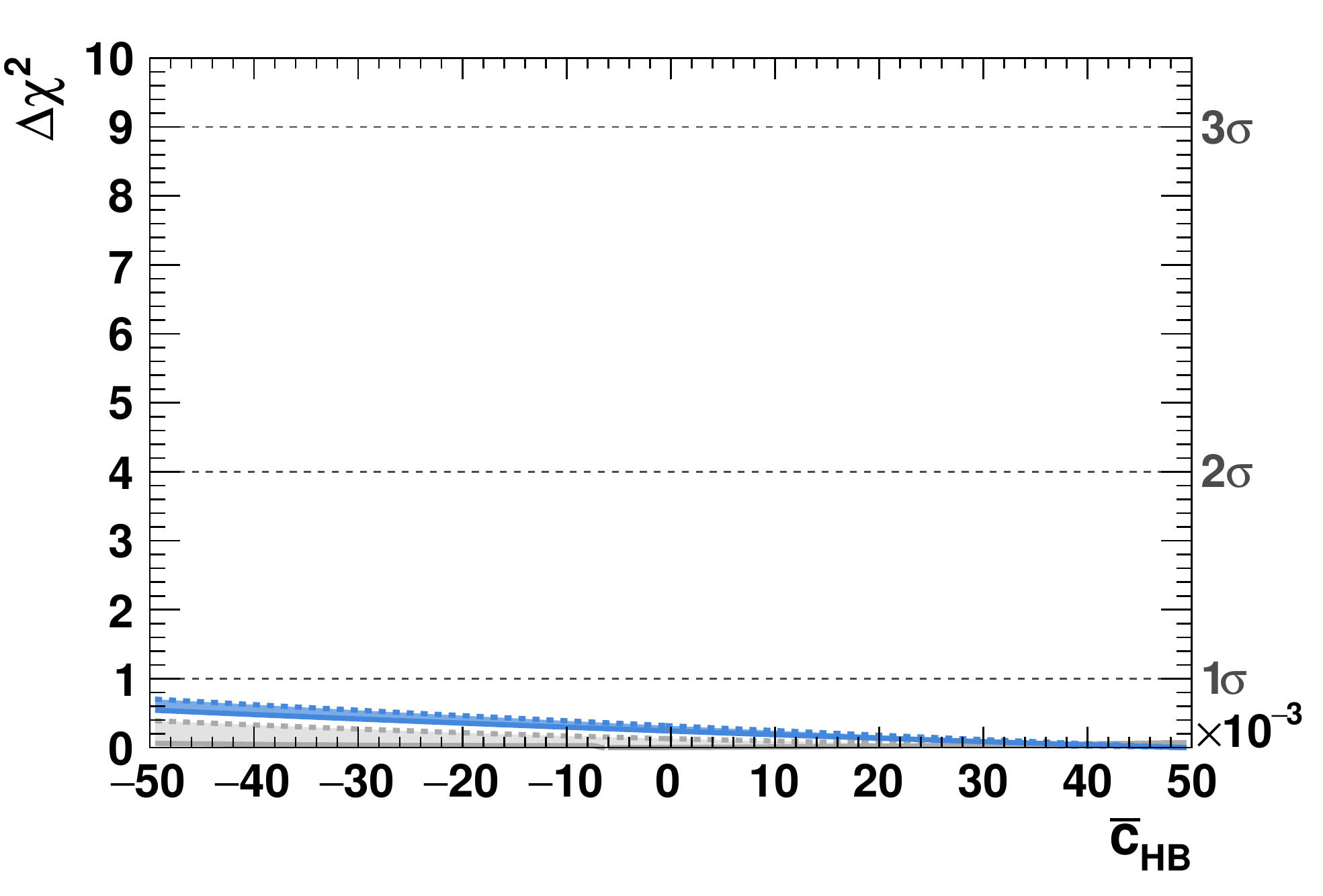}
  \hspace{0.7cm}
  \includegraphics[width=0.44\textwidth]{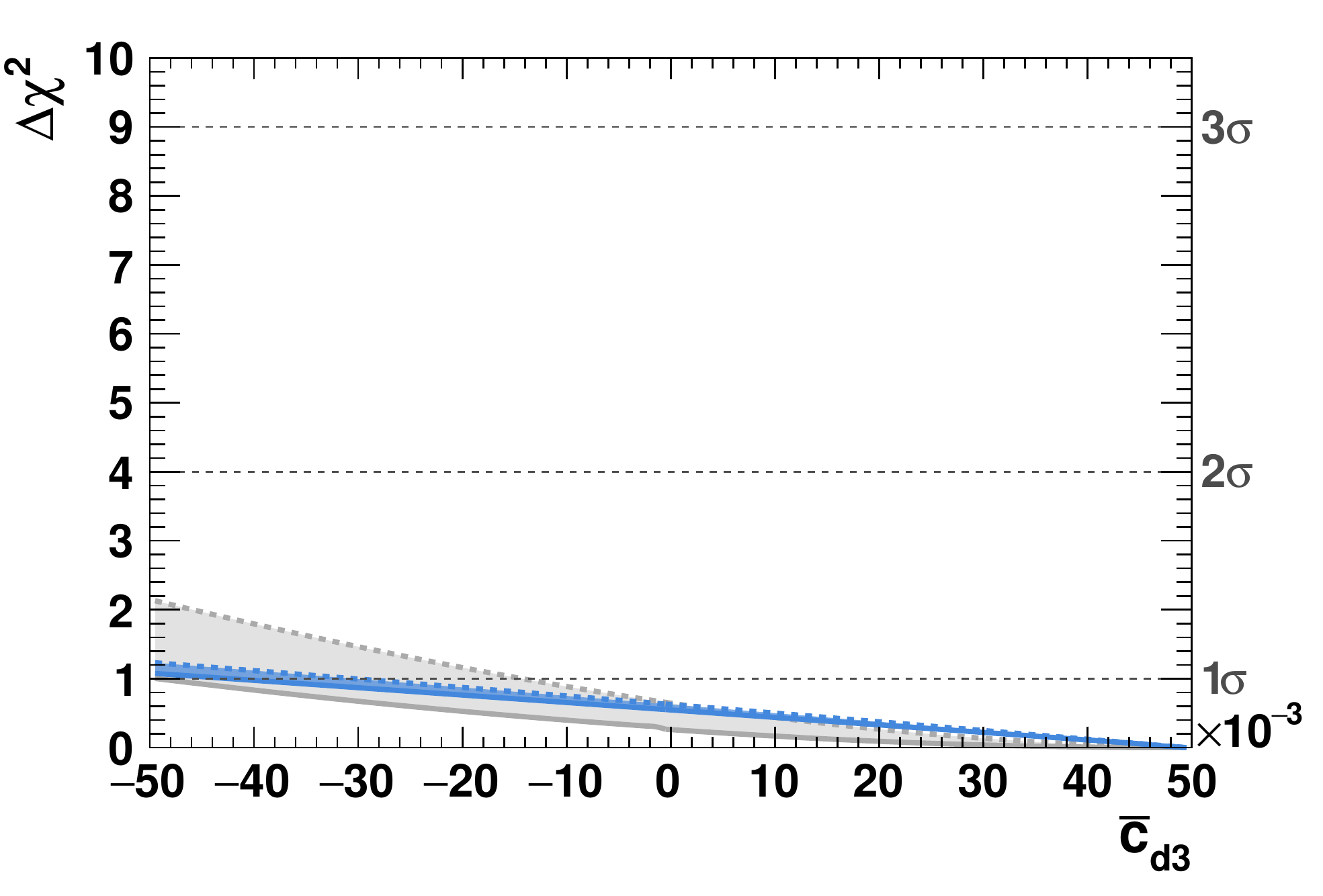}  
 \end{center}
 \caption{Confronting the Lagrangian Eq.~\eqref{eq:silh} with the 8
   TeV LHC Run 1 measurements. Solid lines correspond to a fit with theoretical uncertainties included,
   dashed lines show results without theoretical uncertainties, the band shows the impact of these. 
   Grey lines and bands denote the individual constraints on a given parameter, and blue refers to the marginalised results.  
   For details see the main text.
   \label{fig:run1}}
\end{figure*}
%%%%%%%%%%%%%%%%%%%%%%%%%%%%%%%%%%%%%%
%
The results are shown in Fig.~\ref{fig:run1} and are in
good agreement with the results obtained in Refs.~\cite{Ellis:2014jta,Aad:2015tna}. 
Numerical values are given below in Tab.~\ref{tab:8tevresults}.
Small differences can be understood from working under different assumptions 
(specifically the strict linearisation of dimension six effects) 
as well as including more analyses. It should be noted that our 
choice of limiting the range of Wilson coefficient values 
(necessary for the positive definiteness of differential distributions)
is necessitated by our extrapolation and inclusion of differential distributions. 
Consequently, we cannot set a limit on many operators in the light
of Run 1 measurements within our approximations. 
However, the direct comparison to the Figs.~\ref{excl:int} and \ref{excl:diff} 
will allow us to see how these can be improved when going to higher 
centre-of-mass energy and luminosity.
Relaxing these constraints will lead to increased Wilson coefficient intervals 
for the marginalised scans over the 8 TeV signal strength measurements 
(for a recent fit without limited coefficient ranges see 
Ref.~\cite{Falkowski:2015jaa}).

The fit converges with a minimum value of $\chi^2$ of 87.7 for 70 degrees of freedom $(n_{\text{dof}})$,
corresponding to a $p$-value of about 0.07. Without theory uncertainties the value of 
$\chi^2$ increases to 96.5. 
The goodness-of-fit is slightly worse than the result of a $\chi^2$ test of the SM hypothesis, 
which gives a minimum value of $\chi^2 / n_{\text{dof}} = 91.3 / 78 = 1.17$, or a $p$-value of 0.14.
The smaller $p$-value for the dimension-six fit with respect to the SM result 
can be understood because of the addition of free parameters not needed to describe the data,  
in other words, some dimension-six coefficients are not constrained by the current data. 
Two coefficients, $\bar{c}_{g}$ and $\bar{c}_{\gamma}$, can be reliably constrained 
at 95\% confidence level (CL) within the range of Wilson coefficient values considered. 
We find the allowed 95\% CL ranges 
\begin{eqnarray}
\bar{c}_{g} \in [\, -0.64, \, 0.43 \, ] \times 10^{-4} \nonumber \\ 
\bar{c}_{\gamma} \in [\, -7.8, \, 4.3 \, ] \times 10^{-4} \, .
\end{eqnarray}
These constraints are somewhat tighter than the ones obtained by the 
ATLAS collaboration, 
$\bar{c}_{g} \in [\, -0.7, \, 1.3 \, ] \times 10^{-4}$ and 
$\bar{c}_{\gamma} \in [\, -7.4, \, 5.7 \, ] \times 10^{-4}$~\cite{Aad:2015tna}, 
because the ATLAS values are derived using only the ATLAS $H\to\gamma\gamma$
measurement. 

Let us compare these limits to the SM to get
an estimate of how big these constraints are if we move away from the
bar convention. The limits on, e.g., $\bar{c}_g\lesssim 0.4\times
10^{-4}$ can be compared for instance against the effective $ggH$
operator that arises from integrating out the top quark in the limit
$m_t\to \infty$. The effective operator for this limit, using low
energy effective theorems
\cite{Ellis:1975ap,Shifman:1979eb,Kniehl:1995tn} reads
\begin{multline}
\frac{\alpha_s}{12\pi} G_{\mu\nu}^a G^{a\mu\nu} \log ( 1+ H/v) \\  \simeq
\frac{ \alpha_s}{12\pi v} G_{\mu\nu}^a G^{a\mu\nu}  H + \dots\,
\end{multline}
Matching this operator onto SILH convention of Eq.~\eqref{eq:silh}, we
obtain $|\bar c_g({\text{effective SM}})|\simeq 0.2\times 10^{-3}$. 
So in this sense, new physics is constrained to a ${\cal{O}}(10\%)$ deviation relative
to the SM from inclusive observables. The relative deviations in the
tails of the Higgs transverse momentum distributions that are induced by this operator 
can easily be as big as factors of two (see
e.g. \cite{Cacciapaglia:2014rla,Azatov:2014jga,Englert:2014ffa}),
which highlights the necessity to resolve this deviation with energy
or momentum dependent observables during Run 2 and the high luminosity
phase to best constrain the presence of non-resonant physics using
high momentum transfers.

%%%%%%%%%%%%%%%%%%%%%%%%%%%%%%%%%%%%%%%
\section{Projections for 14 TeV and the High Luminosity Phase}
\label{sec:analysis}
%%%%%%%%%%%%%%%%%%%%%%%%%%%%%%%%%%%%%%
Throughout our analysis we normalise our results to the recommendation
of the Higgs cross section working group~\cite{Dittmaier:2011ti,Dittmaier:2012vm, Heinemeyer:2013tqa}. 
Predicted rates are using the narrow width approximation of Eq.~\eqref{eq:nw}. 
We construct pseudo-measurements to asses the sensitivity of the LHC with a centre-of-mass energy of 14~TeV 
to the set of operators considered in this work. 
The theoretically predicted number of events for a 
specific final state $N_{\text{th}}$ is obtained by multiplying by additional branching ratios if
necessary and the luminosity $L$ of the particular analysis:
\begin{multline}
  N_{\text{th}} = \sigma(H+X) \times \text{BR}(H\to YY) \\ \times L \times \text{BR}(X,Y\to \text{{final~state}}) \, .
\end{multline}
This number is then multiplied by the
efficiency to measure the production channel $\epsilon_p$ and the
efficiency to measure the decay products $\epsilon_d$, to obtain the measured number of events
\begin{equation}
N_{\text{ev}} = \epsilon_p \epsilon_d N_{\text{th}}.
\end{equation}
The relative statistical uncertainty for a given pseudo-measurement is 
estimated to be $\sqrt{N_{\text{ev}}}$.
For the efficiency to reconstruct a specific final state, we rely on experimental results from Run 1, where available. 
The efficiencies used are $\epsilon_{p,\mathrm{t\bar{t}h}} = 0.10$~\cite{Aad:2015gra, Khachatryan:2014qaa,Moretti:2015vaa,Buckley:2015vsa}, 
$\epsilon_{p,\text{ZH}} = 0.12$, 
$\epsilon_{p,\text{WH}} = 0.04$, 
$\epsilon_{p,\text{VBF}} =  0.30$~\cite{Chatrchyan:2013lba,Khachatryan:2014ira,Aad:2014xzb, Aad:2014eha}.
We assume a value of $\epsilon_{p,\text{H+j}} = 0.5$~\cite{Schlaffer:2014osa} (see also \cite{Buschmann:2014sia,Dawson:2014ora,Schmidt:2015cea}) where no experimental results targeting this production 
mode are available so far.
In order to simplify the assumptions and the background estimates, we consider only 
leptonic channels for the $VH$ and $t\bar tH$ production modes. Here only final states with
electrons and muons are used. These are however allowed to originate from $\tau$-decays.
In case of the gluon fusion production mode, analyses targeting different final states have 
different reconstruction efficiencies. We use the following efficiencies for the process $pp\to H$:
$\epsilon_{p,\text{GF}} = 0.4$ for $H\to \gamma \gamma$~\cite{Aad:2014eha, Khachatryan:2014ira}, 
$\epsilon_{p,\text{GF}} = 0.01$ for $H\to \tau^+ \tau^-$~\cite{Aad:2015vsa, Chatrchyan:2014nva}, 
$\epsilon_{p,\text{GF}} = 0.25$ for $H\to 4l$~\cite{Aad:2014eva, Chatrchyan:2013lba}, 
$\epsilon_{p,\text{GF}} = 0.10$ for $H\to 2l 2\nu$~\cite{ATLAS:2014aga, Chatrchyan:2013iaa}, 
$\epsilon_{p,\text{GF}} = 0.10$ for $H\to Z \gamma$~\cite{Aad:2014fia, Chatrchyan:2013vaa}, and 
$\epsilon_{p,\text{GF}} = 0.50$ for $H\to \mu \mu$~\cite{Aad:2014xva, Khachatryan:2014aep}. 
The $H\to b\bar{b}$ decay is not considered for the gluon fusion production mode. 
Taking a conservative approach we assume the same reconstruction efficiencies for measurements at $14$ TeV, 
independent of the Higgs transverse momentum.

In the reconstruction of the Higgs boson we include reconstruction and identification efficiencies of the final state objects:
\begin{description}[labelsep=0.5em,itemsep=0.01cm]
\item[$H\to \mathrm{b\bar{b}}$:] We assume a flat $b$-tagging efficiency of $60\%$, i.e. $\epsilon_{d,b\bar{b}}=0.36$. 
\item[$H\to \gamma \gamma$:] For the identification and reconstruction of isolated photons we assume respectively an efficiency of $85\%$. 
Hence, we find $\epsilon_{d,\gamma\gamma} \simeq 0.72$.
\item[$H\to \tau^+ \tau^-$:] We consider $\tau$-decays into hadrons ($\text{BR}_{\text{had}}=0.648$) or leptons, i.e. an electron ($\text{BR}_{e}=0.178$) or muon ($\text{BR}_{\mu}=0.174$). For the reconstruction efficiency of the hadronic $\tau$ we assume a value of $50\%$ 
and for the electron and muon we use $95\%$. 
Thus, the total reconstruction efficiency is $\epsilon_{d,\tau\tau} \simeq 0.433$.
\item[$H\to ZZ^* \to 4l$:] We consider $Z$ decays into electrons and muons only, also taking into account $\tau$ decays into lighter leptons.
For each lepton we assume a reconstruction efficiency of $95\%$, which gives a total 
reconstruction efficiency of $\epsilon_{d,4l} \simeq 0.815$.
\item[$H\to WW^* \to 2l 2\nu$:] Only lepton decays into electrons and muons are considered and for each visible lepton we include a $95\%$ reconstruction efficiency, 
i.e.\ $\epsilon_{d,2l 2\nu} = 0.9025$
\item[$H\to Z \gamma$:] Again, we include separately an $85\%$ identification and reconstruction efficiency for isolated photons 
and a $95\%$ reconstruction efficiency for each electron and muon. As a result we find $\epsilon_{d,Z\gamma} \simeq 0.767$.
\item[$H\to \mu^+ \mu^-$:] Each muon is assumed to have a reconstruction efficiency of $95\%$, resulting in $\epsilon_{d,\mu\mu}=0.9025$.
\end{description}

Owing to the different selections made in the various experimental analyses, 
each channel has a unique background composition, resulting in different
additional systematic uncertainties on the measurements. 
We approximate those by adding uncorrelated systematic uncertainties for each  
production and decay channel in quadrature. 
The uncertainties used are given in Tab.~\ref{tab:backgroundunc}
and are assumed to be flat in $p_{T,H}$. 
The uncertainties are taken from experimental Run 1
analyses~\cite{Aad:2013wqa, Aad:2014eha, Chatrchyan:2013mxa, 
Aad:2015vsa, Aad:2014eva, ATLAS:2014aga, Aad:2014fia, Aad:2014xva, 
Aad:2015gra, Aad:2015ona, Khachatryan:2014ira, Chatrchyan:2014nva, 
Chatrchyan:2013lba, Chatrchyan:2013iaa, 
Chatrchyan:2013vaa, Khachatryan:2014aep, Khachatryan:2015bnx}, 
where publicly available. In cases where these uncertainties 
are not explicitly given they are approximated to reproduce the
total experimental uncertainties. 
The total uncorrelated uncertainty is obtained by adding the 
systematic uncertainty from background processes to the statistical uncertainty from 
signal events in quadrature.
%
%%%%%%%%%%%%%%%%%%%%%%%%%%%%%%%%%%%%%%%%%%%%%%%%%%%%%%%%
\begin{table}[!t]
\begin{tabular}{| C{2.5cm} | C{1.5cm}  | C{2.5cm} | C{1.5cm}  |}
\hline
\multicolumn{2}{|c|}{production process}     &   \multicolumn{2}{c|}{decay process}  \\
\hline
\hline
$pp\to H$                       &     10          &  $H\to b\bar{b}$  & 25 \\
$pp\to H+\text{j}$           &      30         &  $H\to \gamma \gamma$ &  20 \\
$pp\to H+2\text{j}$         &   100          &  $H\to \tau^+ \tau^-$   & 15 \\
$pp\to HZ$ 	             &   10         &   $H\to 4l$   & 20 \\
$pp\to HW$	             &    50         &  $H\to 2l 2\nu$   & 15 \\
$pp\to t\bar{t}H$             &    30      &  $H\to Z \gamma$  &  150 \\
                                       &                 &  $H\to \mu^+ \mu^-$  & 150 \\
\hline
\end{tabular}     
    \caption{Relative systematic uncertainties due to background processes
    for each production and decay channel in $\%$. \label{tab:backgroundunc}}
\end{table}
%%%%%%%%%%%%%%%%%%%%%%%%%%%%%%%%%%%%%%%%%%%%%%%%%%%%%%%

Beyond identification and reconstruction efficiencies for production channels and Higgs decays, 
each channel is plagued by individual experimental systematic uncertainties. 
We adopt flat systematic uncertainties in $p_{T,H}$ for the individual channels.
The numerical values are based on the results from experimental Run 1 
analyses~\cite{Aad:2013wqa, Aad:2014eha, Chatrchyan:2013mxa, Aad:2015vsa, Aad:2014eva, ATLAS:2014aga, Aad:2014fia, Aad:2014xva, 
Aad:2015gra, Aad:2015ona, Khachatryan:2014ira, Chatrchyan:2014nva, Chatrchyan:2013lba, Chatrchyan:2013iaa, 
Chatrchyan:2013vaa, Khachatryan:2014aep, Khachatryan:2015bnx}, see Tab.~\ref{tab:syst}. 
In channels where no measurement has been performed or no information is publicly available, 
e.g.\ $pp\to H+2\text{j}, H\to Z \gamma$, 
we choose a conservative estimate of systematic uncertainties of 100\%.
In addition to the uncertainties listed in Tab.~\ref{tab:syst}, we include a systematic uncertainty of
30\% for the $H\to 2l 2\nu$ channel for differential cross sections. This uncertainty is due to the 
inability of reconstructing the Higgs transverse momentum accurately. 

During future runs, experimental uncertainties are likely to 
improve with the integrated luminosity. 
Hence for our results at 14~TeV we use the 8~TeV uncertainties as a starting point, 
as displayed in Tabs.~\ref{tab:backgroundunc} and \ref{tab:syst}, 
and rescale them by $L_8/L_{14}$ 
for a given integrated luminosity at 14~TeV $L_{14}$.  
This results in a reduction of statistical and experimental 
systematic uncertainties by a factor of about
$0.3$ for $L_{14}=300~\text{fb}^{-1}$ 
and about $0.1$ for $L_{14}=3000~\text{fb}^{-1}$. 
This simplified procedure has also been adopted by the ATLAS and CMS collaborations
to extrapolate the sensitivity of experimental analyses to future 
runs~\cite{ATLAS-collaboration:1484890,ATLAS:2013hta, CMS-NOTE-2012-006, CMS:2013xfa} 
We use this extrapolation for ease of comparison and reproducibility.

We only consider measurements with more than 5 signal events after the application of all 
efficiencies and a total uncertainty smaller than 100\%.
The pseudo-data are constructed using the SM hypothesis, i.e.\ all Wilson coefficients are set to zero.
We construct expected signal strength measurements for all accessible production and decay modes.
Additionally, differential cross sections as function of the Higgs transverse momentum are simulated
with a bin size of 100~GeV.
In $2\to3$ processes like $ttH$ other differential distributions might
provide higher sensitivity than $p_{T,H}$, but at this point 
we restrict the analysis to include $p_{T,H}$ distributions only, 
as these are likely to be provided as unfolded distributions by the experimental 
collaborations. We leave studies on the sensitivity of additional 
kinematic variables in a global fit to future work~\cite{toapp}.

Comparing our predictions for the uncertainties on the signal strength measurements 
for 14~TeV using an integrated luminosity of $L_{14}=300~\text{fb}^{-1}$ and 
$L_{14}=3000~\text{fb}^{-1}$, with the expectations published by 
ATLAS~\cite{ATLAS-collaboration:1484890,ATLAS:2013hta}
and CMS~\cite{CMS-NOTE-2012-006, CMS:2013xfa}, we find good quantitative agreement 
with the publicly available channels.

%%%%%%%%%%%%%%%%%%%%%%%%%%%%%%%%%%%%%%%%%%%%%%%%%%%%%%%
\begin{table}[tb]
\begin{tabular}{|c|c|c|c|c|c|c|}
\hline
                                       & $t\bar{t}H$ & $HZ$ 	& $HW$	& $H$ incl. & $H+\text{j}$ & $H+2\text{j}$  \\
\hline\hline
  $H\to b\bar{b}$               &  80    	&     25     	&    40     	&    100 	&    	100		&	150	\\
 $H\to \gamma \gamma$ &  60     	&     70     &   30      	&     10	&       10   		&	20	\\
 $H\to \tau^+ \tau^-$        &  100   	&     75     &   75      	&     80	&      	 80		&	30	\\
 $H\to 4l$                         &  70     	&     30     &   30      	&     20   	&   	 20		&	30	\\
 $H\to 2l 2\nu$                 &  70     	&    100    &  100     	&     20	&     	 20		&	30	\\
 $H\to Z \gamma$           &  100    	&    100    &   100    	&     100   	&       100 		&	100	\\
 $H\to \mu^+ \mu^-$       &  100     	&    100    &   100    	&     100   	&       100 		&	100	\\
 \hline
\end{tabular}
    \caption{Relative systematic uncertainties for each production times decay channel in $\%$. 
    \label{tab:syst}}
\end{table}
%%%%%%%%%%%%%%%%%%%%%%%%%%%%%%%%%%%%%%%%%%%%%%%%%%%%%%%

%%%%%%%%%%%%%%%%%%%%%%%%%%%%%%%%%%%%%%%%%%%%%%%%%%%%%%%%
\begin{table}[tb]
\begin{tabular}{| C{2.5cm} | C{1.5cm}  | C{2.5cm} | C{1.5cm}  |}
\hline
\multicolumn{2}{|c|}{production process}     &   \multicolumn{2}{c|}{decay process}  \\
\hline
\hline
$pp\to H$                       &     14.7       &  $H\to b\bar{b}$  & 6.1 \\
$pp\to H+\text{j}$           &      15         &  $H\to \gamma \gamma$ &  5.4 \\
$pp\to H+2\text{j}$         &      15         &  $H\to \tau^+ \tau^-$   & 2.8 \\
$pp\to HZ$ 	             &      5.1         &   $H\to 4l$   & 4.8 \\
$pp\to HW$	             &      3.7         &  $H\to 2l 2\nu$   & 4.8 \\
$pp\to t\bar{t}H$             &     12         &  $H\to Z \gamma$  & 9.4 \\
                              &                &  $H\to \mu^+ \mu^-$  & 2.8  \\  
\hline
\end{tabular}     
    \caption{Theoretical uncertainties for each production and decay channel in $\%$. \label{tab:theo}}
\end{table}
%%%%%%%%%%%%%%%%%%%%%%%%%%%%%%%%%%%%%%%%%%%%%%%%%%%%%%%

%%%%%%%%%%%%%%%%%%%%%%%%%%%%%%%%%%%%%%%%%%%%%%%%%%%%%%%
\begin{table*}[tb]
\begin{tabular}{l | C{3.3cm} | C{3.3cm} | C{3.3cm} | C{3.5cm}}
& individual  &   \multicolumn{3}{c}{ marginalized (all $c_i$ free)}  \\[0.2cm]
\hline
& Run 1 & Run 1 & pseudo-data $\mu$ & pseudo-data $\mu$ and $p_{T,h}$ \\
\hline
\hline
$\bar{c}_{g}$ $[\times 10^{4}]$ &  $[\, -0.32, \, 0.35 \, ]$ &  $[\, -0.64, \, 0.43 \, ]$ &  $[\, -0.84, \,  > 1.000 \, ]$  &  $[\, -0.82, \,  > 1.000 \, ]$  \\
$\bar{c}_{\gamma}$ $[\times 10^{4}]$ &  $[\, -5.5, \, 3.6 \, ]$ &  $[\, -7.8, \, 4.3 \, ]$ &  $[\, < -10.000, \, 7.3 \, ]$ &  $[\, < -10.000, \, 6.6 \, ]$ \\
$\bar{c}_{W}$  &  $[\, < -0.01, \, 0.007 \, ]$ &  $[\, < -0.01, \,  > 0.01 \, ]$  &  $[\, < -0.01, \,  > 0.01 \, ]$  &  $[\, < -0.01, \,  > 0.01 \, ]$  \\
$\bar{c}_{H}$  &  $[\, < -0.05, \,  > 0.05 \, ]$  &  $[\, < -0.05, \,  > 0.05 \, ]$  &  $[\, < -0.05, \,  > 0.05 \, ]$  &  $[\, < -0.05, \,  > 0.05 \, ]$  \\
$\bar{c}_{HW}$  &  $[\, -0.047, \, 0.014 \, ]$ &  $[\, < -0.05, \, 0.035 \, ]$ &  $[\, < -0.05, \,  > 0.05 \, ]$  &  $[\, -0.044, \,  > 0.05 \, ]$  \\
$\bar{c}_{HB}$  &  $[\, < -0.05, \,  > 0.05 \, ]$  &  $[\, < -0.05, \,  > 0.05 \, ]$  &  $[\, < -0.05, \,  > 0.05 \, ]$  &  $[\, < -0.05, \,  > 0.05 \, ]$  \\
$\bar{c}_{u3}$  &  $[\, < -0.05, \,  > 0.05 \, ]$  &  $[\, < -0.05, \,  > 0.05 \, ]$  &  $[\, < -0.05, \,  > 0.05 \, ]$  &  $[\, < -0.05, \,  > 0.05 \, ]$  \\
$\bar{c}_{d3}$  &  $[\, < -0.05, \,  > 0.05 \, ]$  &  $[\, < -0.05, \,  > 0.05 \, ]$  &  $[\, < -0.05, \,  > 0.05 \, ]$  &  $[\, < -0.05, \,  > 0.05 \, ]$  \\
\hline
\end{tabular}
\caption{Constraints at 95\% CL 
on dimension-six operator coefficients (first column) from LHC Run 1 data, 
considering only one operator in the fit (second column) and all operators simultaneously (third column).
The results obtained using pseudo-data are shown in the last two columns, 
with signal strengths measurements only (fourth column) and including differential 
distributions (fifth column). In case no constraints can be derived within the 
parameter ranges considered in this work, the lower and upper limits 
are indicated to lie outside this range.
\label{tab:8tevresults}}
\end{table*}
%%%%%%%%%%%%%%%%%%%%%%%%%%%%%%%%%%%%%%%%%%%%%%%%%%%%%%%

Theory uncertainties included in the fit are listed in
Tab.~\ref{tab:theo} and have been obtained by the Higgs cross section
working group~\cite{Dittmaier:2011ti,Dittmaier:2012vm,
  Heinemeyer:2013tqa} (see also \cite{Fichet:2015xla} about their role in Higgs fits).  We assume the same size of theory
uncertainties for the SM predictions as for calculations using the EFT
framework. 
The theory uncertainties are not scaled with luminosity and
retain the values given in Tab.~\ref{tab:theo} throughout this work. 

Systematic uncertainties are crucial limiting factors of a 
coupling extraction and the scaling we choose in the present
paper are unlikely to be realistic, but provide a clean extrapolation
picture for potential progress over the next decades. 
In summary, the assumptions chosen to get our
estimate are
\begin{itemize}
\item the above luminosity scaling of experimental uncertainties,
\item a clean separation of the measurements of all production and decay
  channels (no cross talk between channels),
\item flat experimental systematic uncertainties as function of $p_{T,H}$,
\item flat theory uncertainties as function of $p_{T,H}$ as quoted in Tab.~\ref{tab:theo}, 
which we assume to be independent of the Wilson coefficients.
\end{itemize}
A more detailed investigation of systematics beyond the approximations 
chosen in this work can provide a guideline 
for future precision efforts, this work is currently ongoing~\cite{toapp}.

%%%%%%%%%%%%%%%%%%%%%%%%%%%%%%%%%%%%%%
\section{Predicted Constraints}
\label{sec:results14}
%%%%%%%%%%%%%%%%%%%%%%%%%%%%%%%%%%%%%%

The projected measurements of the Higgs signal strengths and the Higgs 
transverse momentum ($p_{T,H}$) distributions
are used to test the sensitivity to the dimension six operators that can be obtained 
with the LHC. 
In all fits theory uncertainties are included as nuisance parameters with Gaussian constraints. 
The constraints on individual Wilson coefficients are obtained by a marginalisation over the 
remaining coefficients and the nuisance parameters related to the theory uncertainties. 

In order to test this approach, we first generate pseudo-data for 8~TeV 
following the procedure detailed above. The integrated luminosity is chosen to be $L_8$, 
i.e.\ $25~\text{fb}^{-1}$ per experiment which corresponds to the full Run 1 data. 
With this setting no luminosity scaling of experimental uncertainties is performed. 
Besides statistical uncertainties, the generated 8~TeV data have systematic uncertainties 
corresponding to the values given in Tabs.~\ref{tab:backgroundunc} and~\ref{tab:syst}. 
We compare the constraints obtained with these pseudo-data with the ones obtained from 
the Run 1 analysis in Tab.~\ref{tab:8tevresults}.
Similar to the constraints derived in Sec.~\ref{sec:results8} no reliable constraints 
at 95\% CL on coefficients other than $\bar{c}_{g}$ and $\bar{c}_{\gamma}$ can be derived 
within the parameter ranges considered in this work. 
We observe that the constraints using pseudo-data are considerably weaker than 
the ones from the existing Run 1 measurements. This is no surprise, as the simplified 
approach outlined above can not reflect the complexity of real analyses, where
a number of signal regions are used to disentangle different production modes. 
This picture does not change when including differential distributions 
(last column of Tab.~\ref{tab:8tevresults}) which results 
in slightly better constraints at 8~TeV compared to the fit with signal strengths only. 
We note that although the constraints obtained with pseudo-data are generally weaker, 
they are very similar to the ones using current Run 1 experimental data. We therefore trust 
our method and proceed to derive the expected sensitivity of the LHC.

\begin{figure*}[p]
 \begin{center}
  \includegraphics[width=0.44\textwidth]{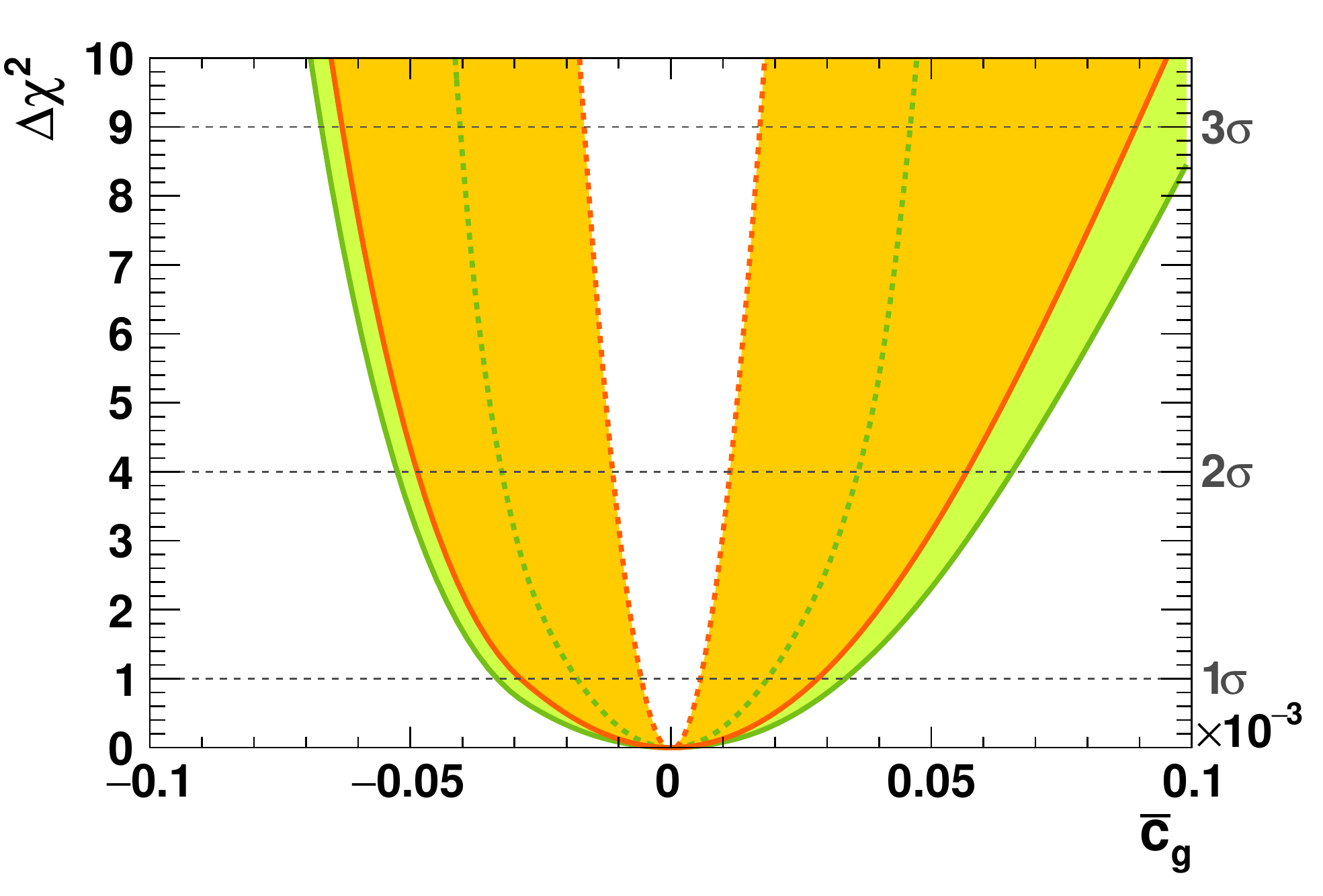}
  \hspace{0.7cm}
  \includegraphics[width=0.44\textwidth ]{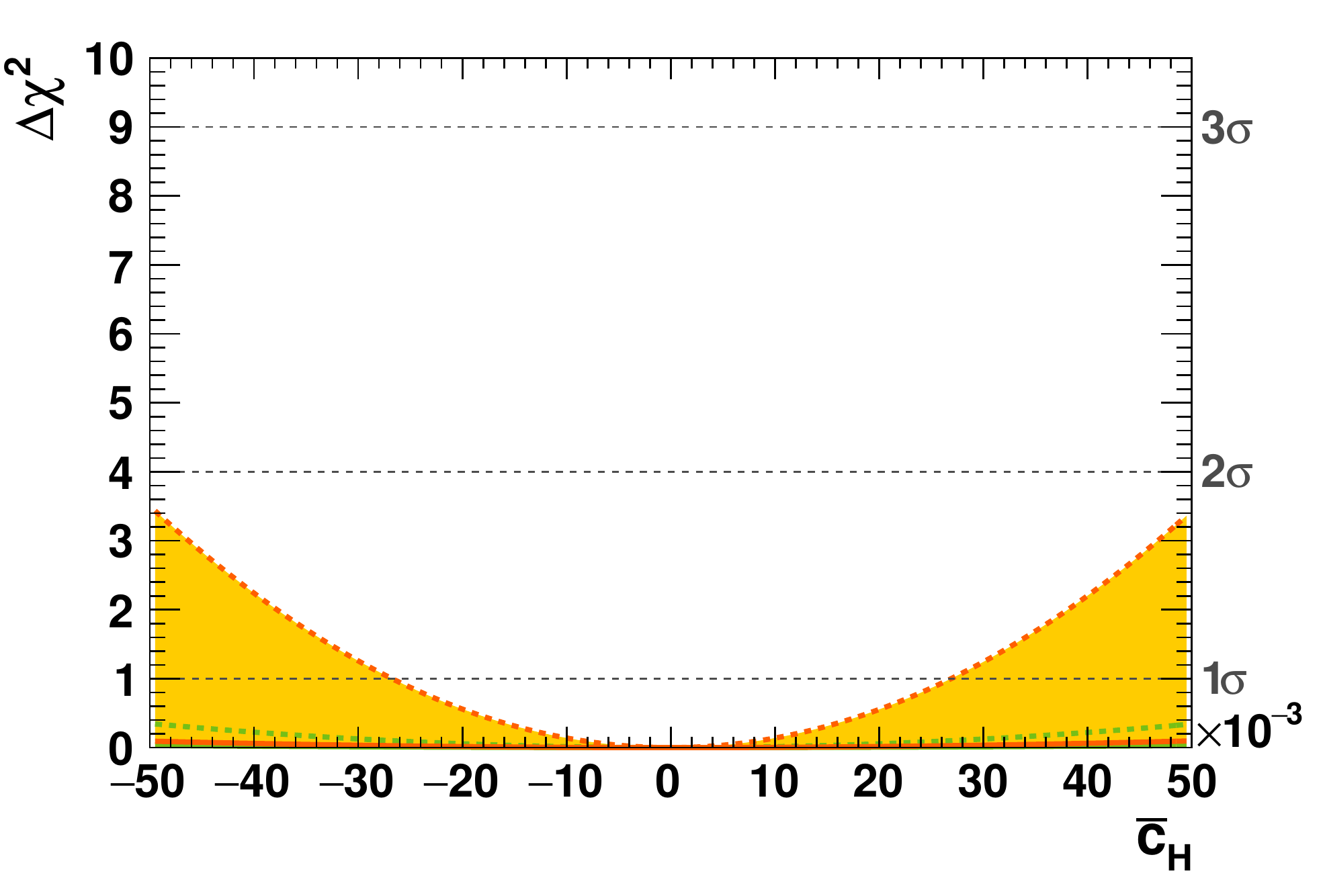}\\[0.2cm]
  \includegraphics[width=0.44\textwidth ]{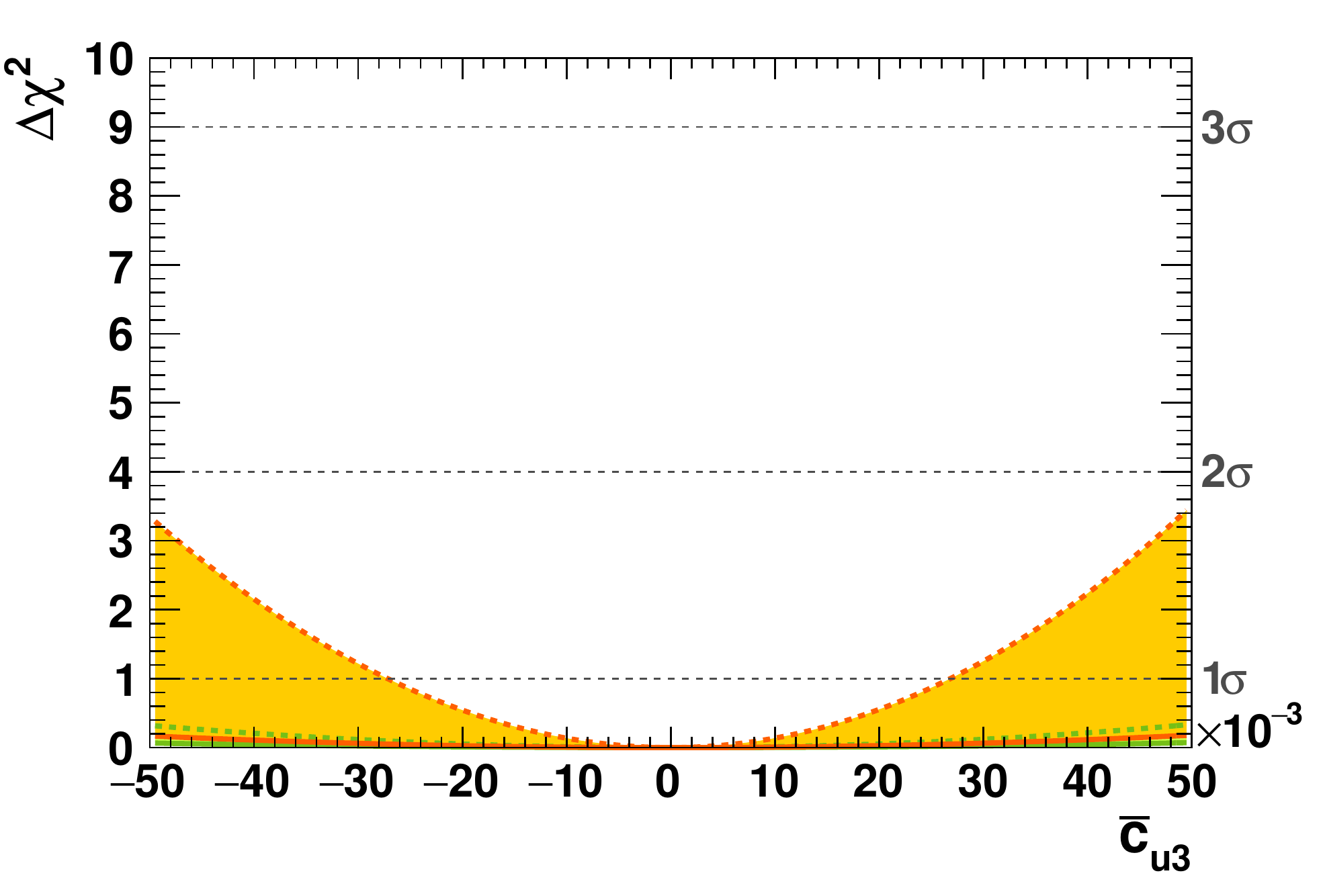}
  \hspace{0.7cm}
  \includegraphics[width=0.44\textwidth]{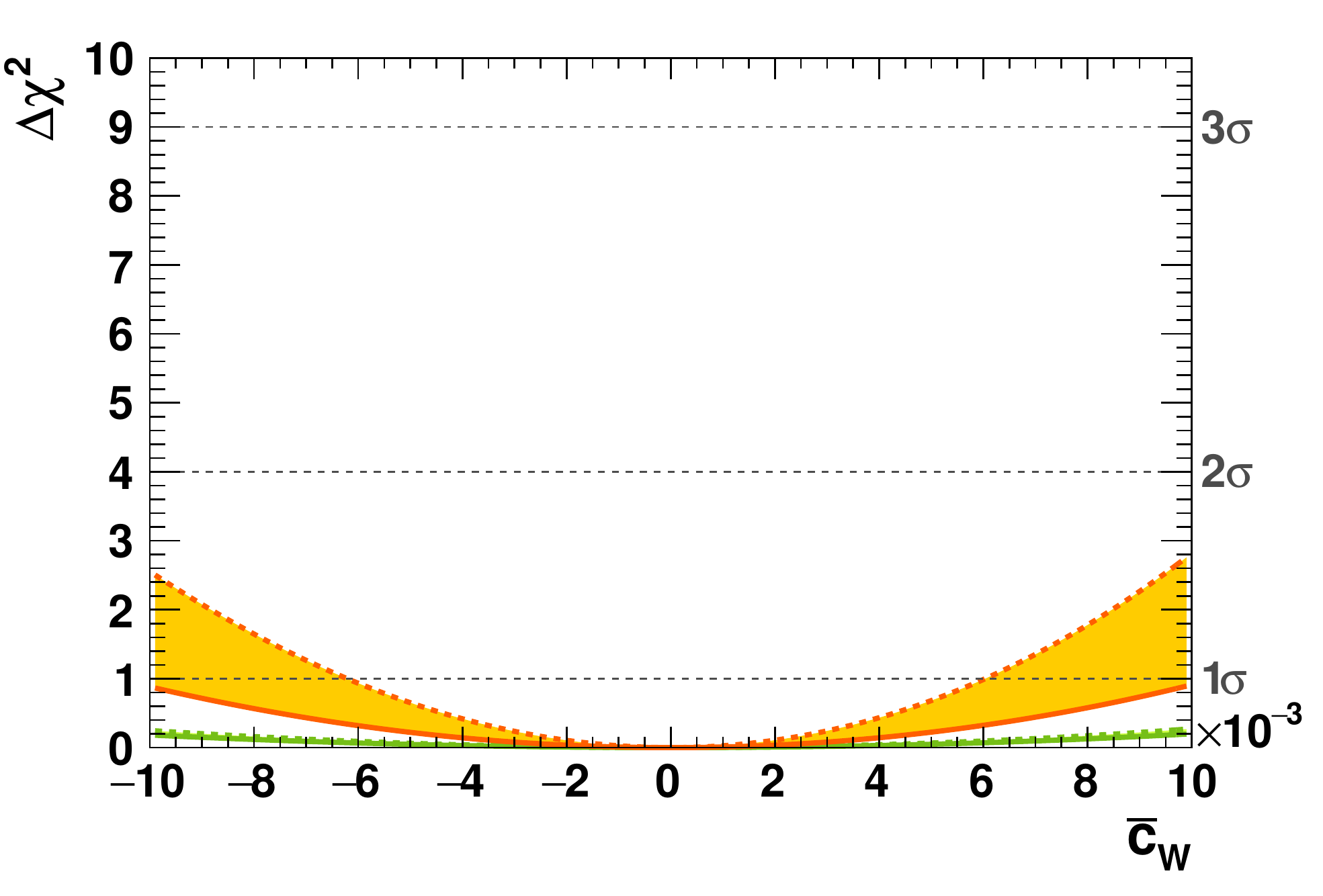} \\[0.2cm]
  \includegraphics[width=0.44\textwidth]{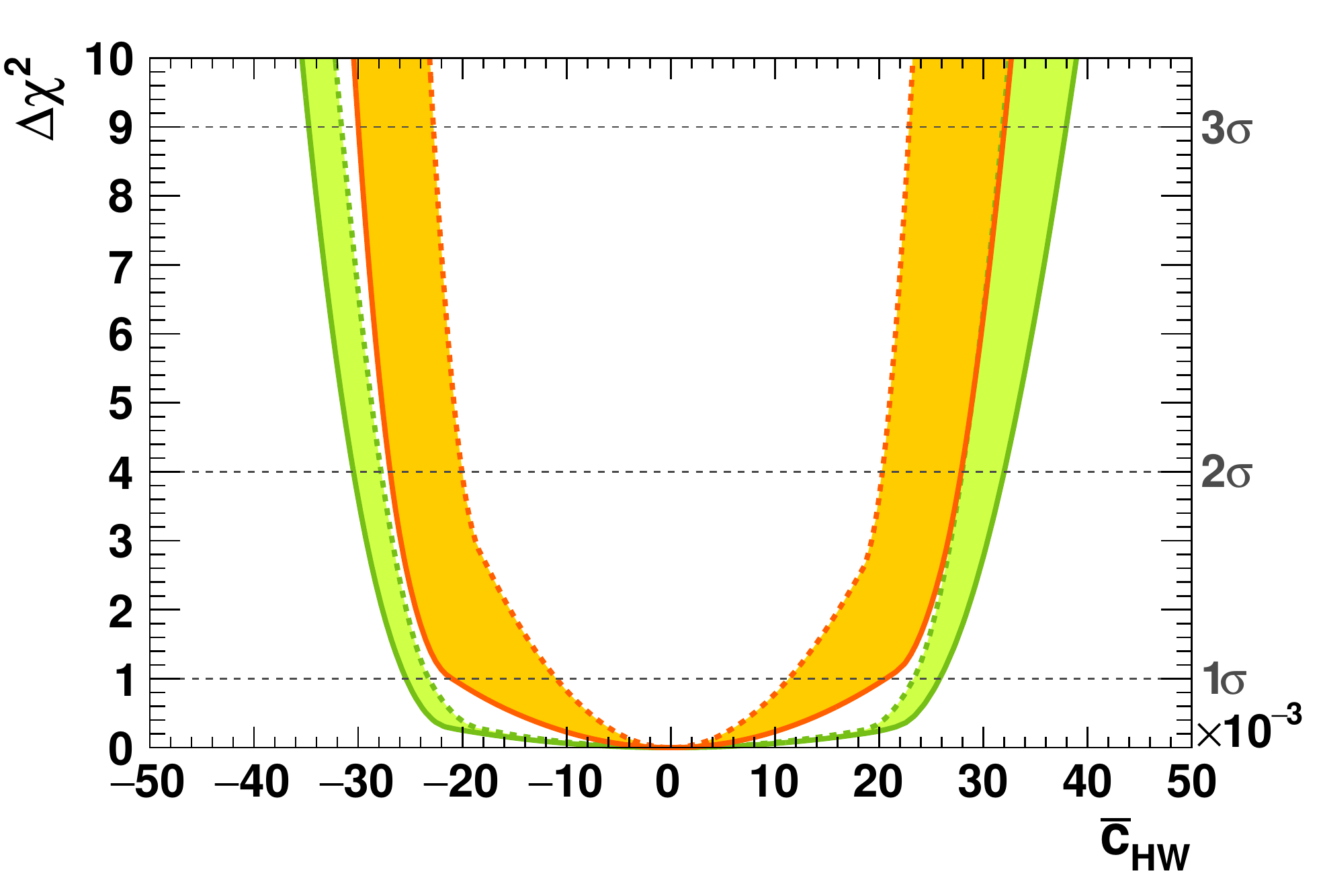}
  \hspace{0.7cm}
  \includegraphics[width=0.44\textwidth]{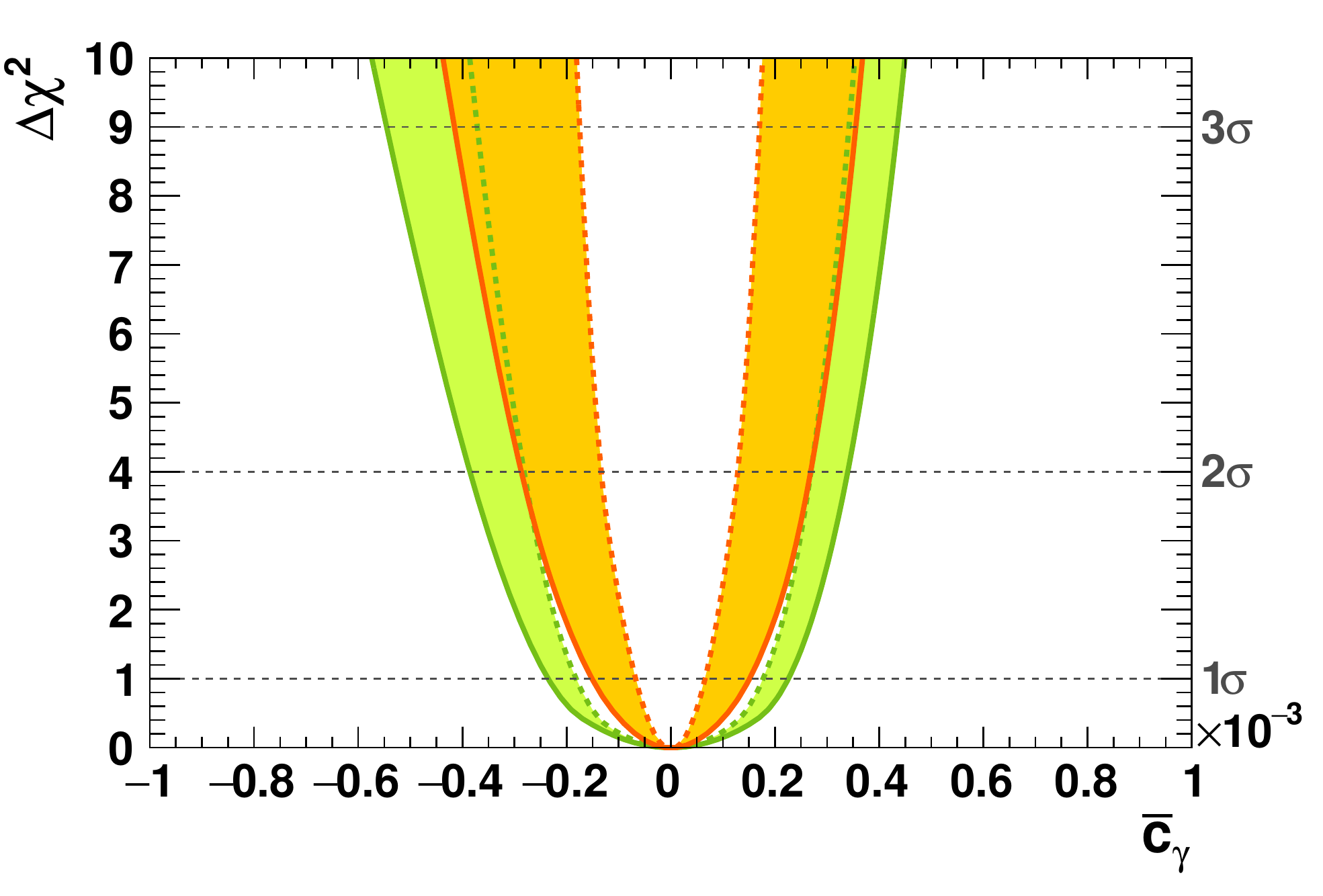}\\[0.2cm]
  \includegraphics[width=0.44\textwidth]{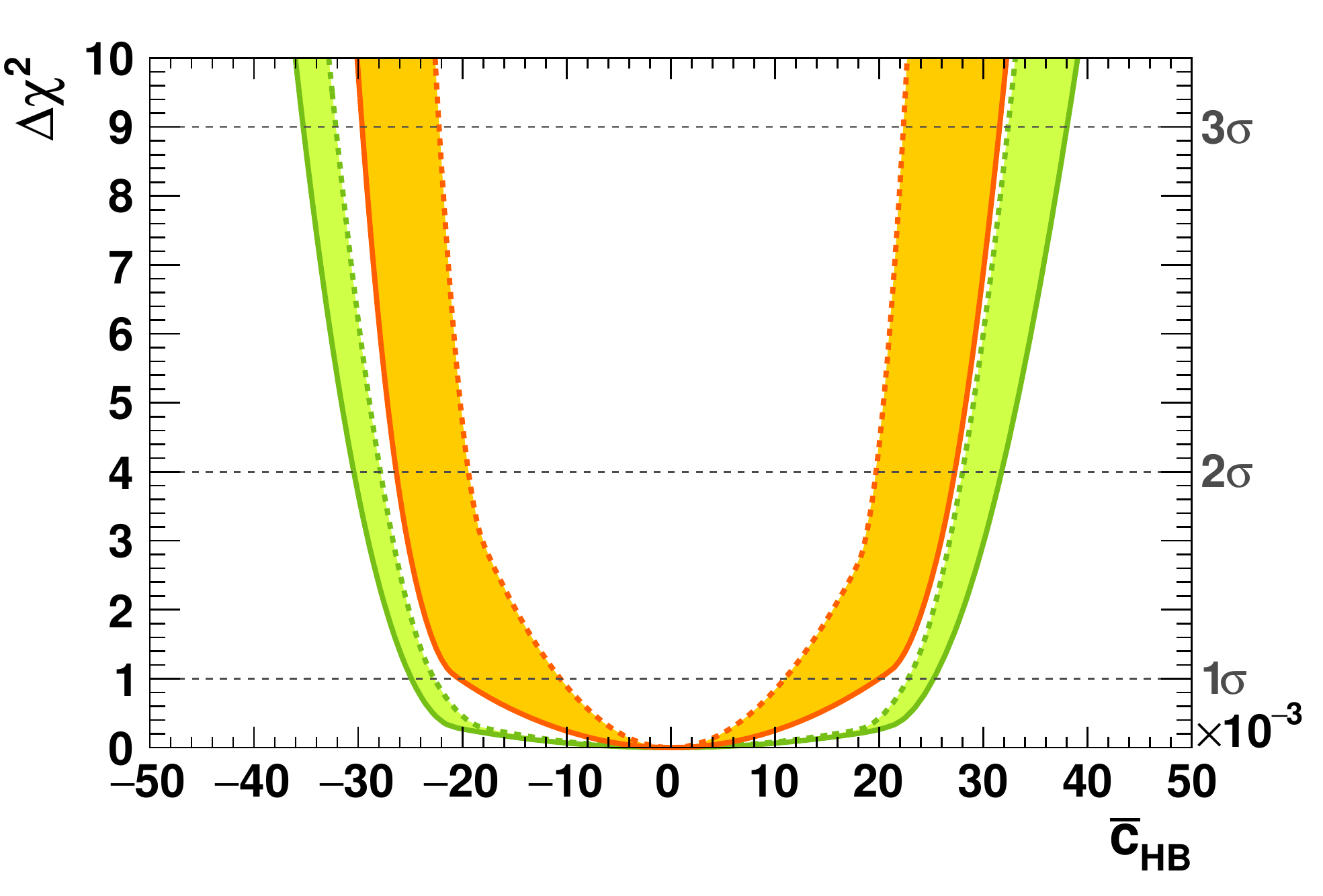}
  \hspace{0.7cm}
  \includegraphics[width=0.44\textwidth]{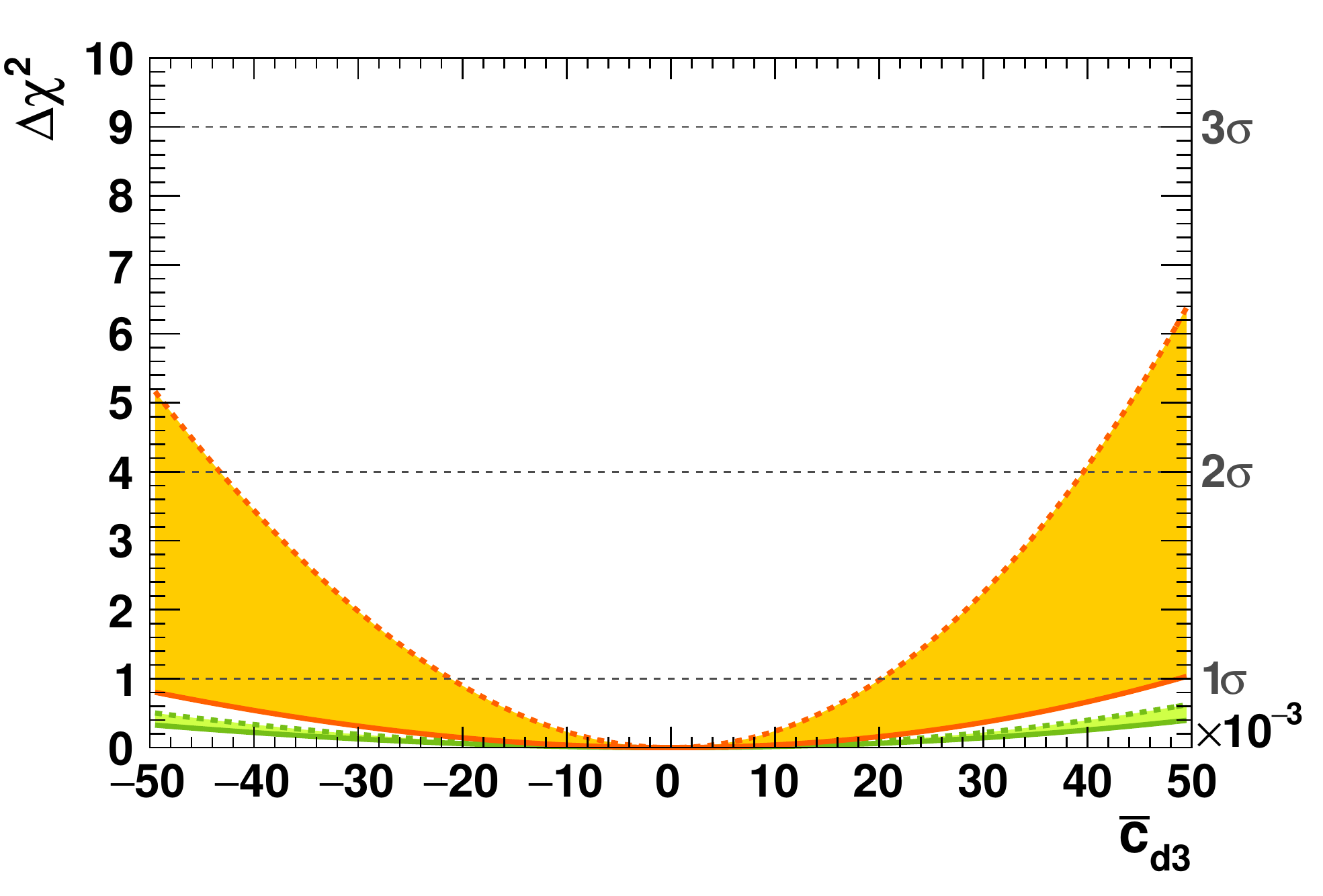}  
 \end{center}
 \caption{Confronting the Lagrangian Eq.~\eqref{eq:silh} with the 14
   TeV LHC measurements with $L=300$ (green) and $3000~\text{fb}^{-1}$ (orange). 
   We only take signal strength measurements into account. 
   Solid lines correspond to a fit with theoretical uncertainties included,
   dashed lines show results without theoretical uncertainties, the band shows the impact of these. 
   For details see the text.
   \label{excl:int}}
\end{figure*}

In Fig.~\ref{excl:int} we show in how far the limits from the LHC Run 1 with 8~TeV
extrapolate to 14 TeV at luminosities of $300~\text{fb}^{-1}$, as
well as after the high luminosity phase with $3000~\text{fb}^{-1}$. 
In these fits we only include expected signal strength measurements. With
these statistics, more production and decay channels become observable
at smaller statistical and systematic uncertainties, which leads to a more constrained fit. 
The fit for the $300~\text{fb}^{-1}$ scenario uses 36 signal strength measurements, 
and 46 measurements are used for the scenario with $3000~\text{fb}^{-1}$. 
All details of the pseudo-data used in performing these extrapolations can be found in the appendix, where also the luminosity-scaling of systematic uncertainties is depicted.
Specifically the constraints on operators that modify associated
Higgs production and weak boson fusion benefit from the increased
centre-of-mass energy and luminosity. In the scenario for the 
high luminosity phase the theoretical uncertainties become dominant 
in some cases.

\begin{figure*}[p]
 \begin{center}
  \includegraphics[width=0.44\textwidth]{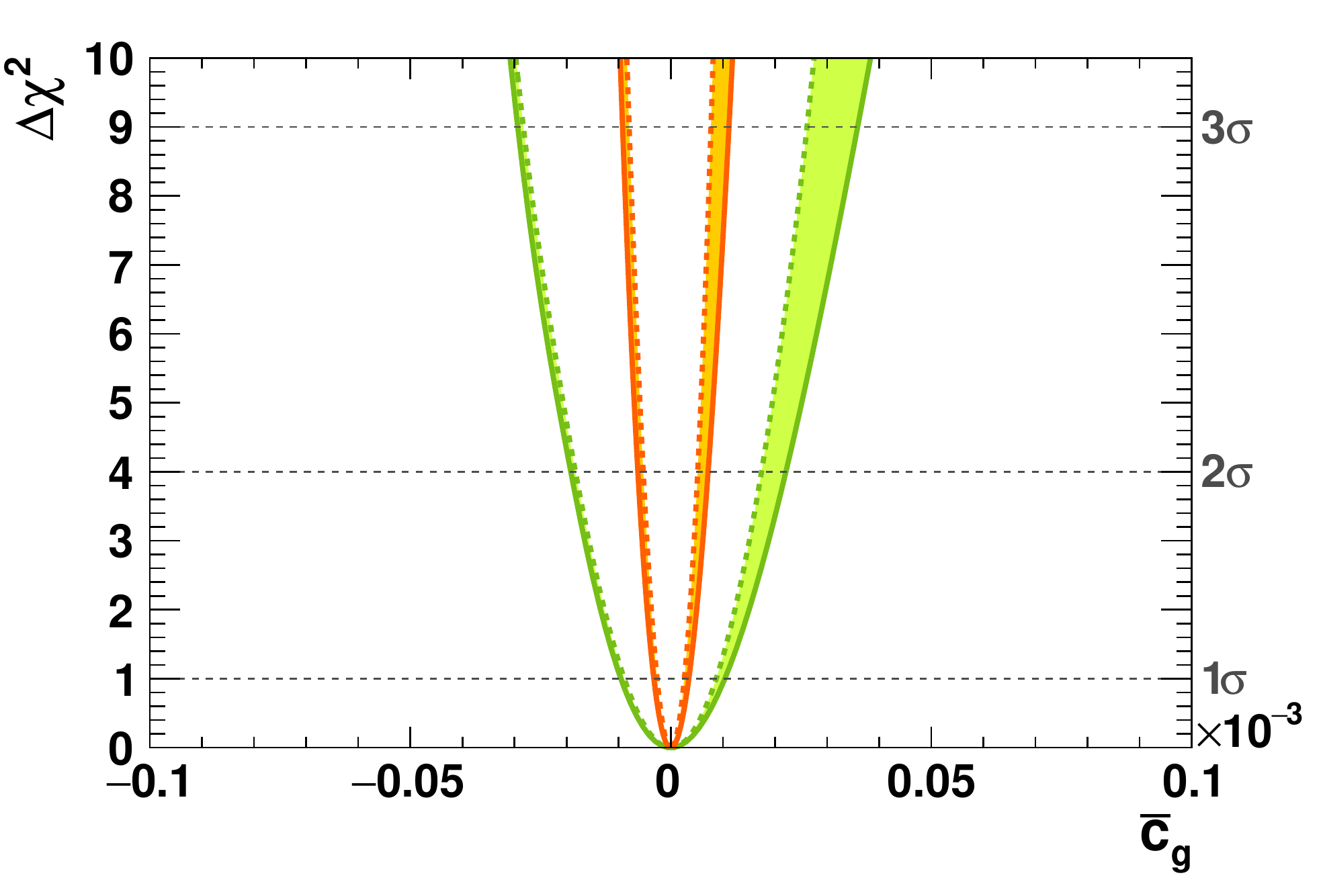}
  \hspace{0.7cm}
  \includegraphics[width=0.44\textwidth ]{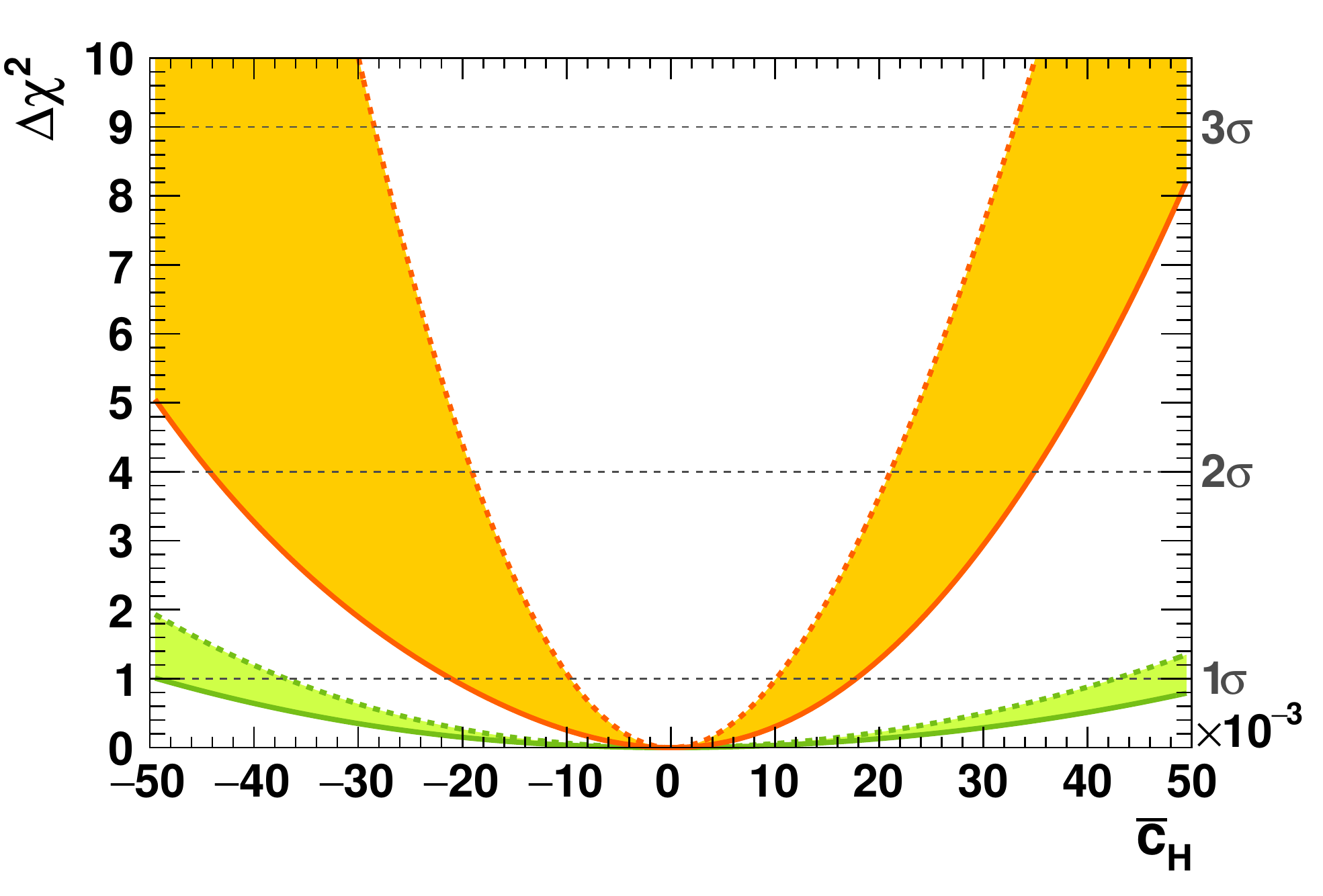}\\[0.2cm]
  \includegraphics[width=0.44\textwidth ]{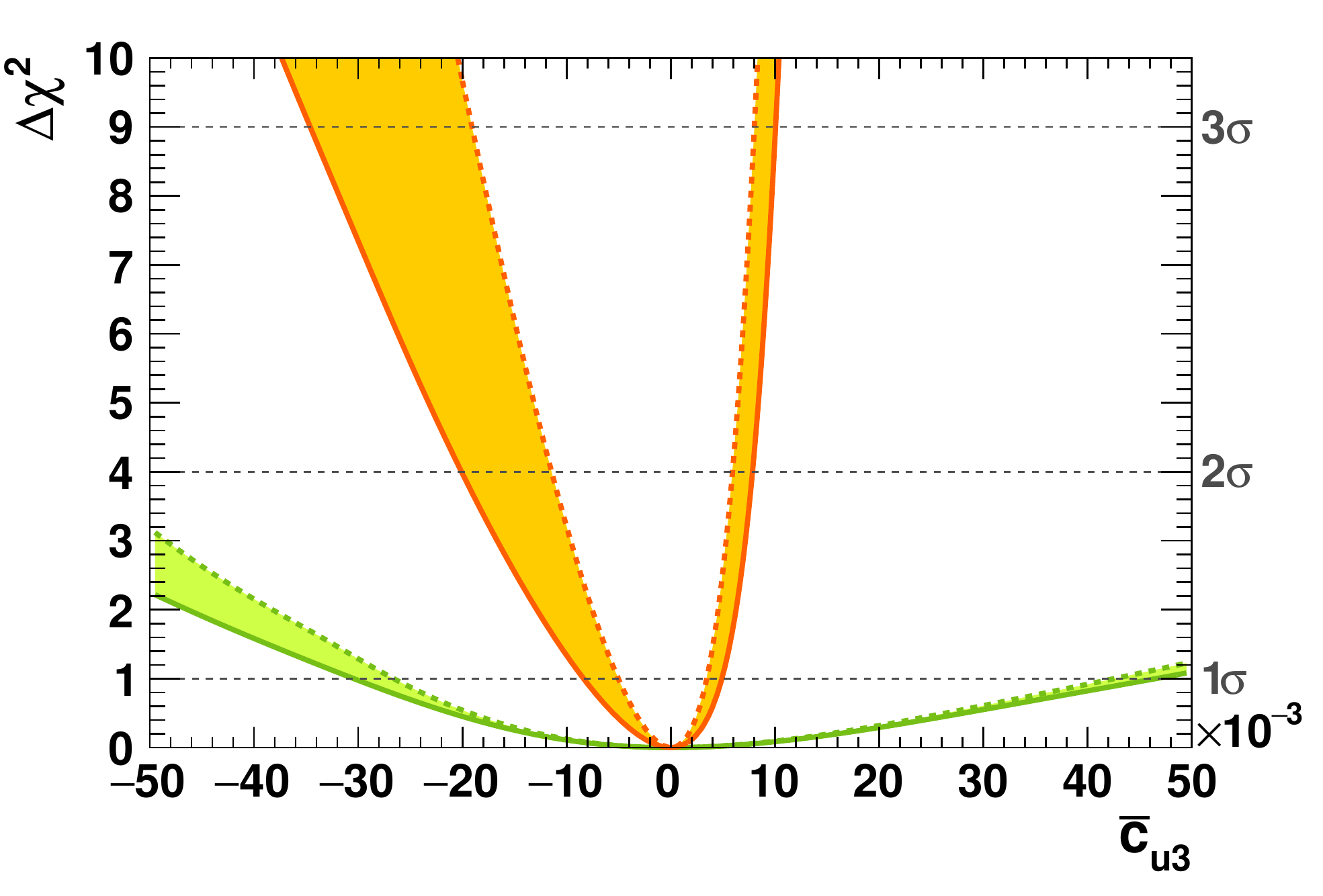}
  \hspace{0.7cm}
  \includegraphics[width=0.44\textwidth]{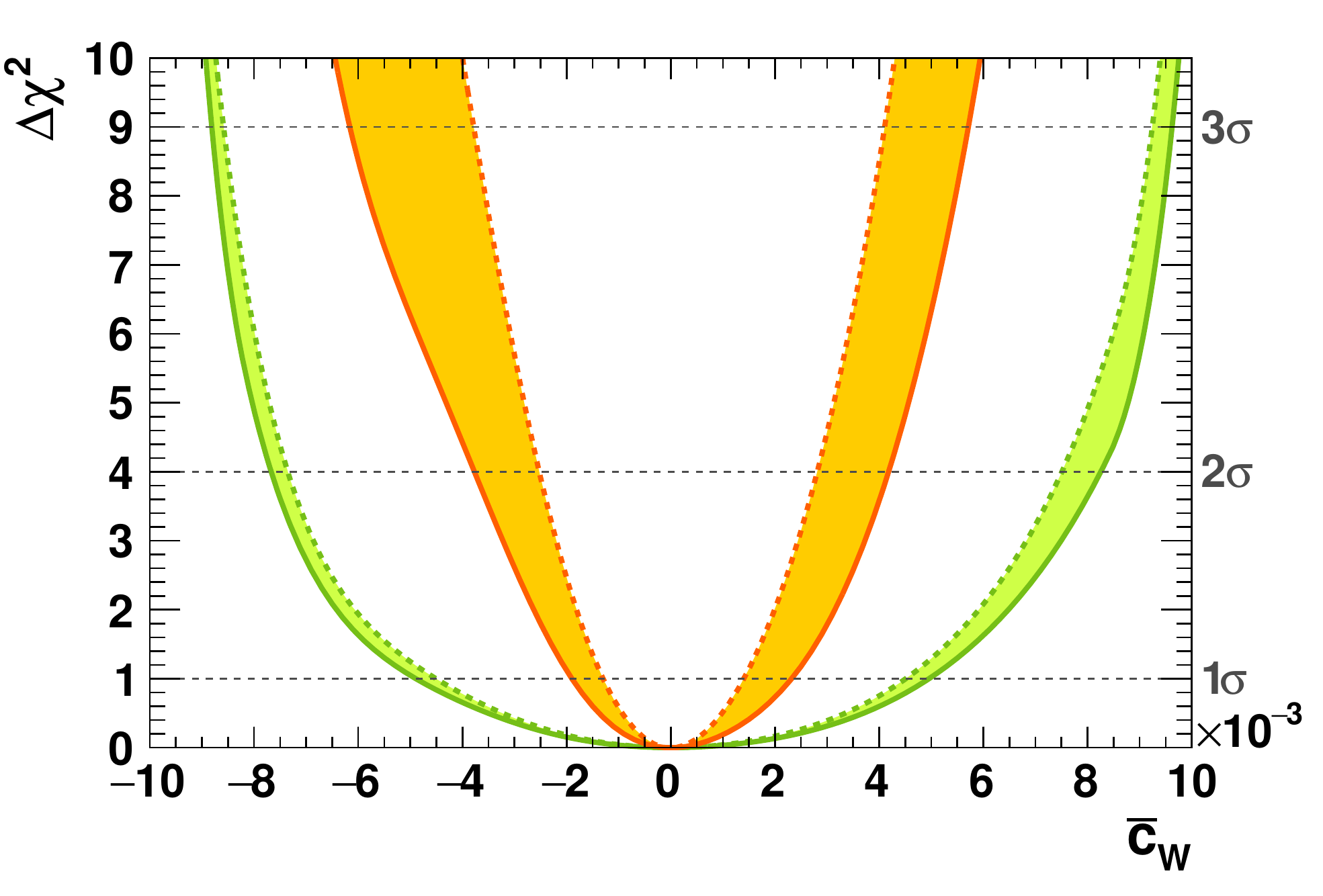} \\[0.2cm]
  \includegraphics[width=0.44\textwidth]{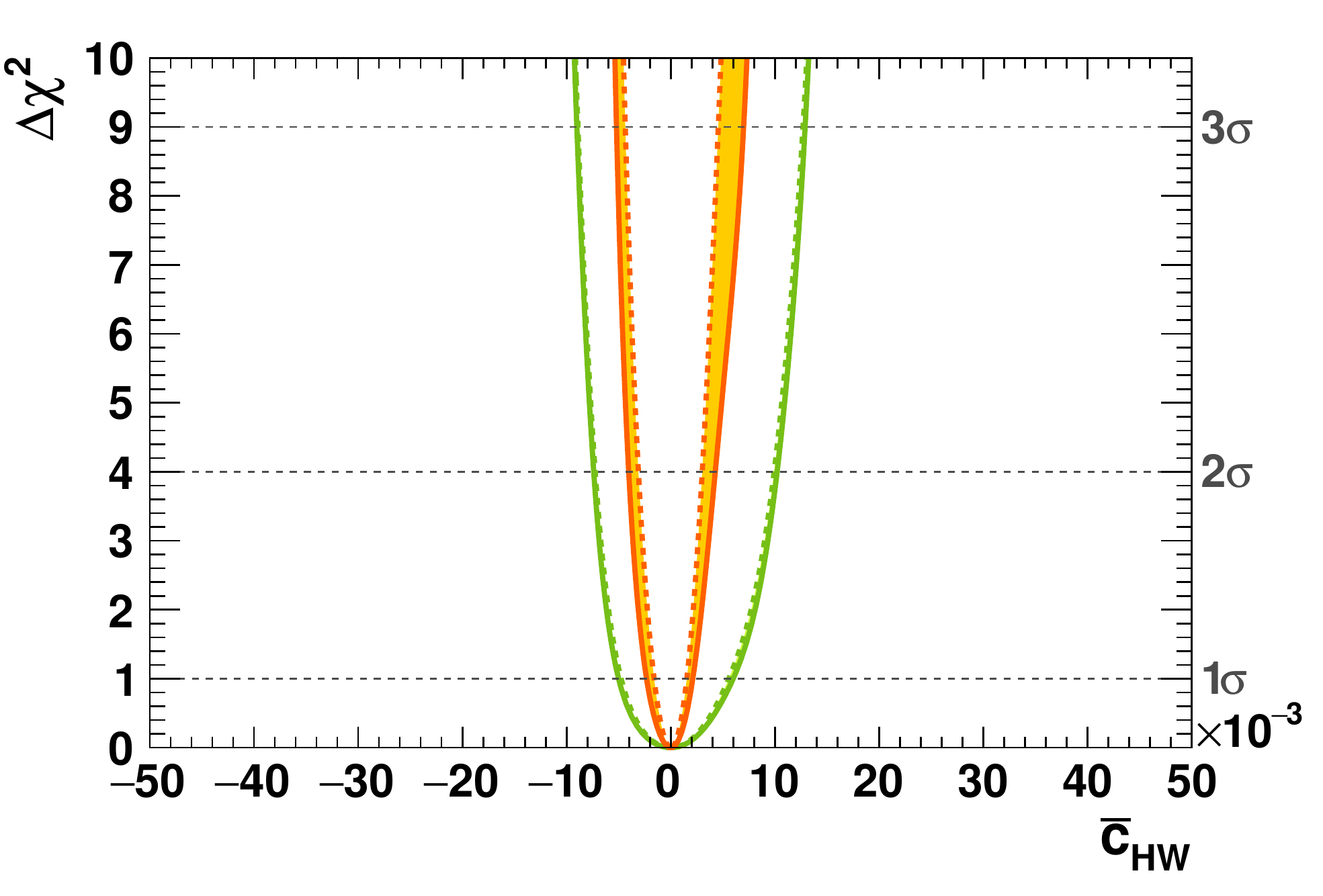}
  \hspace{0.7cm}
  \includegraphics[width=0.44\textwidth]{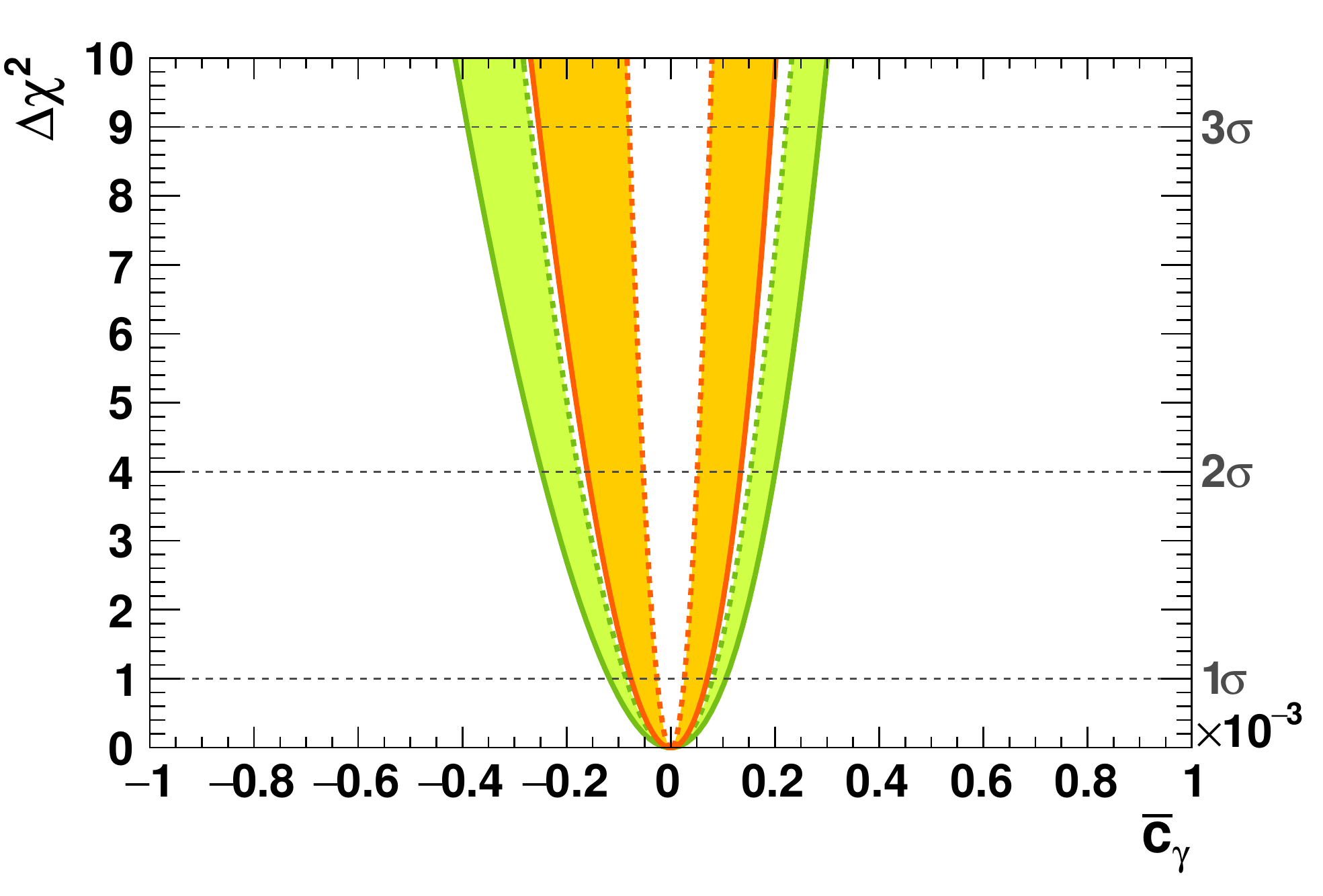}\\[0.2cm]
  \includegraphics[width=0.44\textwidth]{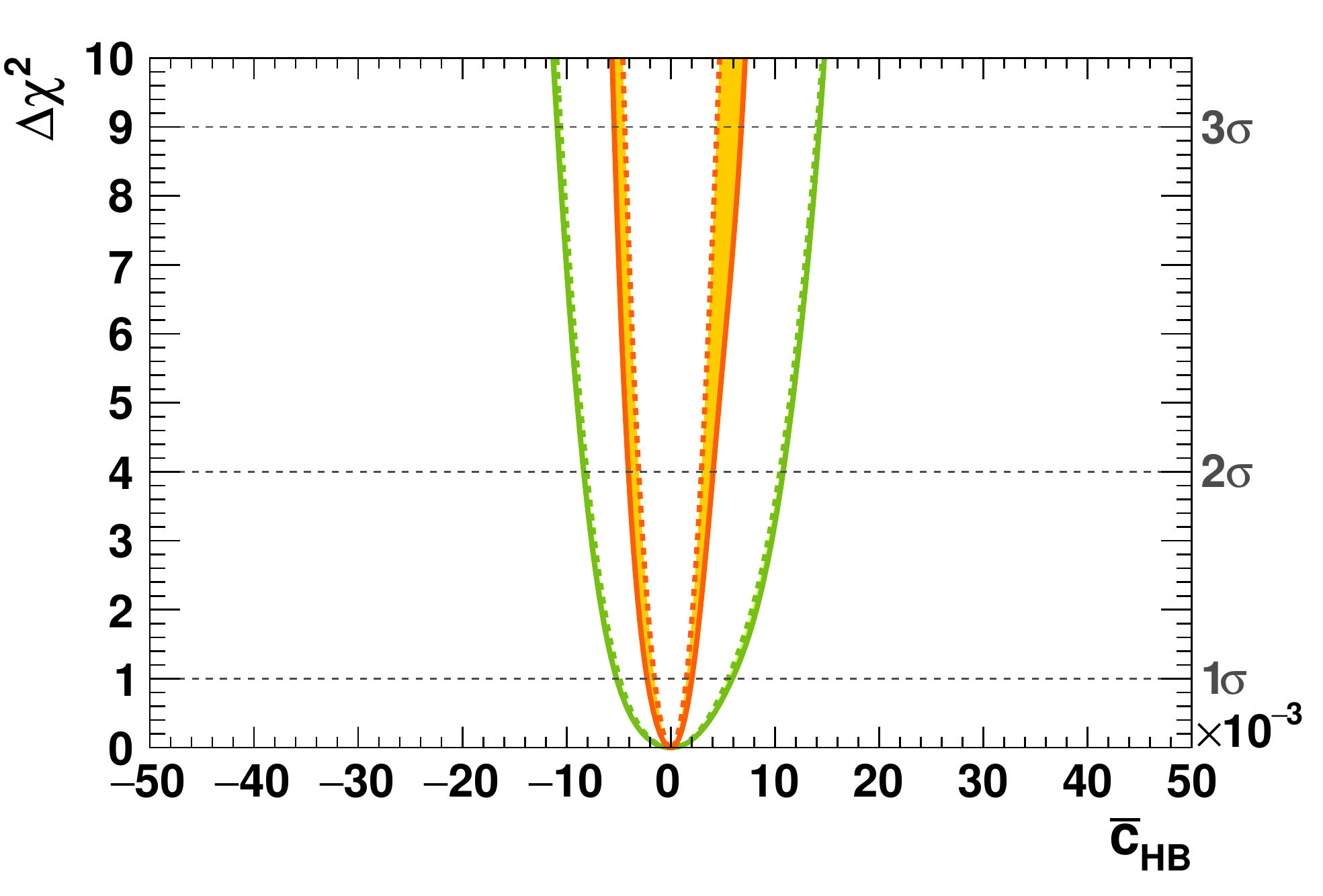}
  \hspace{0.7cm}
  \includegraphics[width=0.44\textwidth]{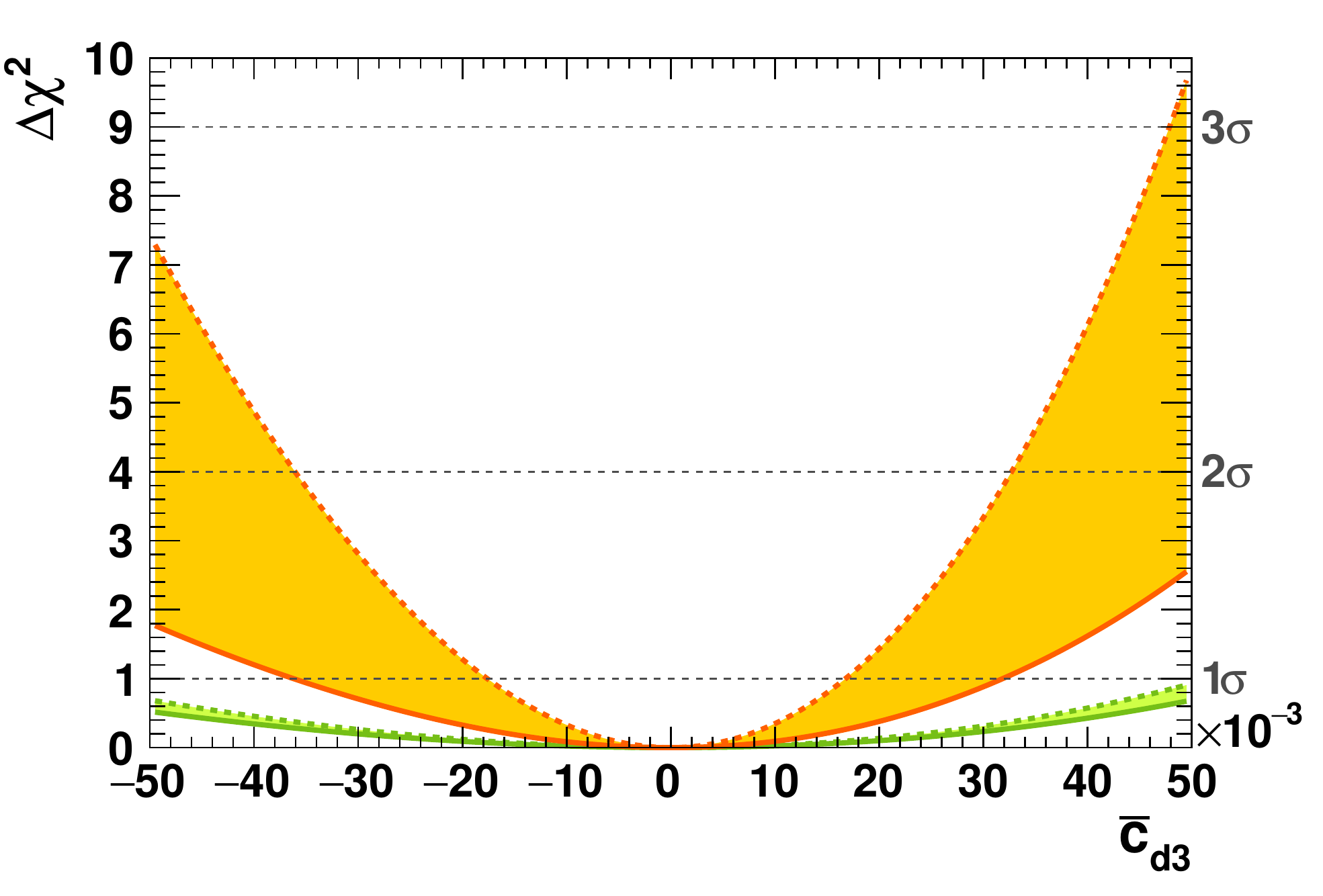}  
 \end{center}
 \caption{Confronting the Lagrangian Eq.~\eqref{eq:silh} with the 14
   TeV LHC measurements with $L=300$ (green) and $3000~\text{fb}^{-1}$ (orange). 
   We include the full $p_{T,H}$ distribution and the signal strength measurement for $pp\to H$ production in the limit setting procedure.
   Solid lines correspond to a fit with theoretical uncertainties included,
   dashed lines show results without theoretical uncertainties, the band shows the impact of these. 
   \label{excl:diff}}
\end{figure*}

\begin{figure*}[t]
  \includegraphics[trim=2cm 10.2cm 1cm 1.5cm, clip, width=0.99\textwidth]{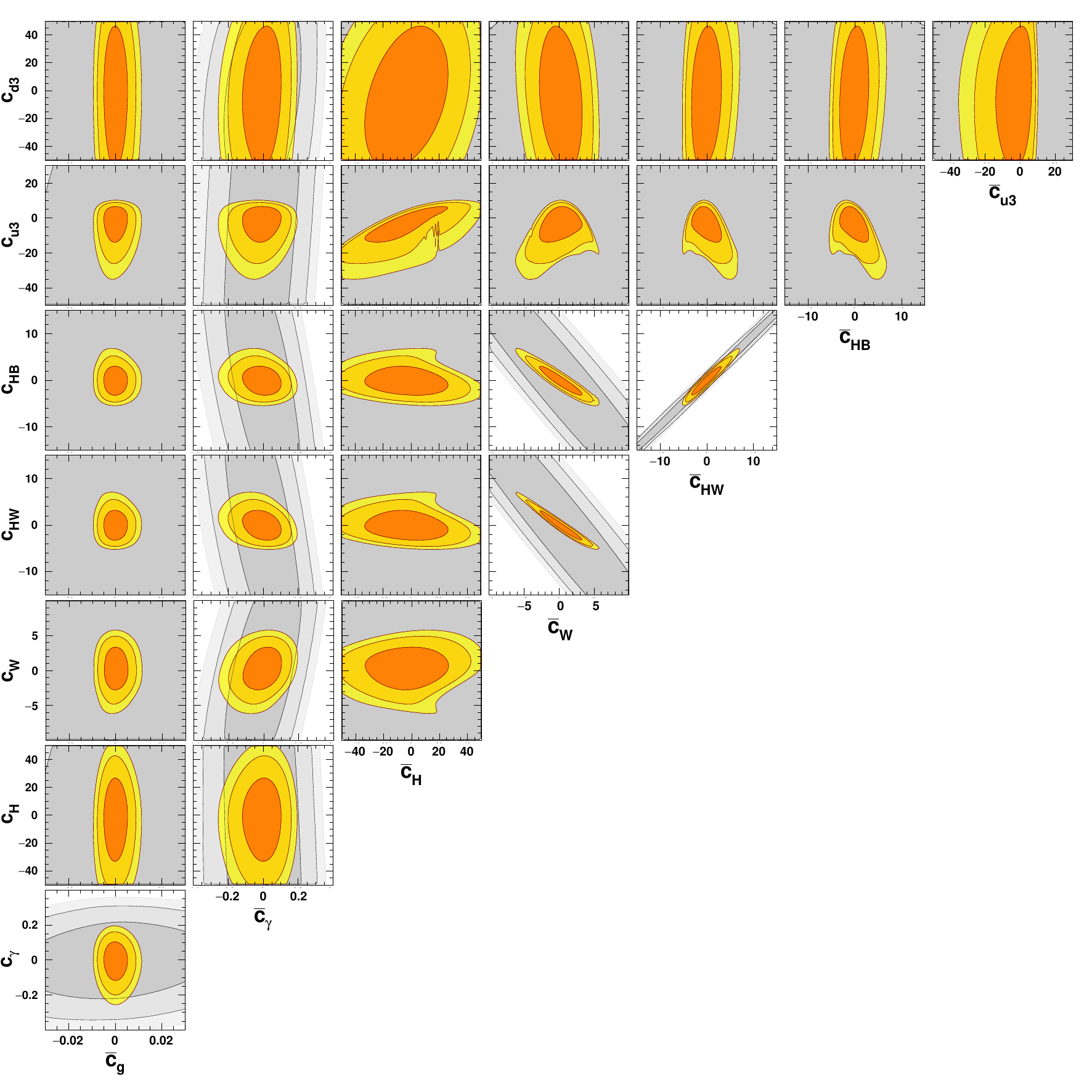}
 \caption{
   \label{excl:2d_3000} Contours of 68\%, 95\% and 99\% CL (from dark to light), obtained from a marginalised fit using the expected signal strength modifiers only (gray) and the
   $p_{T,H}$ measurements (orange) for 14 TeV with a luminosity of $3000~\text{fb}^{-1}$. 
   All coefficients have been multiplied by a factor of $10^{3}$.
   }
\end{figure*}

In a second step, we include the differential $p_{T,H}$ measurements from all
production modes, except $pp\to H$. For the $pp\to H$ production mode
we include six signal strength measurements (see the appendix), 
as no transverse momentum of the Higgs boson
is generated on tree-level.
This results in 82+6 independent measurements included for the fit with $300~\text{fb}^{-1}$
and 117+6 for $3000~\text{fb}^{-1}$. 
In a given production and decay channel, experimental systematic uncertainties 
are included as correlated uncertainties among bins in $p_{T,H}$.
Comparing the above constraints with those expected from including the
differential distributions, Fig.~\ref{excl:diff}, we see a
tremendous improvement. 
The improvement compared to the constraints 
presented in Fig.~\ref{excl:int} is solemnly due to the inclusion of differential distributions,
as no new channels are added in this step. 
We also observe a reduction of the impact of theoretical uncertainties. 
Two-dimensional contours of the expected constraints are shown in Fig.~\ref{excl:2d_3000}
for the scenario with $3000~\text{fb}^{-1}$.
The fits using signal strength measurements only (gray) reveal a series 
of flat directions which cannot be amended by a different operator choice.
Several flat directions are resolved
with the fit using information from the differential $p_{T,H}$ measurements.
While the improvement on the exact numerical constraints 
can be somewhat compromised by larger systematic uncertainties,
the general feature of lifting flat directions still remains~\cite{toapp}.
Even with $3000~\text{fb}^{-1}$ it is not possible to constrain 
$\bar c_{u3}$ and $\bar c_{g}$ or $\bar c_{HW}$ and $\bar c_{HB}$
simultaneously using signal strength modifiers only. 
Using information from the differential $p_{T,H}$ measurements, 
which are are systematically under sufficient control,
effectively allows to constrain all coefficients simultaneously.  
Elements of studying differential
distributions to effective Higgs dimension six framework have been
investigated with similar findings in the
literature~\cite{Corbett:2015ksa,Ellis:2014jta,Banerjee:2015bla}, but, to our
knowledge, Figs.~\ref{excl:diff} and \ref{excl:2d_3000} provide the first consistent fit of
all single-Higgs relevant operators in a fully differential fashion,
in particular with extrapolations to 14 TeV.
The numerical values of the 95\% CL intervals for the different scenarios are
given in Tab.~\ref{tab:results}. 

%%%%%%%%%%%%%%%%%%%%%%%%%%%%%%%%%%%%%%
\begin{table*}[tb]
\begin{tabular}{l | C{3.3cm} | C{3.3cm} | C{3.3cm} | C{3.5cm}}
& \multicolumn{2}{c|}{ signal strengths only}  &   \multicolumn{2}{c}{ with differential $p_{T,H}$ measurements}  \\[0.2cm]
\hline
& LHC-300 & LHC-3000 & LHC-300 & LHC-3000 \\
\hline
\hline
$\bar{c}_{g}$ $[\times 10^{4}]$ &  $[\, -0.53, \, 0.66 \, ]$ &  $[\, -0.49, \, 0.57 \, ]$ &  $[\, -0.19, \, 0.22 \, ]$ &  $[\, -0.06, \, 0.07 \, ]$ \\
$\bar{c}_{\gamma}$ $[\times 10^{4}]$ &  $[\, -3.9, \, 3.4 \, ]$ &  $[\, -2.9, \, 2.7 \, ]$ &  $[\, -2.5, \, 2.0 \, ]$ &  $[\, -1.6, \, 1.3 \, ]$ \\
$\bar{c}_{W}$  &  $[\, < -0.010, \,  > 0.010 \, ]$  &  $[\, < -0.010, \,  > 0.010 \, ]$  &  $[\, -0.008, \, 0.008 \, ]$ &  $[\, -0.004, \, 0.004 \, ]$ \\
$\bar{c}_{H}$  &  $[\, < -0.050, \,  > 0.050 \, ]$  &  $[\, < -0.050, \,  > 0.050 \, ]$  &  $[\, < -0.050, \,  > 0.050 \, ]$  &  $[\, -0.044, \, 0.035 \, ]$ \\
$\bar{c}_{HW}$  &  $[\, -0.030, \, 0.032 \, ]$ &  $[\, -0.027, \, 0.028 \, ]$ &  $[\, -0.007, \, 0.010 \, ]$ &  $[\, -0.004, \, 0.004 \, ]$ \\
$\bar{c}_{HB}$  &  $[\, -0.030, \, 0.032 \, ]$ &  $[\, -0.026, \, 0.027 \, ]$ &  $[\, -0.008, \, 0.011 \, ]$ &  $[\, -0.004, \, 0.004 \, ]$ \\
$\bar{c}_{u3}$  &  $[\, < -0.050, \,  > 0.050 \, ]$  &  $[\, < -0.050, \,  > 0.050 \, ]$  &  $[\, < -0.050, \,  > 0.050 \, ]$  &  $[\, -0.020, \, 0.008 \, ]$ \\
$\bar{c}_{d3}$  &  $[\, < -0.050, \,  > 0.050 \, ]$  &  $[\, < -0.050, \,  > 0.050 \, ]$  &  $[\, < -0.050, \,  > 0.050 \, ]$  &  $[\, < -0.050, \,  > 0.050 \, ]$  \\
\hline
\end{tabular}
\caption{Predicted constraints at 95\% CL 
on dimension-six operator coefficients (first column) for the LHC with 14 TeV with 
an integrated luminosity of $300~\text{fb}^{-1}$ (LHC-300) and $3000~\text{fb}^{-1}$ (LHC-3000). 
In the second and third columns results are given using signal strength measurements only, 
in the last two columns results including differential $p_{T,H}$ measurements are shown.
In case no constraints can be derived within the 
parameter ranges considered in this work, the lower and upper limits 
are indicated to lie outside this range.
\label{tab:results}}
\end{table*}
%%%%%%%%%%%%%%%%%%%%%%%%%%%%%%%%%%%%%%

%%%%%%%%%%%%%%%%%%%%%%%%%%%%%%%%%%%%%%
%%%%%%%%%%%%%%%%%%%%%%%%%%%%%%%%%%%%%%
%
%%%%%%%%%%%%%%%%%%%%%%%%%%%%%%%%%%%%%%
%%%%%%%%%%%%%%%%%%%%%%%%%%%%%%%%%%%%%%

%%%%%%%%%%%%%%%%%%%%%%%%%%%%%%%%%%%%%%
%
%%%%%%%%%%%%%%%%%%%%%%%%%%%%%%%%%%%%%%
%%%%%%%%%%%%%%%%%%%%%%%%%%%%%%%%%%%%%%

A series of dimension six operators, on which no constraints can be
formulated at this stage of the LHC programme or by only including
signal strength measurements, can eventually be constrained with
enough data and differential distributions. The reason behind this is
that differential measurements {\it{ipso facto}} increase the number of
(correlated) measurements by number of bins, leading to a highly
over-constrained system. Also, since the impact of many operators is
most significant in the tails of energy-dependent distribution, the
relative statistical pull is decreased by only considering inclusive
quantities.

\section{Interpretation of constraints}
\label{sec:interpretation}

The whole purpose of interpreting data in terms of an effective field theory is to use this framework as a means of communication between a low-scale measurement at the LHC and a UV model defined at a high scale, out of reach of the LHC. This way, the EFT framework allows us to limit a large class of UV models. 

For a well-defined interpretation using effective operators, we assume that the operators, induced by the UV theory, only directly depend on the SM particle and symmetry content, and we also need to assume that the UV theory is weakly coupled to the SM sector. The last condition is necessary to justify the truncation of the effective Lagrangian at dimension six. After establishing limits on Wilson coefficients of the effective theory, as performed in Secs.~\ref{sec:results8}-\ref{sec:results14}, we can now address the implications for a specific UV model. 

Two popular ways of addressing the Hierarchy problem are composite Higgs models and supersymmetric theories. Let us quickly investigate in how far these constraints are
  relevant once we match the EFT expansion to a concrete UV scenario. 

%%%%%%%%%%%%%%%%%%%%%%%%%%%%%%%%  
\begin{figure}[!t]
\includegraphics[width=0.45\textwidth]{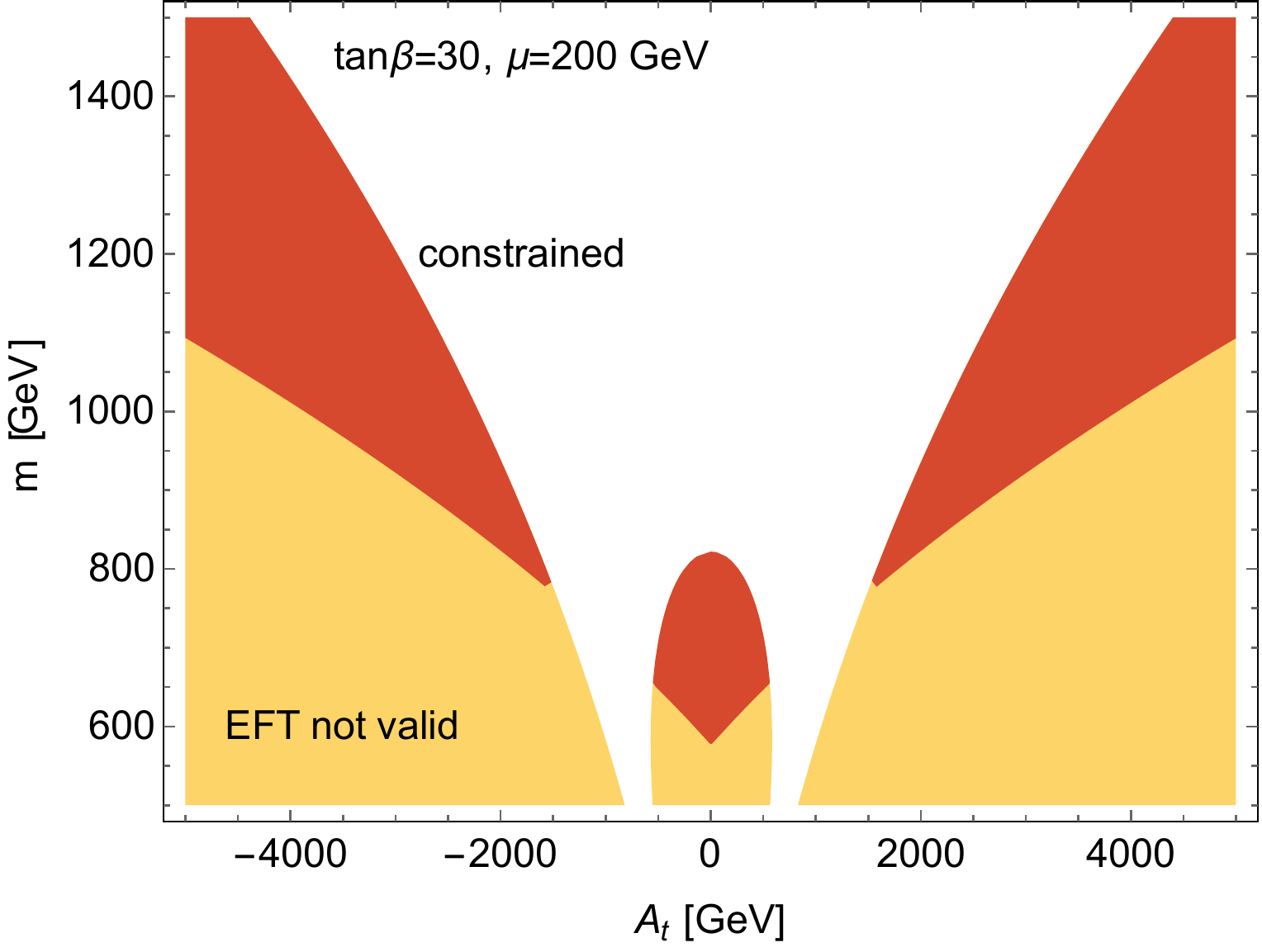}
\caption{\label{fig:mssmmatch} Matching the constraints on $|\bar c_g|\lesssim 5\times 10^{-6}$ of Fig.~\ref{excl:diff} onto stop contributions using Eq.~\eqref{eq:stops} for identified soft masses $m_{\tilde Q}=m_{\tilde{t}}=m$. The EFT approach is valid only for parameter combinations (red area) that result in stop masses that do not directly impact the Higgs transverse momentum distribution directly (e.g. through thresholds). For further details see text.}
\end{figure}  
%%%%%%%%%%%%%%%%%%%%%%%%%%%%%%%%

In the strongly-interacting Higgs case, from the power-counting arguments of Ref.~\cite{Giudice:2007fh,Coleman:1969sm,Callan:1969sn}, one typically expects
\begin{equation}
\overline c_g \sim \frac{ m_W^2 } {16\pi^2 f^2 } \frac{y_t^2}{g_\rho^2}\,,
\end{equation}
where $g_\rho \lesssim 4\pi$ and the compositeness scale is set by $\Lambda \sim g_\rho f$. So our predicted constraint on $\bar c_g$ including information from the differential Higgs $p_T$ distribution translates 
into $\Lambda \gtrsim 2.8 $ TeV, which falls outside the effective kinematic coverage of the Higgs phenomenology at the LHC. This means that new composite physics with a fundamental scale $\Lambda \gtrsim 2.8$ TeV can naively not be probed in the Higgs sector alone. However, new contributions, such as narrow resonances around this mass can be discovered in different channels such as weak-boson fusion \cite{Englert:2015oga} or Drell-Yan production \cite{Pappadopulo:2014qza}.
  
%%%%%%%%%%%%%%%%%%%%%%%%%%%%%%%%
\begin{figure*}[p]
\includegraphics[width=0.35\textwidth]{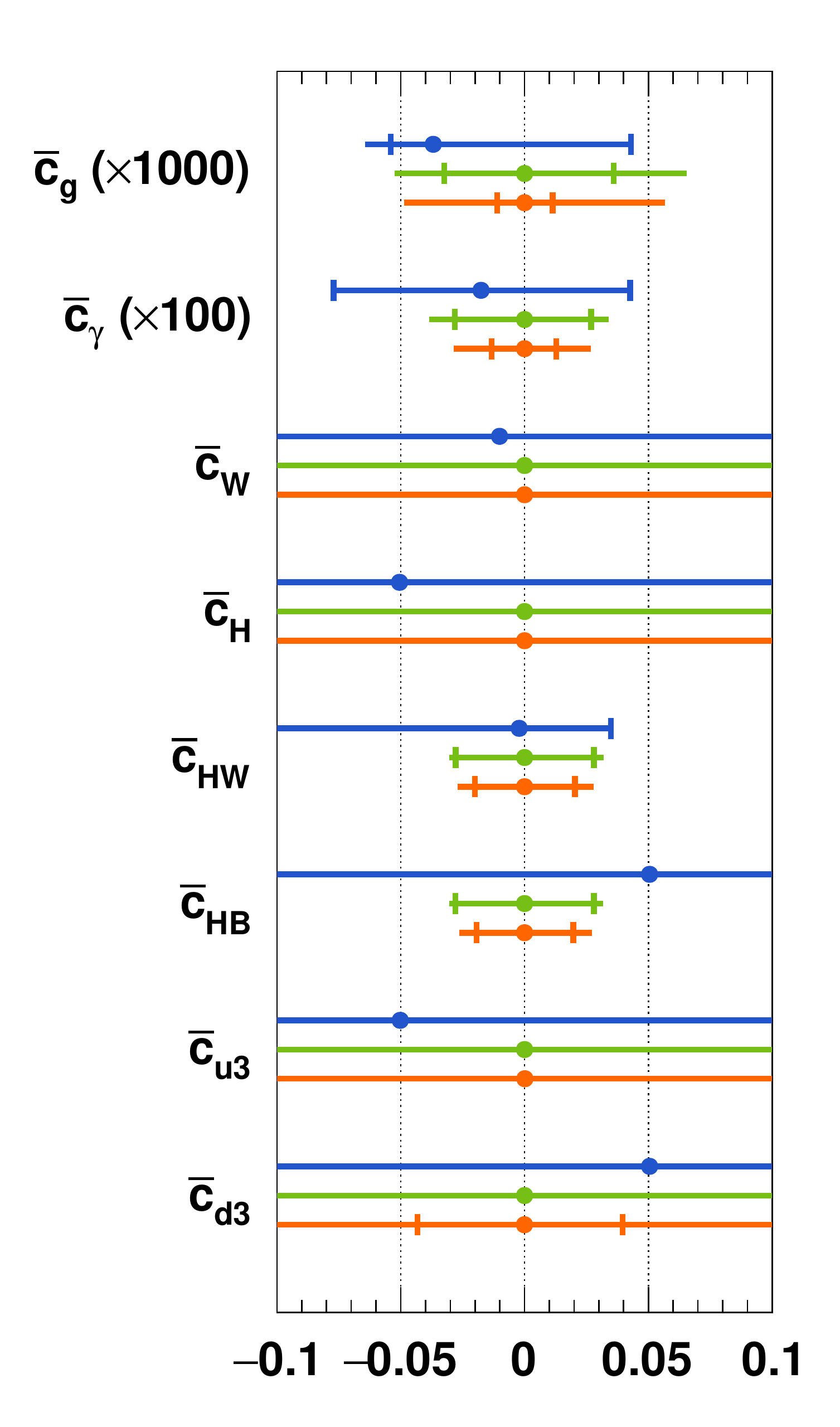}
\includegraphics[width=0.35\textwidth]{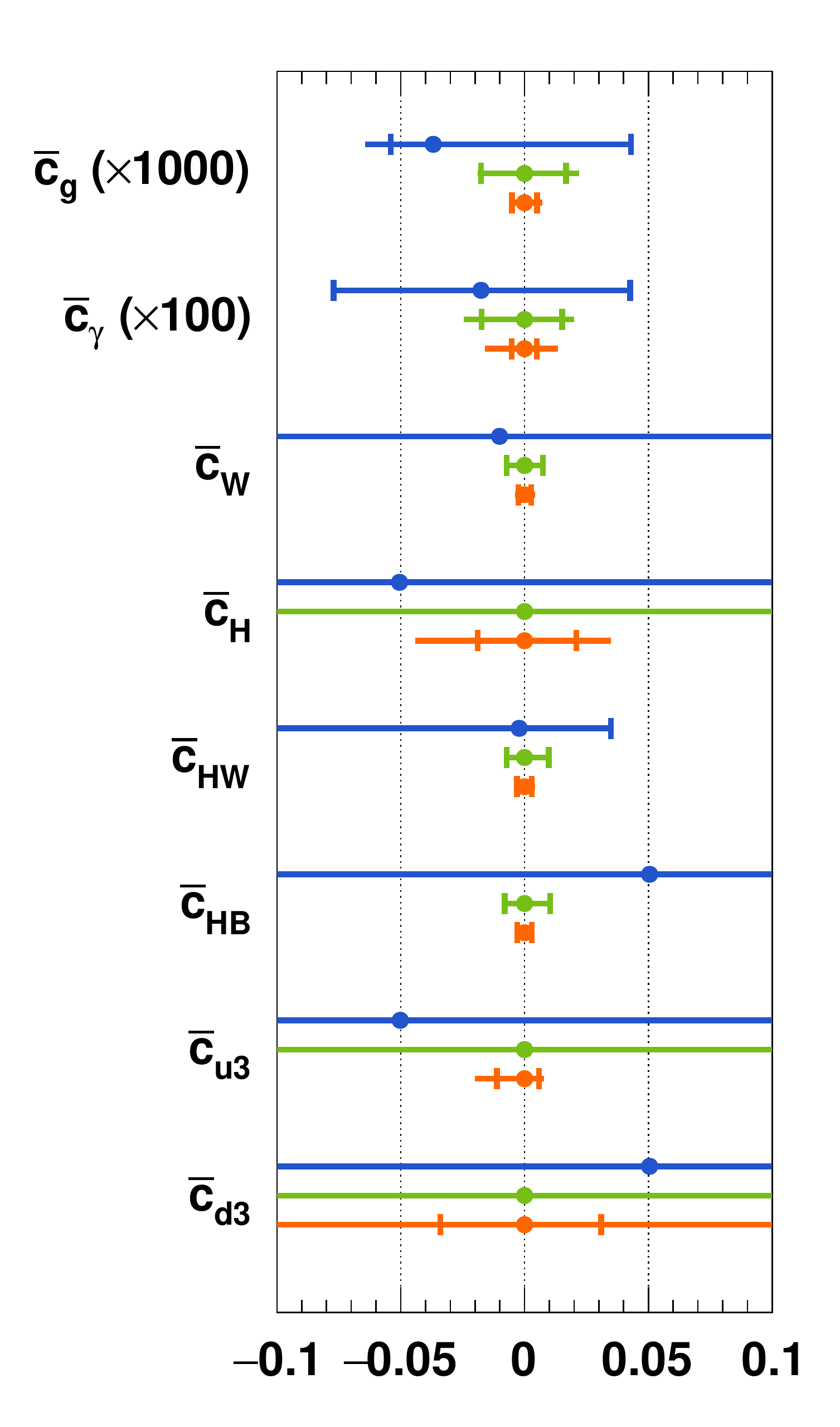}
\caption{\label{fig:summary_hpt} Marginalised 95\% CL constraints for 
the dimension-six operator coefficients for current data (blue), 
the LHC at 14 TeV with an integrated luminosity of $300~\text{fb}^{-1}$ (green) and 
$3000~\text{fb}^{-1}$ (orange). 
The expected constraints are centred around zero by construction, 
since the pseudo-data are generated by using the SM hypothesis. 
The left panel shows the constraints obtained using signal strength measurements only, 
and on the right differential $p_{T,H}$ measurements are included.
The inner error bar depicts the experimental uncertainty, the outer error bar shows the total uncertainty, 
given the assumptions detailed in the text.}
\end{figure*}
%%%%%%%%%%%%%%%%%%%%%%%%%%%%%%%%
  
Matching, say, the MSSM stop contribution on the $\bar c_g$ operator, we have (see e.g. \cite{Grojean:2013nya,Huo:2015nka,Drozd:2015kva} for a more detailed discussion)
\begin{multline}
\label{eq:stops}
\overline c_g = {m_W^2 \over (4\pi)^2} {1\over 24}  \bigg (
{h_t^2 - g_1^2c_{2\beta}/6 \over m_{\tilde Q}^2 } + {h_t^2 + g_1^2 c_{2\beta}/3 \over m_{\tilde{t}_R}^2}
\\ - {h_t^2 X_t^2\over m_{\tilde Q}^2 m_{\tilde{t}_R}^2}\bigg)\,,
\end{multline}
where $h_t \equiv y_t s_\beta$, $X_t \equiv A_t - \mu \cot \beta$ and $m_{\tilde Q}$ and $m_{\tilde{t}_R}$ denote the soft masses of the left and right-handed stops respectively. To ensure the validity of our EFT approach based on differential distributions, we have to make the strong assumption that all supersymmetric particles are heavier than the momentum transfer probed in all processes that are involved in of our fit~\cite{Englert:2014cva,Isidori:2013cga} (see also \cite{Englert:2014ffa,Brehmer:2015rna} for discussions of \hbox{(non-)}resonant signatures in BSM scenarios and EFT). For convenience, we additionally assume that all supersymmetric particles except the lightest stop $\tilde{t}_1$ are very heavy and decouple from $\overline c_g$. The largest value for $p_{T,H}$ we expect to probe during the LHC high-luminosity runs, based on our leading-order theory predictions is 500 GeV in the SM. And we can therefore trust the effective field theory approach for $m_{\tilde{t}_1} > 600$ GeV in our limit setting procedure that inputs SM pseudo-data. For instance, fixing the soft masses $m_{\tilde Q}=m_{\tilde{t}}=m$, $\mu=200~\text{GeV}$ and $\tan\beta=30$ we can understand the constraints on $\overline c_g$ as constraints in the $A_t-m$ plane, Fig.~\ref{fig:mssmmatch}. Similar interpretations are, of course, possible with the other Wilson coefficients.

%%%%%%%%%%%%%%%%%%%%%%%%%%%%%%%%
\section{Discussion, Conclusions and Outlook}
\label{sec:conc}
%%%%%%%%%%%%%%%%%%%%%%%%%%%%%%%%
Even though current measurements as performed by ATLAS and CMS show
good agreement with the SM hypothesis for the small statistics
collected during LHC Run 1, the recently discovered Higgs boson
remains one of the best candidates that could be a harbinger of physics
beyond the SM. If new physics is heavy enough, modifications to the
Higgs boson's phenomenology from integrating out heavy states can
be expressed using effective field theory methods. 

In this paper we
have constructed a scalable fitting framework, based on adapted
versions of {\sc{Gfitter}},~{\sc{Professor}},~{\sc{Vbfnlo}}, and
{\sc{eHdecay}} and have used an abundant list of available single-Higgs LHC
measurements to constrain new physics in the Higgs sector for the
results of Run 1.  In these fits we have adopted the leading order
strongly-interacting light Higgs basis assuming vanishing tree-level
$T$ and $S$ parameters and flavour universality of the new physics
sector. Our results represent the latest incarnation of fits at 8
TeV, and update results from the existing literature. 
The main goal of this work, however, is to provide an
estimate of how these constraints will improve when turning to high
energy collisions at the LHC with large statistics in light of
expected systematics. In this sense our work represents a first step
towards an ultimate Higgs sector fit, which is not limited to
inclusive measurements, but uses highly sensitive differential
distributions throughout.

Using extrapolations to 14~TeV, we find a major improvement of the
expected constraints, in particular when differential information is
included to the limit setting procedure. A summary of the 
current and expected constraints is given in Fig.~\ref{fig:summary_hpt}; these are of immediate
relevance for the expected sensitivity of the Higgs sector to concrete UV physics in the limit of large scale separations and unresolved new physics
at the LHC. 

It is interesting to see that including differential information at
the LHC, we can expect the limits on certain operators to become
competitive with measurements at a future
FCC-ee~\cite{Ellis:2015sca,Craig:2014una}. This is not entirely
unexpected since the high $p_{T,H}$ cross sections, especially for
hadronic channels, are sensitive probes of BSM physics. A major
limiting factor, however, are the involved theoretical uncertainties,
especially when moving to differential distributions at large
statistics. Obviously, electroweak precision constraints provide a
complementary avenue to constrain the presence of higher dimensional
operators~\cite{Efrati:2015eaa,Ellis:2014jta,Berthier:2015oma,Falkowski:2015jaa}
and are guaranteed to improve the sensitivity. We reserve a
dedicated discussion for the future.

%%%%%%%%%%%%%%%%%%%%%%%%%%%%%%%%
%%%%%%%%%%%%%%%%%%%%%%%%%%%%%%%%
\acknowledgments 
We thank A.~Falkowski, M.~Gonzalez-Alonso, A.~Greljo, D.~Marzocca,
T.~Plehn, M.~Trott and T.~You for discussions, and T.~Stefaniak for help
with HiggsSignals. This research was supported in
part by the European Commission through the ``HiggsTools'' Initial
Training Network PITN-GA-2012-316704 and
by the German Research Foundation (DFG) in the Collaborative 
Research Centre (SFB) 676 ``Particles, Strings and the Early Universe'' 
located in Hamburg.
%%%%%%%%%%%%%%%%%%%%%%%%%%%%%%%%
%%%%%%%%%%%%%%%%%%%%%%%%%%%%%%%%

\appendix

\section{Interpolation with {\sc{Professor}}} 
\label{sec:prof} 
Running a Monte-Carlo event generator and a subsequent analysis tool
to fill a bin of a histogram can be thought of as a CPU-expensive
evaluation of a function, $f_\text{MC}(\vec{p})$, at a certain point
in a $P$-dimensional parameter space, $\vec{p}$. The
{\sc{Professor}} method~\cite{Buckley:2009bj} is an approach that reduces the
time to evaluate $f_\text{MC}$ dramatically using $P$-dimensional
polynomial parametrisations.

The key idea is to treat each bin of a histogram as an independent
function of the parameter space as iterated above. The
parametrisations ${f_\text{MC}(\vec{p})}$ all together provide a fast
pseudo generator that yields an approximate response in milliseconds
rather than hours. Further, due to the usage of polynomials, the
response function is steady. These properties make
${f_\text{MC}(\vec{p})}$ suitable for numeric applications.

So far it has been applied with great emphasis and success to the
problem of Monte-Carlo generator tuning --- essentially a numerical
minimisation of a goodness-of-fit measure between real data and
${f_\text{MC}(\vec{p})}$.  When facing the problem of hypothesis
testing of a Monte-Carlo prediction as it is done in this work, the
same principle can be applied.  The difference being that the axes of
the parameter-space in this case are the theoretically well motivated
Wilson coefficients which are to be set limits for.
In its latest incarnation a C++ version of the core functionality of
{\sc{Professor}}, i.e. the parametrisation, has been added. It uses Eigen3 to
perform the SVD and calculate the approximate
${f_\text{MC}(\vec{p})}$.

%%%%%%%%%%%%%%%%%%%%%%%%%%%%%%%%%%%%%%%%%%%%
\begin{figure}[!t]
\includegraphics[width=0.48\textwidth]{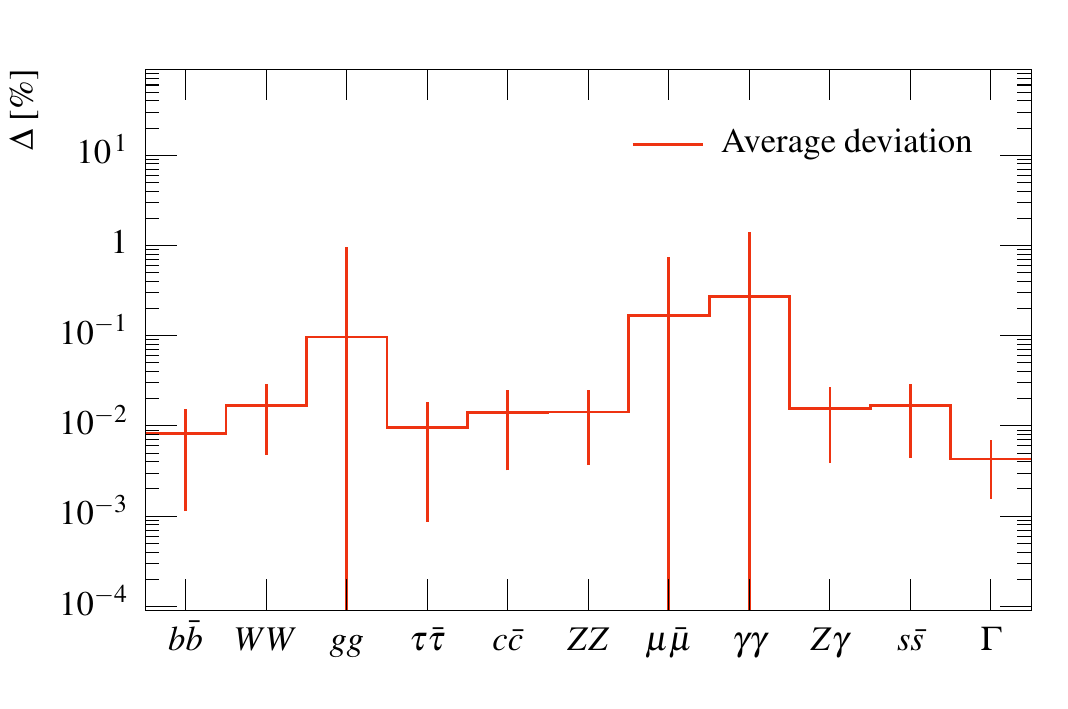}
\caption{\label{fig:profproof} Relative uncertainty of the interpolation for the Higgs branching ratios as well as the total width in the dimension six extension of the SM in percent, including uncertainties. Note that due to the dependence on the total width this interpolation is highly non-linear.}
\end{figure}
%%%%%%%%%%%%%%%%%%%%%%%%%%%%%%%%%%%%%%%%%%%%

The lowest oder polynomial to incorporate parameter correlations is of
second order.  For a certain bin, $b$, at a point $\vec{p}$ in
parameter space, this can be written as:
\begin{multline}
  \label{eq:poly}
  \text{MC}_b(\vec{p}) 
  \approx f^{(b)}(\vec{p})
  \\= \alpha^{(b)}_0 + \sum_i \beta^{(b)}_i \, p_i 
  + \sum_{i \le j} \gamma^{(b)}_{ij} \, p_i \, p_j
\end{multline}
with the to be determined coefficients
${\alpha_0,\beta_i,\gamma_{ij}}$.  The {\sc{Professor}} approach to
determine the latter is to construct an over-constrained system of
equations using the ensemble of bin contents $v^{(b)}_a,\:a\in[1,N]$
obtained when running the MC generator with the parameter settings
$\vec{p}_a = (x_a, y_a),\:a\in[1,N]$ (``anchors'' of the
parametrisation).

With the system being overconstrained, the matrix $\tilde{P}$ can be
(pseudo-) inverted using the functionality of eigen3. With the
pseudoinverse, ${\tilde{P}}^{-1}$, at hand, the coefficients,
$\vec{c}^{(b)}$ can be solved for easily by calculating
${\tilde{P}}^{-1}\cdot \vec{v}_a$ and the approximation
$f^{(b)}(\vec{p})$ can be calculated according to Eq.~\ref{eq:poly}. The fast pseudo-generator is then simply a collection
of coefficients $\vec{c}^{(b)}$ for all bins, $b$, of interest.
\newcommand{\columnfill}{%
  \begin{pmatrix}
    \alpha_0 \\ \beta_x \\ \beta_y \\ \gamma_{xx} \\ \gamma_{xy} \\ \gamma_{yy}
  \end{pmatrix}
}
\begin{align}
  \underbrace{
    \vphantom{\columnfill}
    \begin{pmatrix}
      v_1 \\ v_2 \\ \vdots \\ v_N
    \end{pmatrix}
}_{v^{(b)}_a}
  =
  \underbrace{
    \vphantom{\columnfill}
    \begin{pmatrix}
      1 & x_1 & y_1 & x^2_1 & x_1y_1 & y^2_1 \\
      1 & x_2 & y_2 & x^2_2 & x_2y_2 & y^2_2 \\
      &     &     &  \vdots   &   &          \\
      1 & x_N & y_N & x^2_N & x_Ny_N & y^2_N
    \end{pmatrix}
}_{\tilde{P}}
  \underbrace{
    \begin{pmatrix}
      \alpha_0 \\ \beta_x \\ \beta_y \\ \gamma_{xx} \\ \gamma_{xy} \\ \gamma_{yy}
    \end{pmatrix}
}_{\vec{c}^{(b)}}
\end{align}
To show the accuracy with which this procedure works, we show the interpolation of the Higgs branching ratios in 
Fig.~\ref{fig:profproof}, which due to its non-linear character is the most complicated interpolation involved in this work. We reproduce the branching ratios at the per mille level.

%%%%%%%%%%%%%%%%%%%%%%%%%%%%%%%%%%%%%%
\section{LHC Run 1 measurements} 
\label{sec:run1meas}

The signal strength measurements used in the Run 1 analysis (Sec.~\ref{sec:results8}) are listed in 
tables~\ref{tab:ATLAS_run1} and \ref{tab:CMS_run1}. 
The values of $\mu$ are given with their total uncertainties, where 
statistical, systematic and theoretical uncertainties have been added in quadrature. 
Signal acceptances are given for the production channels gluon fusion (ggH), 
vector boson fusion (VBF), $WH$, $ZH$ and $t\bar{t}H$ production.

%%%%%%%%%%%%%%%%%%%%%%%%%%%%%%%%%%%%%%
\begin{table*}[t]
\footnotesize
\centering
\begin{tabular}{l | C{1.8cm} | C{1.8cm} | C{1cm}  C{1cm}  C{1cm}  C{1cm}  C{1cm}}
Search channel  & energy $\sqrt{s}$ & $\mu$  &  \multicolumn{5}{c}{SM signal composition [in \%]} \\
 & & & ggH & VBF & $WH$ & $ZH$ & $t\bar{t}H$ \\
\hline
\hline
ATLAS $ pp  \to  H  \to  \gamma  \gamma  $ (central high $p_{T}$)~\cite{Aad:2014eha}    &   8 TeV   &   $1.62^{+1.00}_{-0.83}$    &    7.1    &   25.4    &   20.1    &   21.0    &   26.4    \\[0.1cm]
ATLAS $ pp  \to  H  \to  \gamma  \gamma  $ (central low $p_{T}$)~\cite{Aad:2014eha}   &   8 TeV   &   $0.62^{+0.42}_{-0.40}$    &   31.8    &   22.2    &   18.5    &   19.9    &    7.7    \\[0.1cm]
ATLAS $ pp  \to  H  \to  \gamma  \gamma  $ (forward high $p_{T}$)~\cite{Aad:2014eha}    &   8 TeV   &   $1.73^{+1.34}_{-1.18}$    &    7.1    &   26.2    &   23.1    &   23.6    &   20.1    \\[0.1cm]
ATLAS $ pp  \to  H  \to  \gamma  \gamma  $ (forward low $p_{T}$)~\cite{Aad:2014eha}   &   8 TeV   &   $2.03^{+0.57}_{-0.53}$    &   29.0    &   20.9    &   21.2    &   21.9    &    7.1    \\[0.1cm]
ATLAS $ pp  \to  H  \to  \gamma  \gamma  $ ($t\bar{t}$H hadronic)~\cite{Aad:2014eha}    &   8 TeV   &   $-0.84^{+3.23}_{-1.25}$   &    0.1    &    0.1    &    0.2    &    0.4    &   99.1    \\[0.1cm]
ATLAS $ pp  \to  H  \to  \gamma  \gamma  $ ($t\bar{t}$H leptonic)~\cite{Aad:2014eha}    &   8 TeV   &   $2.42^{+3.21}_{-2.07}$    &    0.0    &    0.0    &    2.9    &    1.4    &   95.6    \\[0.1cm]
ATLAS $ pp  \to  H  \to  \gamma  \gamma  $ (VBF loose)~\cite{Aad:2014eha}   &   8 TeV   &   $1.33^{+0.92}_{-0.77}$    &    3.7    &   90.5    &    1.9    &    1.7    &    2.2    \\[0.1cm]
ATLAS $ pp  \to  H  \to  \gamma  \gamma  $ (VBF tight)~\cite{Aad:2014eha}   &   8 TeV   &   $0.68^{+0.67}_{-0.51}$    &    1.4    &   96.3    &    0.3    &    0.4    &    1.7    \\[0.1cm]
ATLAS $ pp  \to  H  \to  \gamma  \gamma  $ ($VH$ dijet)~\cite{Aad:2014eha}    &   8 TeV   &   $0.23^{+1.67}_{-1.39}$    &    1.9    &    2.2    &   46.0    &   49.3    &    0.5    \\[0.1cm]
ATLAS $ pp  \to  H  \to  \gamma  \gamma  $ ($VH$ $E_T^{\mathrm{miss}}$)~\cite{Aad:2014eha}    &   8 TeV   &   $3.51^{+3.30}_{-2.42}$    &    0.2    &    1.1    &   22.0    &   47.6    &   29.2    \\[0.1cm]
ATLAS $ pp  \to  H  \to  \gamma  \gamma  $ ($VH$ $1\ell$)~\cite{Aad:2014eha}    &   8 TeV   &   $0.41^{+1.43}_{-1.06}$    &    0.0    &    0.1    &   80.4    &    8.9    &   10.6    \\[0.1cm]
ATLAS $ pp  \to  H  \to  \tau  \tau  $ (boosted, $\tau_{\mathrm{had}}\tau_{\mathrm{had}}$)~\cite{Aad:2015vsa}    &   7/8 TeV   &   $3.60^{+2.00}_{-1.60}$    &    6.9    &   21.1    &   38.1    &   33.9    &    0.0    \\[0.1cm]
ATLAS $ pp  \to  H  \to  \tau  \tau  $ (VBF, $\tau_{\mathrm{had}}\tau_{\mathrm{had}}$)~\cite{Aad:2015vsa}   &   7/8 TeV   &   $1.40^{+0.90}_{-0.70}$    &    2.6    &   97.4    &    0.0    &    0.0    &    0.0    \\[0.1cm]
ATLAS $ pp  \to  H  \to  \tau  \tau  $ (boosted, $\tau_{\mathrm{lep}}\tau_{\mathrm{had}}$)~\cite{Aad:2015vsa}    &   7/8 TeV   &   $0.90^{+1.00}_{-0.90}$    &    8.5    &   24.6    &   35.6    &   31.4    &    0.0    \\[0.1cm]
ATLAS $ pp  \to  H  \to  \tau  \tau  $ (VBF, $\tau_{\mathrm{lep}}\tau_{\mathrm{had}}$)~\cite{Aad:2015vsa}   &   7/8 TeV   &   $1.00^{+0.60}_{-0.50}$    &    1.3    &   98.7    &    0.0    &    0.0    &    0.0    \\[0.1cm]
ATLAS $ pp  \to  H  \to  \tau  \tau  $ (boosted, $\tau_{\mathrm{lep}}\tau_{\mathrm{lep}}$)~\cite{Aad:2015vsa}   &   7/8 TeV   &   $3.00^{+1.90}_{-1.70}$    &    9.8    &   47.1    &   26.5    &   16.7    &    0.0    \\[0.1cm]
ATLAS $ pp  \to  H  \to  \tau  \tau  $ (VBF, $\tau_{\mathrm{lep}}\tau_{\mathrm{lep}}$)~\cite{Aad:2015vsa}   &   7/8 TeV   &   $1.80^{+1.10}_{-0.90}$    &    1.1    &   98.9    &    0.0    &    0.0    &    0.0    \\[0.1cm]
ATLAS $ pp  \to  H  \to WW \to  \ell  \nu  \ell  \nu  $ (ggH enhanced)~\cite{ATLAS:2014aga, Aad:2015ona}   &   7/8 TeV   &   $1.01^{+0.27}_{-0.25}$    &   55.6    &   11.1    &   11.1    &   11.1    &   11.1    \\[0.1cm]
ATLAS $ pp  \to  H  \to WW \to  \ell  \nu  \ell  \nu  $ (VBF enhanced)~\cite{ATLAS:2014aga, Aad:2015ona}   &   7/8 TeV   &   $1.27^{+0.53}_{-0.45}$    &    2.0    &   98.0    &    0.0    &    0.0    &    0.0    \\[0.1cm]
ATLAS $ pp  \to  H  \to ZZ \to 4 \ell  $ (ggH-like)~\cite{Aad:2014eva}    &   7/8 TeV   &   $1.66^{+0.51}_{-0.44}$    &   22.7    &   18.2    &   18.2    &   18.2    &   22.7    \\[0.1cm]
ATLAS $ pp  \to  H  \to ZZ \to 4 \ell $ (VBF/$VH$-like)~\cite{Aad:2014eva}    &   7/8 TeV   &   $0.26^{+1.64}_{-0.94}$    &    2.2    &   32.6    &   32.6    &   32.6    &    0.0    \\[0.1cm]
ATLAS $ pp  \to t\bar{t} H  \to \mathrm{leptons} $ ($1\ell 2\tau_{\mathrm{had}}$)~\cite{Aad:2015iha}    &   8 TeV   &   $-9.60^{+9.60}_{-9.70}$   &    0.0    &    0.0    &    0.0    &    0.0    &   100.0   \\[0.1cm]
ATLAS $ pp  \to t\bar{t} H  \to \mathrm{leptons} $ ($2\ell 0\tau_{\mathrm{had}}$)~\cite{Aad:2015iha}    &   8 TeV   &   $2.80^{+2.10}_{-1.90}$    &    0.0    &    0.0    &    0.0    &    0.0    &   100.0   \\[0.1cm]
ATLAS $ pp  \to t\bar{t} H  \to \mathrm{leptons} $ ($2\ell 1\tau_{\mathrm{had}}$)~\cite{Aad:2015iha}    &   8 TeV   &   $-0.90^{+3.10}_{-2.00}$   &    0.0    &    0.0    &    0.0    &    0.0    &   100.0   \\[0.1cm]
ATLAS $ pp  \to t\bar{t} H  \to \mathrm{leptons} $ ($3\ell$)~\cite{Aad:2015iha}   &   8 TeV   &   $2.80^{+2.20}_{-1.80}$    &    0.0    &    0.0    &    0.0    &    0.0    &   100.0   \\[0.1cm]
ATLAS $ pp  \to t\bar{t} H  \to \mathrm{leptons} $ ($4\ell$)~\cite{Aad:2015iha}   &   8 TeV   &   $1.80^{+6.90}_{-6.90}$    &    0.0    &    0.0    &    0.0    &    0.0    &   100.0   \\[0.1cm]
ATLAS $ pp  \to t\bar{t} H  \to t\bar{t} b\bar{b}$~\cite{Aad:2015gra}    &   8 TeV   &   $1.50^{+1.10}_{-1.10}$    &   0.0    &   0.0    &   0.0    &   0.0    &   100.0    \\[0.1cm]
ATLAS $ pp  \to V H  \to Vb\bar{b} $ ($0\ell$)~\cite{Aad:2014xzb}   &   7/8 TeV   &   $-0.35^{+0.55}_{-0.52}$   &    0.0    &    0.0    &   13.2    &   86.8    &    0.0    \\[0.1cm]
ATLAS $ pp  \to V H  \to Vb\bar{b} $ ($1\ell$)~\cite{Aad:2014xzb}   &   7/8 TeV   &   $1.17^{+0.66}_{-0.60}$    &    0.0    &    0.0    &   94.4    &    5.6    &    0.0    \\[0.1cm]
ATLAS $ pp  \to V H  \to Vb\bar{b} $ ($2\ell$)~\cite{Aad:2014xzb}   &   7/8 TeV   &   $0.94^{+0.88}_{-0.79}$    &    0.0    &   0.0    &   0.0    &   100.0    &   0.0    \\[0.1cm]
ATLAS $ pp  \to V H  \to VWW $ ($2\ell$)~\cite{Aad:2015ona}   &   7/8 TeV   &   $3.70^{+1.90}_{-1.80}$    &    0.0    &    0.0    &   74.3    &   25.7    &    0.0    \\[0.1cm]
ATLAS $ pp  \to V H  \to VWW $ ($3\ell$)~\cite{Aad:2015ona}   &   7/8 TeV   &   $0.72^{+1.30}_{-1.10}$    &    0.0    &    0.0    &   78.8    &   21.2    &    0.0    \\[0.1cm]
ATLAS $ pp  \to V H  \to VWW $ ($4\ell$)~\cite{Aad:2015ona}   &   7/8 TeV   &   $4.90^{+4.60}_{-3.10}$    &   0.0    &   0.0    &   0.0    &   100.0    &   0.0    \\[0.1cm]
\hline
\end{tabular}
\caption{Signal strengths measurements $\mu$ from the ATLAS 
collaboration used in the Run 1 analysis. 
In the last five columns the signal compositions are given in terms of 
efficiencies for production channels assuming a SM Higgs boson.  
\label{tab:ATLAS_run1}}
\end{table*}
%%%%%%%%%%%%%%%%%%%%%%%%%%%%%%%%%%%%%%

%%%%%%%%%%%%%%%%%%%%%%%%%%%%%%%%%%%%%%
\begin{table*}[tp]
\footnotesize
\centering
\begin{tabular}{l | C{1.8cm} | C{1.8cm} | C{1cm}  C{1cm}  C{1cm}  C{1cm}  C{1cm}}
Search channel  & energy $\sqrt{s}$ & $\mu$  &  \multicolumn{5}{c}{SM signal composition [in \%]} \\
 & & & ggH & VBF & $WH$ & $ZH$ & $t\bar{t}H$ \\
\hline
\hline
CMS $ pp  \to  H  \to  \gamma   \gamma  $ ($t\bar{t}$H multijet)~\cite{Khachatryan:2014ira}   &   8 TeV   &   $1.24^{+4.23}_{-2.70}$    &    0.0    &    0.1    &    0.1    &    0.2    &   99.5    \\[0.1cm]
CMS $ pp  \to  H  \to  \gamma   \gamma  $ ($t\bar{t}$H lepton)~\cite{Khachatryan:2014ira}   &   8 TeV   &   $3.52^{+3.89}_{-2.45}$    &    0.0    &    0.0    &    0.3    &    0.5    &   99.2    \\[0.1cm]
CMS $ pp  \to  H  \to  \gamma   \gamma  $ ($t\bar{t}$H tags)~\cite{Khachatryan:2014ira}   &   7 TeV   &   $0.71^{+6.20}_{-3.56}$    &    0.0    &    0.1    &    0.4    &    0.4    &   99.2    \\[0.1cm]
CMS $ pp  \to  H  \to  \gamma   \gamma  $ (untagged 0)~\cite{Khachatryan:2014ira}   &   7 TeV   &   $1.97^{+1.51}_{-1.25}$    &   12.1    &   18.7    &   23.8    &   24.0    &   21.3    \\[0.1cm]
CMS $ pp  \to  H  \to  \gamma   \gamma  $ (untagged 0)~\cite{Khachatryan:2014ira}   &   8 TeV   &   $0.13^{+1.09}_{-0.74}$    &    6.7    &   16.7    &   20.5    &   18.4    &   37.7    \\[0.1cm]
CMS $ pp  \to  H  \to  \gamma   \gamma  $ (untagged 1)~\cite{Khachatryan:2014ira}   &   7 TeV   &   $1.23^{+0.98}_{-0.88}$    &   30.6    &   17.4    &   20.9    &   19.5    &   11.7    \\[0.1cm]
CMS $ pp  \to  H  \to  \gamma   \gamma  $ (untagged 1)~\cite{Khachatryan:2014ira}   &   8 TeV   &   $0.92^{+0.57}_{-0.49}$    &   13.7    &   20.3    &   21.7    &   22.4    &   21.8    \\[0.1cm]
CMS $ pp  \to  H  \to  \gamma   \gamma  $ (untagged 2)~\cite{Khachatryan:2014ira}   &   7 TeV   &   $1.60^{+1.25}_{-1.17}$    &   30.3    &   16.8    &   20.6    &   20.8    &   11.5    \\[0.1cm]
CMS $ pp  \to  H  \to  \gamma   \gamma  $ (untagged 2)~\cite{Khachatryan:2014ira}   &   8 TeV   &   $1.10^{+0.48}_{-0.44}$    &   22.9    &   18.8    &   21.1    &   20.3    &   16.9    \\[0.1cm]
CMS $ pp  \to  H  \to  \gamma   \gamma  $ (untagged 3)~\cite{Khachatryan:2014ira}   &   7 TeV   &   $2.61^{+1.74}_{-1.65}$    &   30.9    &   16.7    &   21.0    &   19.7    &   11.7    \\[0.1cm]
CMS $ pp  \to  H  \to  \gamma   \gamma  $ (untagged 3)~\cite{Khachatryan:2014ira}   &   8 TeV   &   $0.65^{+0.65}_{-0.89}$    &   23.4    &   17.9    &   20.6    &   20.7    &   17.3    \\[0.1cm]
CMS $ pp  \to  H  \to  \gamma   \gamma  $ (untagged 4)~\cite{Khachatryan:2014ira}   &   8 TeV   &   $1.46^{+1.29}_{-1.24}$    &   28.5    &   17.6    &   20.6    &   19.5    &   13.8    \\[0.1cm]
CMS $ pp  \to  H  \to  \gamma   \gamma  $ (VBF dijet 0)~\cite{Khachatryan:2014ira}    &   7 TeV   &   $4.85^{+2.17}_{-1.76}$    &    1.8    &   94.9    &    0.7    &    0.9    &    1.7    \\[0.1cm]
CMS $ pp  \to  H  \to  \gamma   \gamma  $ (VBF dijet 0)~\cite{Khachatryan:2014ira}    &   8 TeV   &   $0.82^{+0.75}_{-0.58}$    &    1.3    &   96.1    &    0.5    &    0.4    &    1.7    \\[0.1cm]
CMS $ pp  \to  H  \to  \gamma   \gamma  $ (VBF dijet 1)~\cite{Khachatryan:2014ira}    &   7 TeV   &   $2.60^{+2.16}_{-1.76}$    &    4.2    &   81.2    &    3.4    &    3.5    &    7.7    \\[0.1cm]
CMS $ pp  \to  H  \to  \gamma   \gamma  $ (VBF dijet 1)~\cite{Khachatryan:2014ira}    &   8 TeV   &   $-0.21^{+0.75}_{-0.69}$   &    2.3    &   91.4    &    1.6    &    0.9    &    3.7    \\[0.1cm]
CMS $ pp  \to  H  \to  \gamma   \gamma  $ (VBF dijet 2)~\cite{Khachatryan:2014ira}    &   8 TeV   &   $2.60^{+1.33}_{-0.99}$    &    3.8    &   72.8    &    4.0    &    4.0    &   15.4    \\[0.1cm]
CMS $ pp  \to  H  \to  \gamma   \gamma  $ ($VH$ dijet)~\cite{Khachatryan:2014ira}   &   7 TeV   &   $7.86^{+8.86}_{-6.40}$    &    1.0    &    1.3    &   42.8    &   41.1    &   13.8    \\[0.1cm]
CMS $ pp  \to  H  \to  \gamma   \gamma  $ ($VH$ dijet)~\cite{Khachatryan:2014ira}   &   8 TeV   &   $0.39^{+2.16}_{-1.48}$    &    0.9    &    1.5    &   40.3    &   40.1    &   17.3    \\[0.1cm]
CMS $ pp  \to  H  \to  \gamma   \gamma  $ ($VH$ $E_T^{\mathrm{miss}}$)~\cite{Khachatryan:2014ira}   &   7 TeV   &   $4.32^{+6.72}_{-4.15}$    &    0.1    &    0.3    &   23.8    &   44.2    &   31.6    \\[0.1cm]
CMS $ pp  \to  H  \to  \gamma   \gamma  $ ($VH$ $E_T^{\mathrm{miss}}$)~\cite{Khachatryan:2014ira}   &   8 TeV   &   $0.08^{+1.86}_{-1.28}$    &    0.3    &    0.7    &   20.1    &   35.6    &   43.3    \\[0.1cm]
CMS $ pp  \to  H  \to  \gamma   \gamma  $ ($VH$ loose)~\cite{Khachatryan:2014ira}   &   7 TeV   &   $3.10^{+8.29}_{-5.34}$    &    0.1    &    0.5    &   70.2    &   23.3    &    5.9    \\[0.1cm]
CMS $ pp  \to  H  \to  \gamma   \gamma  $ ($VH$ loose)~\cite{Khachatryan:2014ira}   &   8 TeV   &   $1.24^{+3.69}_{-2.62}$    &    0.1    &    0.4    &   66.3    &   24.7    &    8.5    \\[0.1cm]
CMS $ pp  \to  H  \to  \gamma   \gamma  $ ($VH$ tight)~\cite{Khachatryan:2014ira}   &   8 TeV   &   $-0.34^{+1.30}_{-0.63}$   &    0.0    &    0.1    &   57.2    &   24.4    &   18.4    \\[0.1cm]
CMS $ pp  \to  H  \to  \mu  \mu $ ~\cite{Khachatryan:2014aep}    &   7/8 TeV   &   $2.90^{+2.80}_{-2.70}$    &   20.0    &   20.0    &   20.0    &   20.0    &   20.0    \\[0.1cm]
CMS $ pp  \to  H  \to  \tau  \tau $ (0 jet)~\cite{Chatrchyan:2014nva}   &   7/8 TeV   &   $0.40^{+0.73}_{-1.13}$    &   70.2    &    8.8    &   10.5    &   10.5    &    0.0    \\[0.1cm]
CMS $ pp  \to  H  \to  \tau  \tau $ (1 jet)~\cite{Chatrchyan:2014nva}   &   7/8 TeV   &   $1.06^{+0.47}_{-0.47}$    &   12.8    &   31.0    &   28.1    &   28.1    &    0.0    \\[0.1cm]
CMS $ pp  \to  H  \to WW \to 2 \ell 2 \nu  $ (0/1 jet)~\cite{Chatrchyan:2013iaa}   &   7/8 TeV   &   $0.74^{+0.22}_{-0.20}$    &   19.0    &   31.3    &   24.9    &   24.9    &    0.0    \\[0.1cm]
CMS $ pp  \to  H  \to WW \to 2 \ell 2 \nu  $ (VBF)~\cite{Chatrchyan:2013iaa}   &   7/8 TeV   &   $0.60^{+0.57}_{-0.46}$    &    2.0    &   98.0    &    0.0    &    0.0    &    0.0    \\[0.1cm]
CMS $ pp  \to  H  \to ZZ \to 4 \ell  $ (0/1 jet)~\cite{Khachatryan:2014jba,Chatrchyan:2013mxa}    &   7/8 TeV   &   $0.88^{+0.34}_{-0.27}$    &   41.7    &   58.3    &    0.0    &    0.0    &    0.0    \\[0.1cm]
CMS $ pp  \to  H  \to ZZ \to 4 \ell  $ (2 jet)~\cite{Khachatryan:2014jba,Chatrchyan:2013mxa}    &   7/8 TeV   &   $1.55^{+0.95}_{-0.66}$    &   16.7    &   83.3    &    0.0    &    0.0    &    0.0    \\[0.1cm]
CMS $ pp  \to t\bar{t} H  \to 2  \ell  $ (same sign)~\cite{Khachatryan:2014qaa}   &   8 TeV   &   $5.30^{+2.10}_{-1.80}$    &    0.0    &    0.0    &    0.0    &    0.0    &   100.0   \\[0.1cm]
CMS $ pp  \to t\bar{t} H  \to 3  \ell $ ~\cite{Khachatryan:2014qaa}   &   8 TeV   &   $3.10^{+2.40}_{-2.00}$    &    0.0    &    0.0    &    0.0    &    0.0    &   100.0   \\[0.1cm]
CMS $ pp  \to t\bar{t} H  \to 4  \ell $ ~\cite{Khachatryan:2014qaa}   &   8 TeV   &   $-4.70^{+5.00}_{-1.30}$   &    0.0    &    0.0    &    0.0    &    0.0    &   100.0   \\[0.1cm]
CMS $ pp  \to t\bar{t} H  \to t\bar{t} b\bar{b}$~\cite{Khachatryan:2014qaa}   &   7/8 TeV   &   $0.70^{+1.90}_{-1.90}$    &   0.0    &   0.0    &   0.0    &   0.0    &   100.0    \\[0.1cm]
CMS $ pp  \to t\bar{t} H  \to t\bar{t} \gamma\gamma$~\cite{Khachatryan:2014qaa}   &   8 TeV   &   $2.70^{+2.60}_{-1.80}$    &   0.0    &   0.0    &   0.0    &   0.0    &   100.0    \\[0.1cm]
CMS $ pp  \to t\bar{t} H  \to t\bar{t} \tau\tau$~\cite{Khachatryan:2014qaa}   &   7/8 TeV   &   $-1.30^{+6.30}_{-5.50}$   &   0.0    &   0.0    &   0.0    &   0.0    &   100.0    \\[0.1cm]
CMS $ pp  \to  H  \to  \tau  \tau  $ (VBF)~\cite{Chatrchyan:2014nva}    &   7/8 TeV   &   $0.93^{+0.41}_{-0.41}$    &   14.3    &   85.7    &    0.0    &    0.0    &    0.0    \\[0.1cm]
CMS $ pp  \to W H  \to \ell\nu b \bar{b}$ ~\cite{Chatrchyan:2013zna}    &   7/8 TeV   &   $1.10^{+0.90}_{-0.90}$    &   0.0    &   0.0    &   100.0    &   0.0    &   0.0    \\[0.1cm]
CMS $ pp  \to Z H  \to 2\ell b\bar{b}$ ~\cite{Chatrchyan:2013zna}    &   7/8 TeV   &   $0.80^{+1.00}_{-1.00}$    &   0.0    &   0.0    &   0.0    &   100.0    &   0.0    \\[0.1cm]
CMS $ pp  \to Z H  \to \nu\nu b\bar{b}$ ~\cite{Chatrchyan:2013zna}    &   7/8 TeV   &   $1.00^{+0.80}_{-0.80}$    &   0.0    &   0.0    &   0.0    &   100.0    &   0.0    \\[0.1cm]
CMS $ pp  \to V H  \to  \tau  \tau $ ~\cite{Chatrchyan:2014nva}   &   7/8 TeV   &   $0.98^{+1.68}_{-1.50}$    &    0.0    &    0.0    &   50.0    &   50.0    &    0.0    \\[0.1cm]
CMS $ pp  \to V H  \to WW \to 2 \ell 2 \nu  $~\cite{Chatrchyan:2013iaa}    &   7/8 TeV   &   $0.39^{+1.97}_{-1.87}$    &    3.6    &    3.6    &   46.4    &   46.4    &    0.0    \\[0.1cm]
CMS $ pp  \to V H  \to VWW $ (hadronic V)~\cite{CMS-PAS-HIG-13-017}   &   7/8 TeV   &   $1.00^{+2.00}_{-2.00}$    &    4.2    &    3.5    &   49.1    &   43.2    &    0.0    \\[0.1cm]
CMS $ pp  \to  W H  \to WW \to 3 \ell 3 \nu  $~\cite{Chatrchyan:2013iaa}    &   7/8 TeV   &   $0.56^{+1.27}_{-0.95}$    &   0.0    &   0.0    &   100.0    &   0.0    &   0.0    \\[0.1cm]
\hline
\end{tabular}
\caption{Signal strengths measurements $\mu$ from the CMS 
collaboration used in the Run 1 analysis. 
In the last five columns the signal compositions are given in terms of 
efficiencies for production channels assuming a SM Higgs boson.  
\label{tab:CMS_run1}}
\end{table*}
%%%%%%%%%%%%%%%%%%%%%%%%%%%%%%%%%%%%%%

%%%%%%%%%%%%%%%%%%%%%%%%%%%%%%%%
\section{Pseudo-data used for 14 TeV Studies} 
\label{sec:pseudodata}

\begin{figure*}[p]
\begin{center}
\begin{tabular}{ l p{0.262\textwidth} p{0.22\textwidth} p{0.22\textwidth} p{0.22\textwidth} }
& \hspace{2.5cm}$\mathbf{H\to\gamma\gamma}$  &  
\hspace{1.0cm}$\mathbf{H\to Z Z^{*} \to 4\ell}$  &
\hspace{0.4cm}$\mathbf{H\to W W^{*} \to 2\ell 2\nu}$  &  
\hspace{0.7cm} $\mathbf{H\to \tau^+ \tau^-}$ \\
& \multicolumn{4}{c}{\multirow{8}{*}{\includegraphics[trim=1.7cm 10.0cm 2.2cm 1.6cm, clip, width=0.9\textwidth]{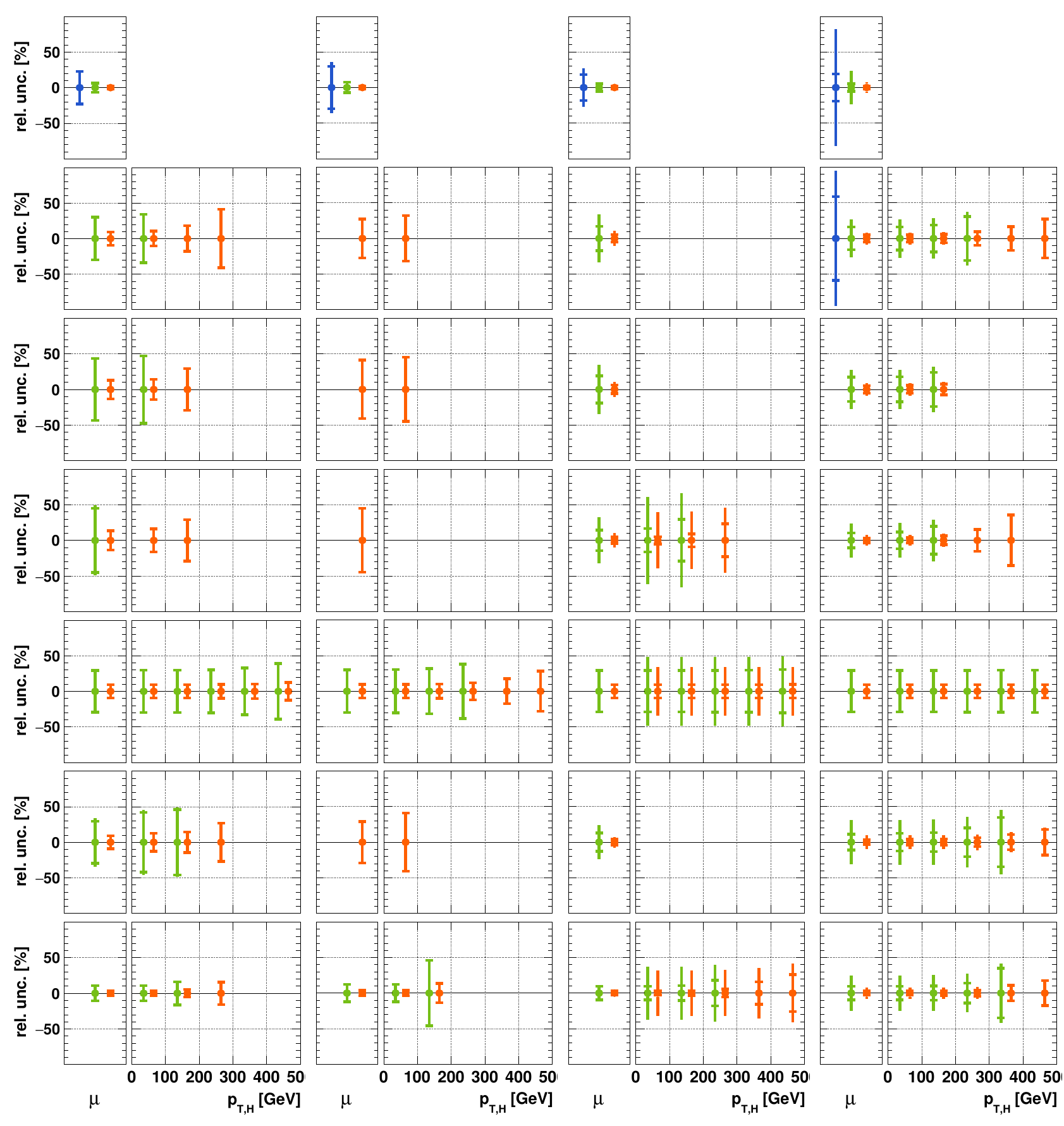}}} \\[0.9cm]
\bf{GF} \\[1.9cm] 
$\mathbf{W^-H}$ \\[1.9cm]
$\mathbf{W^+H}$ \\[1.9cm]
$\mathbf{ZH}$ \\[1.9cm]
\bf{VBF} \\[1.9cm]
$\mathbf{t\bar{t}H}$ \\[1.9cm]
$\mathbf{H+\text{j}}$\\[1.4cm]
\end{tabular}
\end{center}
\caption{\label{fig:pseudodata1} Pseudo-data for the production modes 
 gluon fusion $pp\to H$ (GF), $W^-H$, $W^+H$, $ZH$ 
 VBF, $t\bar{t}H$ and $pp\to H+\text{j}$ (from top to bottom)
 and for the decay channels $H\to\gamma\gamma$, $H\to ZZ^* \to 4l$, $H\to WW^* \to 2l2\nu$ and 
 $H\to \tau^+\tau^-$ (from left to right).
 The pseudo-data for 8~TeV with $L=25~\text{fb}^{-1}$ are shown in blue, 
 the 14~TeV scenarios with $L=300~\text{fb}^{-1}$ and $L=3000~\text{fb}^{-1}$ are shown in green and orange. 
 The inner error bar illustrates the statistical uncertainty and the outer error bar
 shows the total uncertainty. 
 }
\end{figure*}

\begin{figure*}[p]
\begin{center}
\begin{tabular}{ l p{0.262\textwidth} p{0.22\textwidth} p{0.22\textwidth} }
& \hspace{2.3cm}$\mathbf{H\to b\bar{b}}$  &  
\hspace{1.2cm}$\mathbf{H\to \mu^+\mu^-}$  &
\hspace{1.2cm}$\mathbf{H \to Z \gamma}$  \\
& %\multicolumn{3}{c}{\multirow{8}{*}{\includegraphics[trim=1.7cm 10.0cm 2.2cm 1.6cm, clip, width=0.9\textwidth]{material/plots/meas2.pdf}}} \\[0.9cm]
\multicolumn{3}{c}{\multirow{8}{*}{\includegraphics[trim=1.7cm 10.0cm 6.3cm 1.6cm, clip, width=0.685\textwidth]{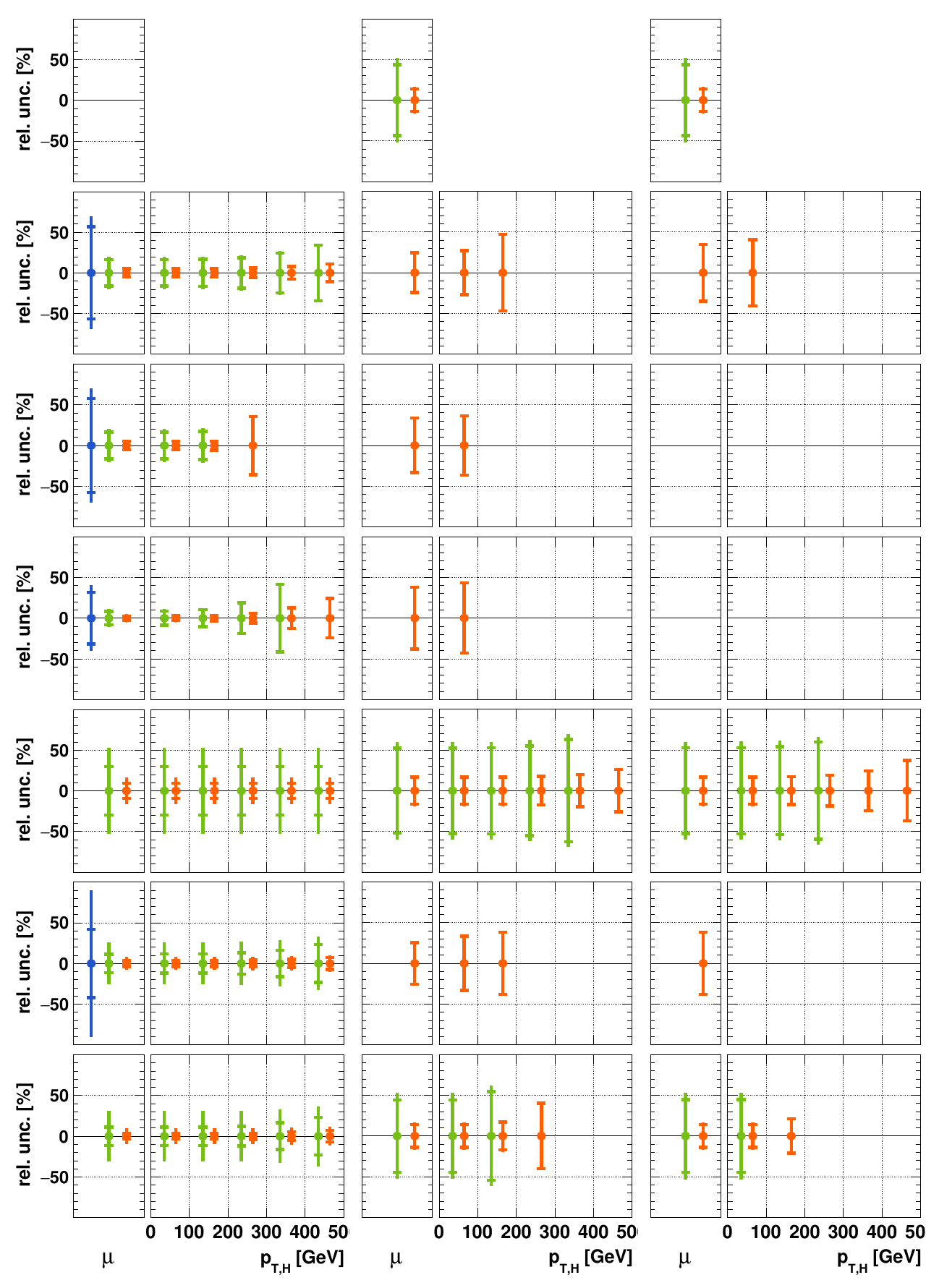}}} \\[0.9cm]
\bf{GF} \\[1.9cm] 
$\mathbf{W^-H}$ \\[1.9cm]
$\mathbf{W^+H}$ \\[1.9cm]
$\mathbf{ZH}$ \\[1.9cm]
\bf{VBF} \\[1.9cm]
$\mathbf{t\bar{t}H}$ \\[1.9cm]
$\mathbf{H+\text{j}}$\\[1.4cm]
\end{tabular}
\end{center}
\caption{\label{fig:pseudodata2} Pseudo-data for the production modes 
 gluon fusion ($pp\to H)$, $W^-H$, $W^+H$, VBF, $t\bar{t}H$ and $pp\to H+\text{j}$ (from top to bottom)
 and for the decay channels $H\to b\bar{b}$, $H\to \mu^+\mu^-$ and $H\to Z \gamma$ 
 (from left to right).
 The pseudo-data for 8~TeV with $L=25~\text{fb}^{-1}$ are shown in blue, 
 the 14~TeV scenarios with $L=300~\text{fb}^{-1}$ and $L=3000~\text{fb}^{-1}$ are shown in green and orange.
 The inner error bar illustrates the statistical uncertainty and the outer error bar
 shows the total uncertainty. 
 }
\end{figure*}

A summary is given in Figures~\ref{fig:pseudodata1} 
and \ref{fig:pseudodata2} of the pseudo-data used for the evaluation 
of the LHC's sensitivity to the set of dimension six operators 
considered in this work. The pseudo-data is obtained using the 
assumptions detailed in Sec.~\ref{sec:analysis}. 
The pseudo-data for 8~TeV has been generated to validate 
the assumptions made.

%%%%%%%%%%%%%%%%%%%%%%%%%%%%%%%%%
%%%%%%%%%%%%%%%%%%%%%%%%%%%%%%%%%
%%%%%%%%%%%%%%%%%%%%%%%%%%%%%%%%%
%%%%%%%%%%%%%%%%%%%%%%%%%%%%%%%%%

%%%%%%%%%%%%%%%%%%%%%%%%%%%%%%%%

\clearpage 

%%%%%%%%%%%%%%%%%%%%%%%%%%%%%%%%
\bibliography{references}
%%%%%%%%%%%%%%%%%%%%%%%%%%%%%%%%

\end{document}